\documentclass[10pt]{article}

\usepackage{amsmath}
\usepackage{amssymb}

\usepackage{graphicx}

\usepackage{cite}

\usepackage{color} 

\usepackage[labelfont=bf,labelsep=period,justification=raggedright]{caption}

\makeatletter
\renewcommand{\@biblabel}[1]{\quad#1.}
\makeatother

\date{}

\begin{document}

\begin{flushleft}
{\Large
\textbf{Marker Genes for Anatomical Regions in the Brain: Insights from the Allen Gene Expression 
Atlas}
}
\\
Pascal Grange, 
Partha P. Mitra 
\\
\bf Cold Spring Harbor Laboratory, One Bungtown Road, Cold Spring Harbor, New York 11724, USA
\\
$\ast$ E-mail: pascal.grange@polytechnique.org
\end{flushleft}

\section*{Abstract}

Quantitative criteria are proposed to identify genes (and sets of genes) whose expression marks
a specific brain region (or a set of brain regions).
 Gene-expression energies, obtained for thousands of mouse genes by numerization of {\it{in situ}} 
      hybridization images in the Allen Gene Expression Atlas, are used to test these 
methods in the mouse brain.  Individual genes
      are ranked  using integrals of their
      expression energies across brain regions.  The
      ranking is generalized to sets of genes and the problem of optimal
      markers of a classical region receives a
      linear-algebraic solution. Moreover, the 
goodness of the fitting of the expression profile of a gene 
to the profile of a brain region is closely related to 
the co-expression of genes. The geometric interpretation 
of this fact leads to a quantitative criterion to detect markers
of pairs of brain regions. Local properties of the gene-expression profiles are
also used to detect genes that separate a given grain region from its environment.


\tableofcontents

\section{Introduction}

Neuroanatomy is experiencing 
a renaissance under the influence of molecular biology and computational methods.
Brain regions can be delineated on
stained sections of brain tissue. The set of boundaries between brain regions defined
on sections can be registered in order to obtain a three-dimensional atlas.
 Conflicts 
exist between the various nomenclatures of brain regions. The present paper will consider
brain regions defined by classical anatomy as in the Allen Reference Atlas \cite{AllenAtlas}.
 Gene-expression energies are positive quantities defined
at every point in the brain (or rather at every cubic voxel of side equal to the resolution, 
which is 200 microns in the present paper).
With contemporary techniques of {\emph{in situ}} hybridization, such data were produced by the Allen Institute for thousands
of genes in the mouse brain \cite{AllenAtlasMol, images}. This makes the ISH data much higher-dimensional 
than classical neuroanatomy. Given an anatomical atlas, it is therefore natural to ask if the patterns formed by the
expression energy of single 
genes and/or sets of genes can delineate and/or separate brain regions.\\

The structure of the paper is as follows. We will first 
formalize the notion of marker genes by defining
quantitative criteria that allow to rank individual genes by 
computing scores. The 
{\emph{localization score}} measures how much of the expression energy of a gene 
is contained in the region of interest. The {\emph{fitting score}} measures how close the 
expression-energy profile is to the characteristic function of the region. 
The associated rankings of genes are computed. The genes ranked as the top few markers make sense optically,
but there are conflicts between the two rankings. The two criteria are 
then used to rank sets of genes as markers of brain regions. The localization
score gives rise to a generalized eigenvalue problem, and the solutions can have much higher 
localization scores than individual genes, but they are difficult to interpret because the
sets of genes are very large and weighted by coefficients of alternating signs.
The fitting score of sets of genes gives rise to sparse sets of markers.
 These two scores are easy to compute and to generalize, but they are both global in nature 
and they penalize genes that fit well the centermost part of a brain region, have low expression around the region, 
and high expression in remote parts of the brain. But such genes are interesting to detect, as
they separate brain regions from their environment. A local fitting criterion is proposed,
using the eikonal function in order to formalize the situation described above. 
Markers of pairs of regions are also investigated. They may be of special interest to evolution, especially
for pairs of regions that are not equally well-identified in other species.

\section{Methods and models}
\subsection{Gene expression energies and classical neuroanatomy} 
The gene expression energies we analyzed were drawn from the Allen
Gene Expression Atlas \cite{AllenAtlasMol}. The steps taken in an
automatized pipeline to produce those data for each gene can be
summarized as follows:\\ 1. Colorimetric {\it{in situ}}
    hybridization;\\ 2. Automatic processing of the resulting
    images. Find tissue area eliminating artifacts, look for
cell-shaped objects of size $\simeq 10-30$ microns to minimize
artefacts;\\ 3. Aggregation of the raw pixel data into a grid.\\

The mouse brain is partitioned into cubic voxels of side 200 microns
(the whole brain consists of $\simeq 50,000$ voxels). For every voxel
$v$, the {\it{expression energy}} of the gene $g$ is defined as a
weighted sum of the greyscale-value intensities $I$ evaluated at the
pixels $p$ intersecting the voxel:
$$E(v,g) := \frac{\sum_{p\in v} M( p ) I(p)}{\sum_{p\in v} 1},$$ where
$M( p )$ is a Boolean mask worked out at step 2 with value 1 if the pixel is expressing and 0 if it is non-expressing.\\ 
Partitions of the
brain (or of the left-hemisphere) of various degrees of coraseness in
terms of classically-defined neuroanatomical regions were also
published in the Allen Reference Atlas (\cite{AllenAtlas}, see also the white paper
 {\ttfamily{http://mouse.brain-map.org/documentation/index.html}} for the definition of expression energies).\\

The present analysis is focused on a subset of the genes for which sagittal and coronal 
data are available from the Allen data. We computed the correlation coefficients
between sagittal and coronal data and selected the genes in the top-three 
quartiles of correlation (this makes for 3041 genes) for further analysis. Of course the quantitative methods 
can be tested against larger datasets or different reference atlases, but the 
genes we selected are already numerous enough to motivate the use of 
computational methods to detect markers.\\

\subsection{Individual genes: localization scores}
 Given a brain region of interest, say the cerebral cortex (call it $\omega$), define
 the localization score of a gene $g$ as the fraction of the $L^2$
 norm of its expression energy contained in the region:
$$\lambda_\omega(g) = \frac{\int_\omega E(v,g)^2 dv}{\int_\Omega E(v,g)^2 dv},$$
where $\Omega$ is the whole brain.\\   

 We computed the localization score of every gene in every region, for
a given annotation of the brain. These numbers induce a ranking of
genes as markers of each region of the brain, the better markers
having higher localization scores. A perfect marker of $\omega$ according to this criterion would have 
a score of 1. Going from a region to another region, one has to be careful when comparing 
 the values of the localization scores: as the volumes of the brain regions vary
across the annotation, the localization score is biased by the sizes
of the region. We need a reference in order to estimate how good a localization score 
is compared to what could be expected for a given brain region. We can use two references:\\
$\bullet$ {\bf{Uniform reference.}}  Consider an {\emph{indifferent}} marker that would be
  expressed uniformly across the brain. Its localization score in
  region $\omega$ is simply the relative volume of the region:
$$\lambda^{\mathrm{unif}}_\omega = \frac{\int_\omega dv }{\int_\Omega dv } = \frac{{\mathrm{Vol}}\;\omega}{{\mathrm{Vol}}\;\Omega}.$$
A gene is a better marker of $\omega$ than expected from a uniform expression if its score 
$\lambda_\omega(g)$ is larger than $\lambda^{\mathrm{unif}}_\omega$.\\
$\bullet$ {\bf{Average (data-driven) reference.}} A more realistic reference is given by the gene-expression profile averaged across all the genes:
$$E^{\mathrm{average}}(v) = \frac{1}{G}\sum_{g=1}^G E(v,g).$$
The corresponding localization score in a given region $\omega$ is:
$$\lambda^{\mathrm{average}}_\omega = \frac{\int_\omega E^{\mathrm{average}}(v)^2 dv }{\int_\Omega E^{\mathrm{average}}(v)^2 dv }.$$ 
A gene is a better marker of $\omega$ than expected from an average expression if its score 
$\lambda_\omega(g)$ is larger than $\lambda^{\mathrm{average}}_\omega$.\\

The values of these references, and the rankings of genes for $\omega$ taken from 
the list of 12 largest regions in the left hemisphere, are presented in the results section 
 and in appendices. 

\subsection{Individual genes: fitting scores}
The criterion defined above does not take into account the repartition
of the signal inside the region of interest: the localization score
for a given gene in region $\omega$ is invariant under a
transformation that moves the whole expression energy into a single
voxel within $\omega$, leaving all other voxels in $\omega$ with a
zero signal. It is therefore desirable to have another ranking of genes as markers, that compares
the gene-expression profiles to characteristic functions of brain regions.

This criterion compares the shape of the expression energy profile of
a gene and the shape of the region of interest.  The fitting score
$\phi_\omega(g)$ of gene $g$ in region $\omega$ is defined as follows:
$$\phi_\omega(g) = 1 - \frac{ 1 }{ 2 }\int_\omega
\left(\tilde{E}_{\mathrm{norm}}(v,g)^2 - \chi_\omega( v)\right)^2 dv =
\int_\omega\tilde{E}_{\mathrm{norm}}(v,g) \chi_\omega( v)dv.$$ where
$\chi_{\omega}$ is the characteristic function of $\omega$ normalized in the $L^2$ sense, and
$\tilde{E}_{\mathrm{norm}}$ is a normalized version of the expression
energy (the columns of the matrix $\tilde{E}_{\mathrm{norm}}$ are the
columns of the matrix $E$, normalized in the $L^2$ sense):
$$\tilde{E}_{\mathrm{norm}}(v,g) = \frac{E(v,g)}{\sqrt{\int_\Omega
    E(v,g )^2 dv}},\;\;\;\; \chi_\omega =
\frac{{\mathbf{1}}_\omega}{\sqrt{{\mathrm{Vol}}(\omega)}}.$$ A perfect
marker of the region $\omega$ would be a gene with fitting score equal
to 1. The geometric interpretation of this coefficient is as the
cosine of the angle between the unitary vectors $\tilde{E}_g$ and
$\chi_\omega$ in voxel space.  A perfect marker of region $\omega$ is
a gene whose expression profile is colinear with the characteristic
function of region $\omega$. Again, this error function can be
evaluated for all the genes in the dataset, and induces a ranking of
genes (see the results section and appendices).

\subsection{Sets of genes: optimal localization scores as a generalized eigenvalue problem}

Consider the problem of optimizing the localization score of a set of
genes, whose collective expression energy is taken to be a linear
combination of the expression energies in our dataset:
         $$E_\alpha( v ) : = \sum_{g = 1}^{G} \alpha_g E( v, g ).$$
where $G=3041$ is the number of genes in our dataset. The previous
analysis corresponded to vectors $\alpha$ with only one non-zero
coordinate.\\ 
The localization score in the brain region $\omega$ of a set
of genes is naturally written as
         $$\lambda_\omega( \alpha ) = \frac{\int_\omega \left( \sum_g
  \alpha_g E( v, g )\right)^2 dv}{\int_\Omega \left( \sum_g \alpha_g
  E( v, g )\right)^2 dv} = \frac{\alpha^t J^\omega \alpha}{\alpha^t
  J^\Omega \alpha},$$ where the quadratic forms $J^\omega$ and
$J^\Omega$ have coefficients given by scalar products of gene
expression profiles across $\omega$ and the whole brain:
         $$J^\omega_{g,h}= \int_\omega E(
v,g)E(v,h)dv,\;\;\;\;J^\Omega_{g,h}= \int_\Omega E( v,g)E(v,h)dv.$$

        We can fix an overall dilation invariance by fixing the value
        of the quadratic form in the denominator, and maximizing the
        localization factor boils down to a maximization of one
        quadratic form under a quadratic constraint.
         $$ {\mathrm{max}}_{\alpha\in{\mathbf{R}}^G}\lambda_\omega(
         \alpha ) ={\mathrm{max}}_{\alpha\in{\mathbf{R}}^G, \alpha^t
           J^\Omega \alpha = 1}\alpha^t J^\omega\alpha.$$ Introducing
         the Lagrange multiplier $\chi$ associated to the constraint,
         we are led to maximizing the following quantity under
         $\alpha$:
         $$L_{\omega, \sigma}( \alpha ) = \alpha^t J^\omega\alpha - \sigma( \alpha^t J^\Omega \alpha - 1).$$
         The stationarity condition reads as a generalized eigenvalue problem,
        $$J^\omega \alpha = \sigma J^\Omega \alpha,$$ and the Lagrange
         multiplier is the largest generalized eigenvalue. Maximizing the generalized 
         localization score is therefore equivalent to finding the
         largest generalized eigenvalue corresponding to the quadratic
         forms $J^\omega$ and $J^\Omega$.\\

 Of course the alternating signs of the coefficients make these sets
difficult to interpret.  But these algebraic sums provide absolute
bests that one could not beat by taking combinations of genes with
positive coefficients. The negative coefficients allow to offset the
contribution of some genes outside the region of interest.\\

\subsection{Sets of genes: optimized fitting scores for sparse sets of genes}


As the optimal set of of markers is very hard to interpret due to alternating 
signs of components,
we can take advantage of the simple quadratic structure
of the error function used to compute fitting scores in order
to obtain sets of markers with positive coefficients. Optimization 
of a quadratic form under positivity constraint is all we need to 
 compute the optimal sets of markers.
Let us write down the fitting error function for a set of genes 
and expand it in powers of the coefficients:
\begin{align}
{\mathrm{ErrFit}}_\omega(c) &= \int_\Omega\left( \sum_g c_gE_g(v) - \chi_\omega(v)\right)^2dv\\
  &= \sum_{g,h}c_g c_h J_{gh}-2\sum_g c_g f_g + 1,
\end{align}
where $\omega$ and $\Omega$ respectively denote the brain region of interest 
and the whole brain. The problem of finding thje best-fitting 
set of genes therefore boils down to the following quadratic programming problem under positivity constraints:
$$c^{\mathrm{opt}} = {\mathrm{argmin}}_{c\in\mathbf{R}_+^G}{\mathrm{ErrFit}}_\omega(c)= {\mathrm{argmin}}_{c\in\mathbf{R}_+^G}\left(\frac{1}{2}c^t J c - f^t c\right),$$
with the following notations for the quadratic form $J$ and the vector $f$:
$$J_{gh}= \int_\Omega E(v,g)E(v,h)dv,$$
$$f_g = \int_\Omega E(v,g) \chi_\omega(v) dv.$$
The set of genes with strictly positive coefficients corresponds to the set 
of inactive constraints. It happens to be much sparser than the vector encoding the
 generalized eigenvector for the cortex localization problem (see figure (\ref{fig:genCortex})).\\ 
   
However, lots of secondary minima are guaranteed to exist when larger and larger sets of genes
are taken into account, and coefficients $c$ of very different norms can be hard to 
use to construct markers out of digitized data, as the absolute intensity of genes is quite heterogeneous, 
and a gene with low absolute intensity can happen to be weighted by a large coefficient, thus 
amplifying noise rather than contributing to a realistic marker.\\
But we can take advantage of the expression of the fitting score in
terms of the scalar product between the gene expression profile 
and the characteristic function of the brain region:
$${\mathrm{ErrFit}}_\omega(c) = 2\left( 1- \int_\Omega\sum_g c_g  E_g(v) \chi_\omega(v) dv \right),$$
$$c^{\mathrm{opt}} = {\mathrm{argmax}_{c\in \{0,1\}^G}}\int_\Omega\sum_g c_g  E_g(v) \chi_\omega(v) dv.$$
 Another approach to the optimization problem consists in looking for sets of genes such that the 
co-expression between the sum of the expression energies of those genes and the 
characteristic function is larger than that of any individual genes. This can happen,
for instance if the characteristic function in voxel space equals the sum of two 
genes, whose expression energies are two independent vectors in voxel space: the cosine 
of the angle between
 any of these two vectors with the characteristic function is strictly smaller than one, but the 
cosine of the angle between the sum and the characteristic function equals one.\\

This is a finite problem, even though the number of subsets is extremely large. We impose a maximum
 $G_{\mathrm{max}}$ on the number of genes we want to accept, and adopt a bootstrapping approach: we repeatedly draw random 
subsets of size $G_{\mathrm{max}}$ from our set of genes, and keep the subsets that beat the record fitting score
(this record is initialized at the highest fitting score for an individual gene).

\subsection{Separation properties}
The methods described so far are global in nature in the sense that the error functions 
involve sums of expression energies over the whole brain. This corresponds to
evaluating how a brain region is singled out with respect to the rest of the brain. No attention
is paid to the position of the voxels that contribute to the error functions: a voxel with high expression 
in the cerebellum will penalize a gene as a merker of the striatum, no more but but no less than if it was in the
ventral pallidum. However it may be interesting to detect genes that separate 
some brain region from its environment, without necessarily highlighting these brain regions
in an exclusive way. The description of such a situation implies a more {\emph{local}}
error function.\\

However, when looking at the expression profile of a gene in the 
 neighbourhood of a particular brain regions, one can sometimes
notice that the region is well-separated from the rest of the brain, because the expression
is high in voxels close to the center of the region, and locally declines around the center.
At large distances from the center, the details of the gene-expression profile matter much less, as long
as this pattern of decreasing expression from center to boundary is detected. Such genes have good {\emph{separation}}
 properties.\\

The separation property we described above corresponds to the situation where the gene-expression 
pattern looks like a plateau around the center of the region $\omega$, and gradually fades away when the boundary
of the region is crossed. Of course the notion of center of a brain region needs to be defined
more precisely. So does the notion of distance to a brain region. The
eikonal distance to the boundary of the region is a geometric quantity
that is well adapted to this problem, as it measures the minimal distance traveled by light emitted from the boundary of the region \cite{LLMec}.
In order to control how far from the center of a region a voxel is, one can therefore
solve the eikonal equation with boundary conditions on the boundary of the region:
$$|\nabla h_\omega | = 1,$$
$$ h_\omega|_{\partial\omega}= 0.$$
 The eikonal distance has been used used to place injections in 
the brain in a way that preserves the boundaries of regions defined by the Allen Atlas \cite{injections,MBA,ExprWiring}. 
It is also a useful tool to evaluate the misalignment of skulls and skull variability in stereotactic 
protocols \cite{stereotax}. The equation is solved using level-set methods \cite{Sethian}.\\

We define a model function $\xi_\omega$ that detects the most central 
part of the region $\omega$, using the eikonal function as a measure 
of centrality. The function is positive It is a plateau in the central part of  $\omega$,
and fades away across voxels that are more peripheric to $\omega$. More specifically, let
us define the {\emph{eikonal radius}} $\rho_\omega$ of the region $\omega$ as the maximum value of the 
eikonal function inside the region:
$$\rho_\omega := {\mathrm{max}}_{v\in \omega} h_\omega( v ).$$
Let us first apply a mask to the eikonal function, with negative signs outside
the region and positive signs inside:
$$h^{\mathrm{signed}}_\omega(v):= h_\omega( v ) \times \left(\mathbf{1}(v\in\omega) - \mathbf{1}( v\notin\omega )\right).$$

Our model function $\xi_\omega$ equals one around the center of the region, 
where $h^{\mathrm{signed}}_\omega$ is positive and larger than a specified fraction of the eikonal radius, given 
by a certain fraction $\iota$ of the eikonal radius. It equals zero where $h^{\mathrm{signed}}_\omega$ is negative and 
larger in absolute value than another specified fraction of the eikonal radius. The values are interpolated between
these two regions according to the values of the eikonal function.\\

Having defined this local characteristic function $\xi_\omega$ around the brain region of interest, one
can treat the support of $\xi_\omega$ as we treated the whole brain $\Omega$ in the previous computations,
and adapt the various quantitative criteria by making the following substitution:
$${\mathrm{Global}}\longleftrightarrow \;{\mathrm{Local}},$$
$$\Omega = {\mathrm{whole\;brain}} \longleftrightarrow\; \Omega = {\mathrm{Supp}}\xi_\omega,$$
$$ \chi_\omega  \longleftrightarrow\;\xi_\omega,$$
$$ E(v,g )\longleftrightarrow\; E(v,g ) \mathbf{1}( v\in{\mathrm{Supp}}\xi_\omega ).$$
This substitution expresses the fact that the expression energy outside the support of the local characteristic function 
of $\omega$ can be very singular of very intense without affecting the separation properties.


\subsection{Co-markers for pairs of brain regions}

The various quantitative criteria can be repeated for reunions of
brain regions. For instance one can look for a marker of the two brain
regions $\omega_A$ and $\omega_B$. Ideally one would like the
expression profile of a marker gene to look like the sum of the two
characteristic functions of regions $A$ and $B$, normalized in the
$L^2$ sense\footnote{One can as well look for genes that separate regions $A$ and $B$ from their respective environments,
by considering the local characteristice functions worked out using the eikonal functions with boundary conditions at the boundaries of $A$ and $B$, rather than the characteristic functions.}. But it may be interesting to allow the two characteristic
functions to be weighted by coefficients, in order to detect genes
whose expression looks like two bumps, one centered around $A$, one centered around $B$, with possibly different
intensities.\\

 Consider the two characteristic functions $\chi_A$ and $\chi_B$,
 normalized in the $L^2$ sense, and a linear combination thereof with
 positive coefficients, normalized in the same way. The coefficients
 of the linear coimbination can be interpreted geometrically in terms
 of a single parameter, which is an angle between 0 and $\pi/2$. Let
 us denote it by $\theta$:
$$\int_\Omega \chi_A( v )^2 dv = 1,\;{\mathrm{Supp}}\chi_A = \omega_A,$$
$$\int_\Omega \chi_B( v )^2 dv = 1,\;{\mathrm{Supp}}\chi_B = \omega_B,$$
$$\chi := \alpha \chi_A +\beta \chi_B,\;\alpha \geq 0, \; \beta \geq 0,\; \int_\Omega \chi^2 = \alpha^2 + \beta^2 = 1,$$
$$ \alpha = \cos\theta,\; \beta = \sin\theta, \; 0 \leq \theta \leq\frac{\pi}{2},$$
 where we have used the
 fact that the functions $\chi_A$ and $\chi_B$ are orthogonal, because
 they have disjoint supporty.  Geometrically, the function $\chi$ we
 are trying to fit is the sum of two unit orthogonal vectors in voxel
 space, that sits on the intersection of the unit circle and the first 
quadrant in the two-plane spanned by these two vectors. We can compute the fitting error for each gene
 at fixed angle $\theta$, but it can be optimized wrt the angle:
\begin{align}
{\mathrm{ErrFit}}_{A,B}( g,\theta) &=\int_{\Omega}\left(E( v,g )-(\cos\theta\chi_A( v )+\sin\theta \chi_B(v))\right)^2 dv\\
 &= 2\left( 1  - \cos \theta \int_\Omega E( v,g )\chi_A( v )dv - \sin\theta \int_\Omega E( v,g )\chi_B(v)dv\right).
\end{align}
This optimization step corresponds to the fact that the angle between a fixed vector 
in voxel space (corresponding to a gene), can have a lower angle with a two-plane than with any of the 
vectors of an orthonormal basis of the two-plane.
The optimal angle $\theta^\ast$ is given by the equation:
$$\frac{\partial}{\partial\theta}{\mathrm{ErrFit}}_{A,B}( g,\theta^\ast)= 0,$$
$${\mathrm{i.e.}}\; -\sin\theta^\ast \int_\Omega E( v,g )\chi_A( v )dv + \cos\theta^\ast \int_\Omega E( v,g )\chi_B(v)dv = 0,$$
$${\mathrm{i.e.}}\; \theta^\ast = \arctan\left(\frac{\int_\Omega E( v,g )\chi_B(v)}{\int_\Omega E( v,g )\chi_A(v)}\right).$$

The value of the error function at the optimal angle (meaning the linear combination of the two characteristic function with positive coefficients that is best fit by gene $g$) is then evaluated in terms of the 
scalar product between the gene expression and the two characteristic function  $\chi_A$ and $\chi_B$: 
$$\cos\theta^\ast = \frac{\int_\Omega E( v,g )\chi_A(v)dv}{\sqrt{ \left( \int_\Omega E( v,g )\chi_A(v)dv\right)^2
     + \left( \int_\Omega E( v,g )\chi_B(v)dv\right)^2}},$$
$$\sin\theta^\ast = \frac{\int_\Omega E( v,g )\chi_B(v)dv}{\sqrt{ \left( \int_\Omega E( v,g )\chi_A(v)dv\right)^2
     + \left( \int_\Omega E( v,g )\chi_B(v)dv\right)^2}},$$
so that 
 $$\phi_{A,B}(g,\theta) = 1 - \frac{1}{2}{\mathrm{ErrFit}}_{A,B}( g,\theta^\ast) = \sqrt{
  \left(\int_\Omega E( v,g )\chi_A(v)dv\right)^2 + \left(\int_\Omega E(
  v,g )\chi_B(v)dv\right)^2}.$$ 
This score is comprised between 0 and
1, as the fitting score evaluated for the fitting of a single region
by a gene.  So, given a non-hierarchical atlas $\mathcal{A}$, one can
find better fittings for pairs of regions in the atlas than for single
regions.\\
These scores can be computed for all pairs of regions in a given non-hierarchical atlas.
Of course there is no reason why the top co-marker of regions $A$ and $B$ should be especially
more impressive than the best marker of $A$ or $B$. By the look of the expressions of 
 the coefficients $\cos\theta^\ast$ and $\sin\theta^\ast$, it is clear that in the case where 
$\int_\Omega E(v,g )\chi_B(v)dv$ is much smaller than $\int_\Omega E(v,g )\chi_A(v)dv$, its score 
as a co-marker of regions $A$ and $B$ is slightly larger than its score as a marker of $A$, but 
most of the expression will of course be in the $A$. The value of $\tan \theta^\ast$ controls the balance between the 
expression energies in the two regions. The closer it is to 1, the better co-marker we have. Asking for a value of exactly 
one would amount to trying to fit the sum of the characteristic functions of regions $A$ and $B$ without 
Once the genes have been ranked as co-markers of $A$ and $B$, one can filter out 
the genes for which $\tan \theta^\ast$ is out of a tolerance zone around 1. This is a balance constraint. The genes
at the top of the ranking that do not satisfy it are rather markers of the region ($A$ or $B$) 
that has the highest coefficient. The genes that satisfy it are the co-markers we are after, and they are penalized 
by the localization and fitting criteria, both for region $A$ and for region $B$:
$${\mathrm{Balance}}\;{\mathrm{constraint}}_\tau \equiv |\tan \theta^\ast - 1|\leq \tau.$$


\section{Results and discussion}

\subsection{Rankings of genes}

\begin{figure}
\centering
\begin{minipage}{0.48\textwidth} 
\includegraphics[width=3in,keepaspectratio]{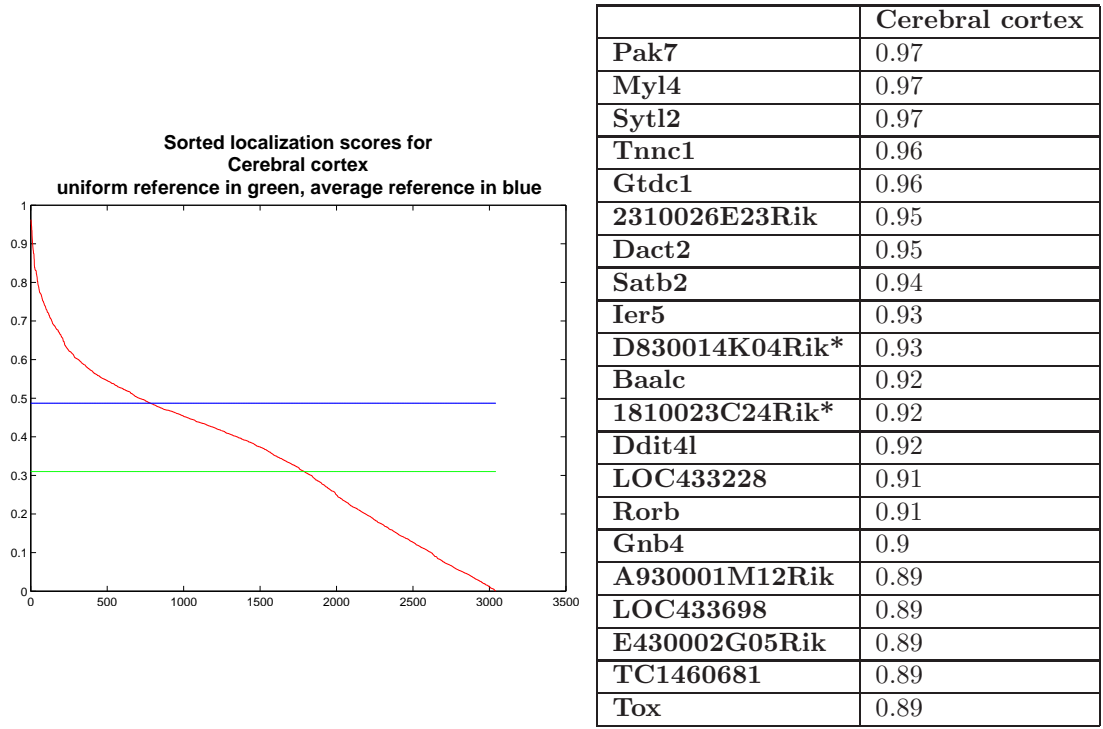}
\end{minipage}   
\begin{minipage}{0.46\textwidth} 
\noindent
\begin{tabular}{|l|l|}
\hline
&\textbf{Cerebral cortex}\\\hline
\textbf{Pak7}&0.97\\\hline
\textbf{Myl4}&0.97\\\hline
\textbf{Sytl2}&0.97\\\hline
\textbf{Tnnc1}&0.96\\\hline
\textbf{Gtdc1}&0.96\\\hline
\textbf{2310026E23Rik}&0.95\\\hline
\textbf{Dact2}&0.95\\\hline
\textbf{Satb2}&0.94\\\hline
\textbf{Ier5}&0.93\\\hline
\textbf{D830014K04Rik*}&0.93\\\hline
\textbf{Baalc}&0.92\\\hline
\textbf{1810023C24Rik*}&0.92\\\hline
\textbf{Ddit4l}&0.92\\\hline
\textbf{LOC433228}&0.91\\\hline
\textbf{Rorb}&0.91\\\hline
\textbf{Gnb4}&0.9\\\hline
\textbf{A930001M12Rik}&0.89\\\hline
\textbf{LOC433698}&0.89\\\hline
\textbf{E430002G05Rik}&0.89\\\hline
\textbf{TC1460681}&0.89\\\hline
\textbf{Tox}&0.89\\\hline
\end{tabular}

\end{minipage} 
\caption{Plot of sorted localization scores in the cerebral cortex (left), with the list of the
first few genes with highest localization scores in the cerebral cortex (right).}
\label{fig:localizationScoresCortex}
\end{figure}

\begin{figure}
 \includegraphics[width=3in,keepaspectratio]{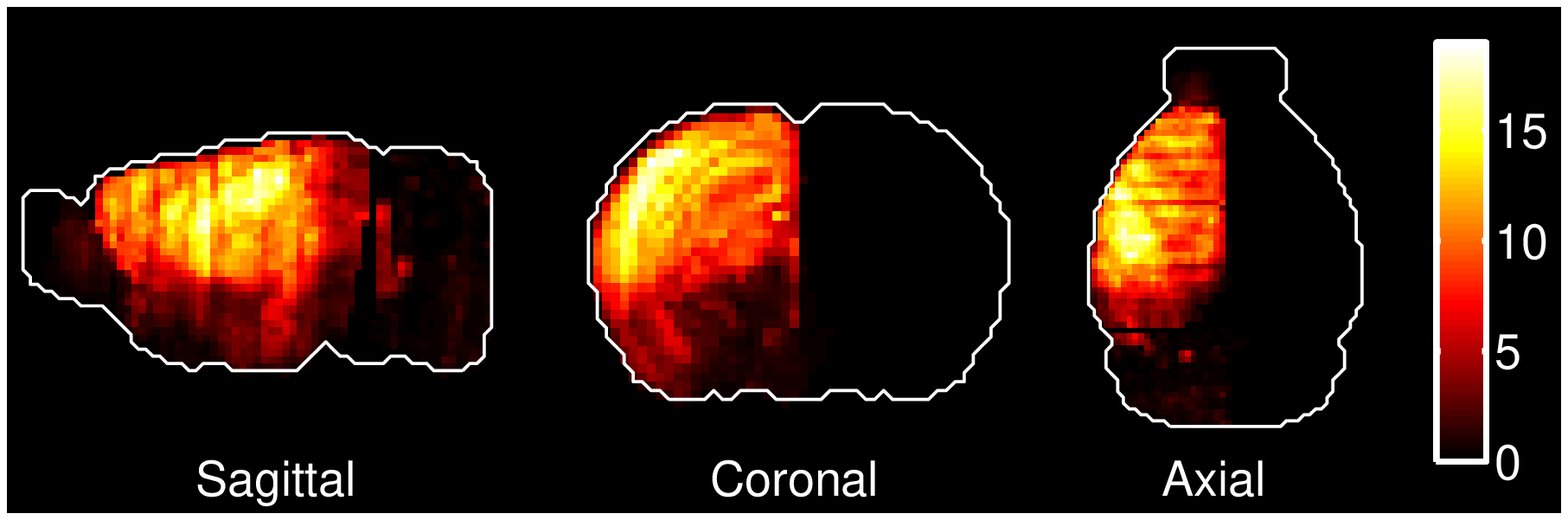} 
\includegraphics[width=3in,keepaspectratio]{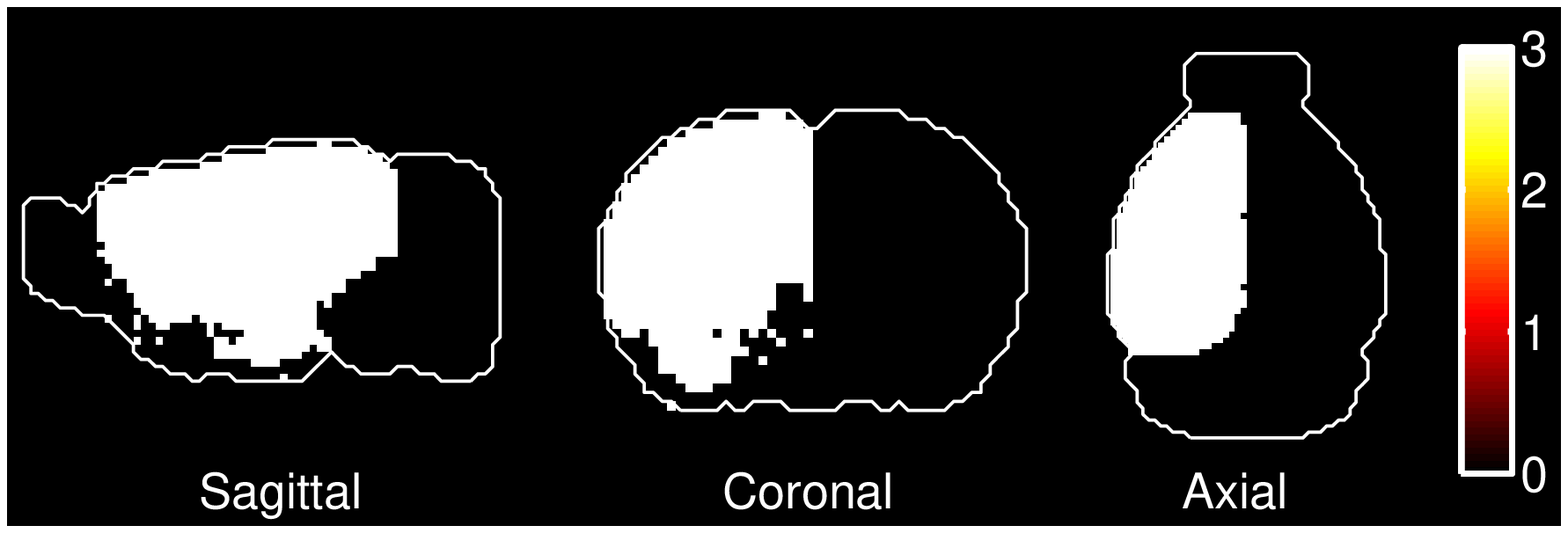}\\
\caption{Heat map of the maximum-intensity projection of Pak7 (left), the best marker of the 
cortex in the sense of localization scores, compared to the characterictic function of the cerebral cortex (right).} 
\label{fig:markerLocCortex}
\end{figure}

\begin{figure}
\centering
\begin{minipage}{0.48\textwidth} 
\includegraphics[width=3in,keepaspectratio]{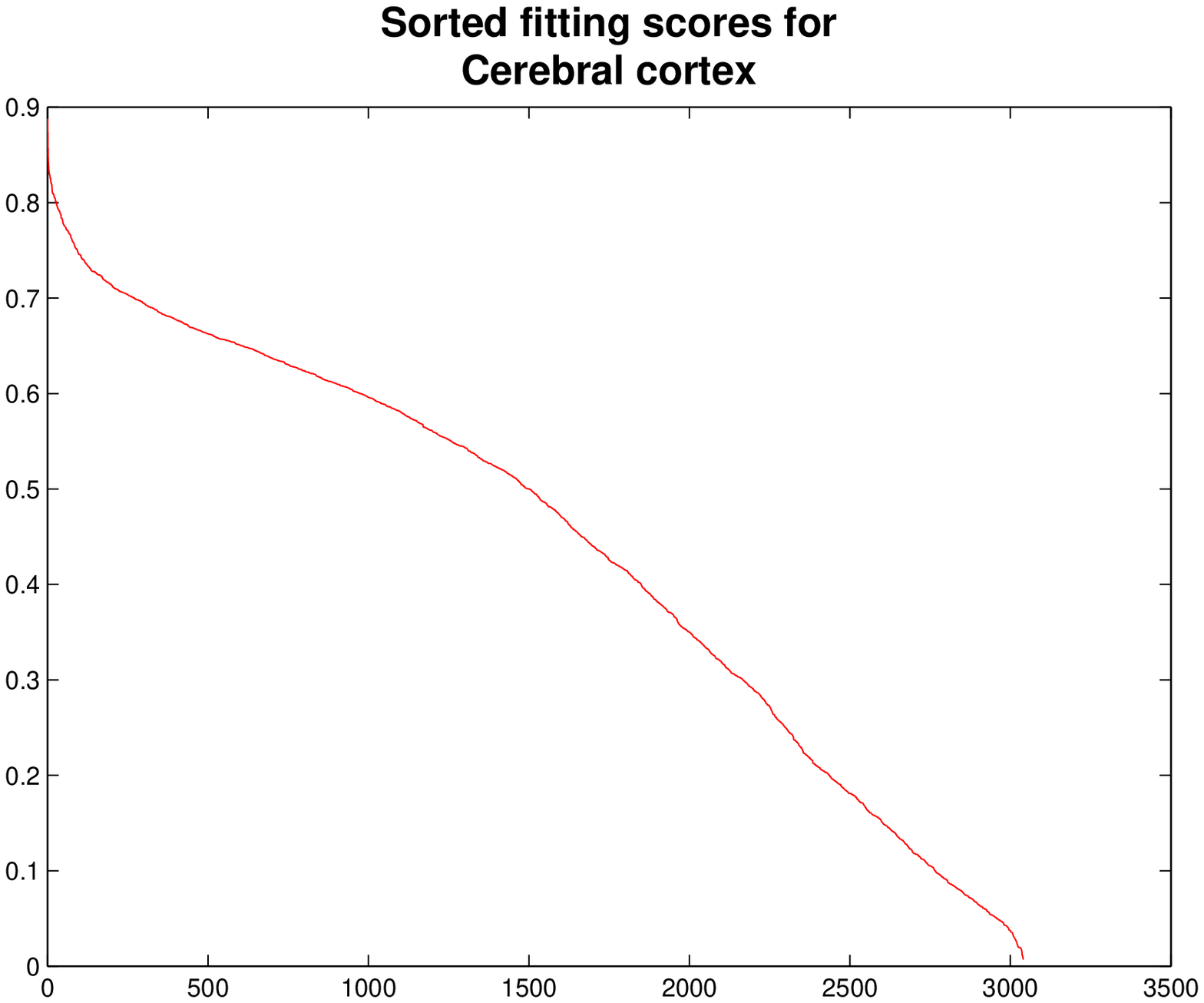}
\end{minipage}   
\begin{minipage}{0.46\textwidth} 
\noindent
\begin{tabular}{|l|l|}
\hline
&\textbf{Cerebral cortex}\\\hline
\textbf{Satb2}&0.89\\\hline
\textbf{Kcnh7}&0.85\\\hline
\textbf{Ephb6}&0.84\\\hline
\textbf{3110035E14Rik}&0.83\\\hline
\textbf{Homer1}&0.83\\\hline
\textbf{Fhl2}&0.83\\\hline
\textbf{Pak7}&0.83\\\hline
\textbf{Klf10}&0.83\\\hline
\textbf{Dusp3}&0.83\\\hline
\textbf{Cckbr}&0.83\\\hline
\textbf{1110008P14Rik}&0.82\\\hline
\textbf{Tbr1}&0.82\\\hline
\textbf{Igsf9b}&0.82\\\hline
\textbf{Stx1a}&0.82\\\hline
\textbf{A230097P14Rik*}&0.81\\\hline
\textbf{D430041D05Rik}&0.81\\\hline
\textbf{Mal2}&0.81\\\hline
\textbf{Igfbp6}&0.81\\\hline
\textbf{Nlk}&0.81\\\hline
\textbf{Arc}&0.81\\\hline
\end{tabular}

\end{minipage} 
\caption{Plot of sorted fitting scores in the cerebral cortex (left), with the list of the
first few genes with highest fitting scores in the cerebral cortex (right).}
\label{fig:fittingScoresCortex}
\end{figure}

\begin{figure}
\begin{center}
\includegraphics[width=5in]{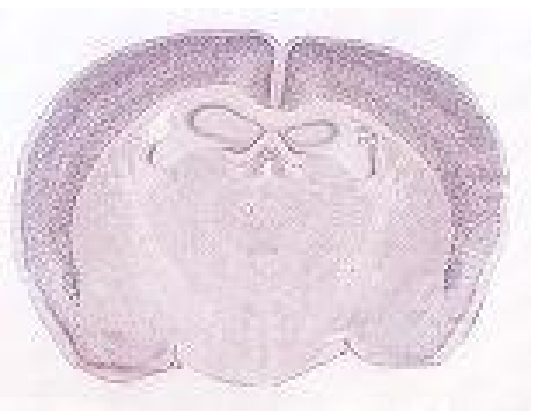}
\end{center}
\caption{
{\bf A coronal section of the ISH of Satb2.} Satb2 has
the highest localization score in the cerebral cortex. The concentration of 
blue precipitate in the region is manifest. 
}
\label{fig:Satb2Coronal}
\end{figure}

\begin{figure}
\begin{center}
\includegraphics[width=5in]{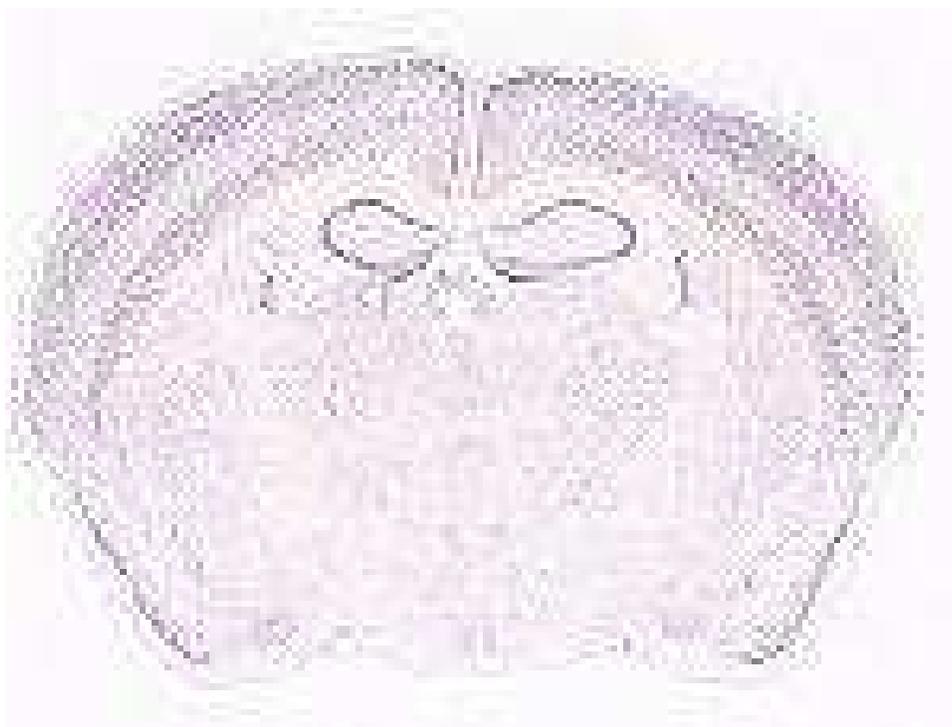}
\end{center}
\caption{
{\bf A coronal section of the ISH of Pak7.} Pak7 has
the highest fitting score in the cerebral cortex. The concentration of 
blue precipitate in the region is manifest. 
}
\label{fig:Pak7Coronal}
\end{figure}

A plot of the sorted localization scores of individual genes
is shown on figure ~\ref{fig:localizationScoresCortex} for the cerebral cortex, as well as a table of the best few marker genes. Tables for all the 
other brain regions in the coarsest Allen Reference Atlas are included in an appendix. 
The maximum-intensity projections of the best marker of the cerebral cortex in the left hemisphere,
 and compared to those of the characteristic function of the cerebral cortex are shown on 
figure \ref{fig:markerLocCortex}.
The sorted fitting scores and the list of top genes for the cerebral cortex are shown on figure (\ref{fig:fittingScoresCortex}).
\begin{figure}
 \includegraphics[width=3in,keepaspectratio]{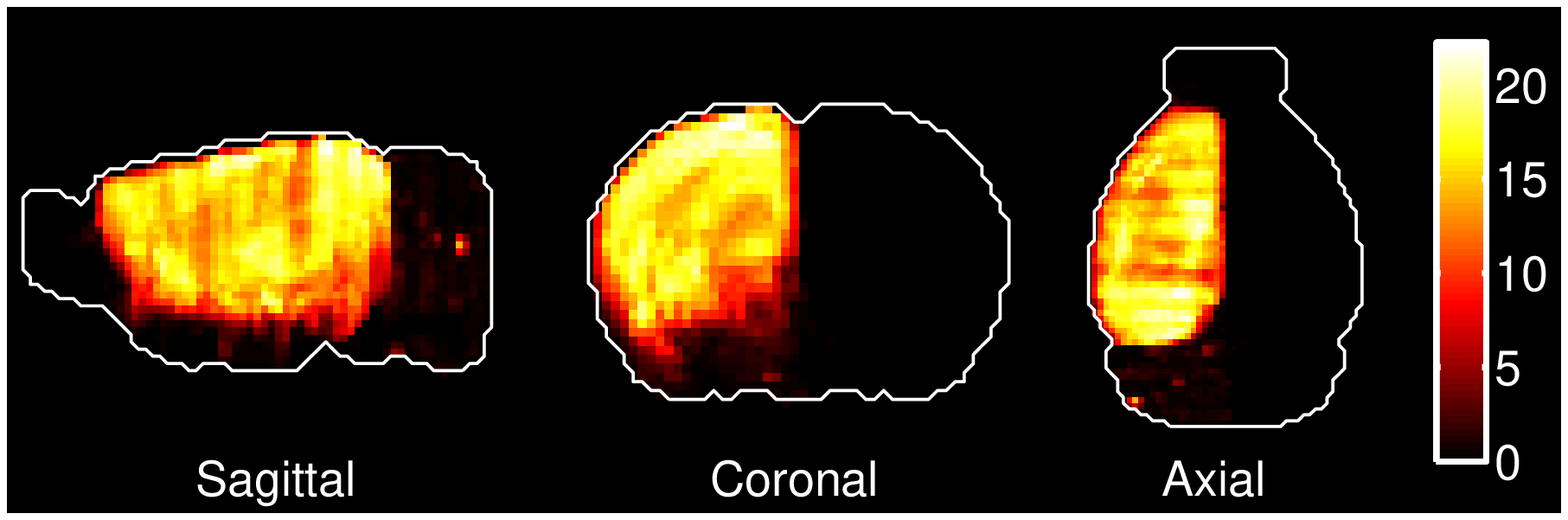} 
\includegraphics[width=3in,keepaspectratio]{regionProfile1.eps}\\
\caption{Heat map of the maximum-intensity projection of Satb2 (left), best marker of the 
cortex in the sense of fitting scores, compared to the characterictic function of the cerebral cortex (right).}
\label{fig:markerFittingCortex}
\end{figure}
A coronal section of the ISH data for Satb2 is shown on figure (\ref{fig:Satb2Coronal}). The cerebral cortex 
indeed appears strikingly on the section. However, Satb2 is not the absolute best gene according to 
the localization criterion, which is Pak7, but it is still among the best 10 genes by localization scores.
 A coronal of the ISH data for Pak7 is shown on figure (\ref{fig:Pak7Coronal}). Maximal-intensity projections
of the registered 3D data on figures (\ref{fig:markerLocCortex}) and (\ref{fig:markerFittingCortex}) show indeed
that the expression energy of Satb2 is more evenly distributed across cortex than the one on Pak7, which makes Satb2 closer 
to the characteristic function of the cerebral cortex.

\subsection{Rankings of regions}
\begin{figure}
\centering
\includegraphics[width=4in,keepaspectratio]{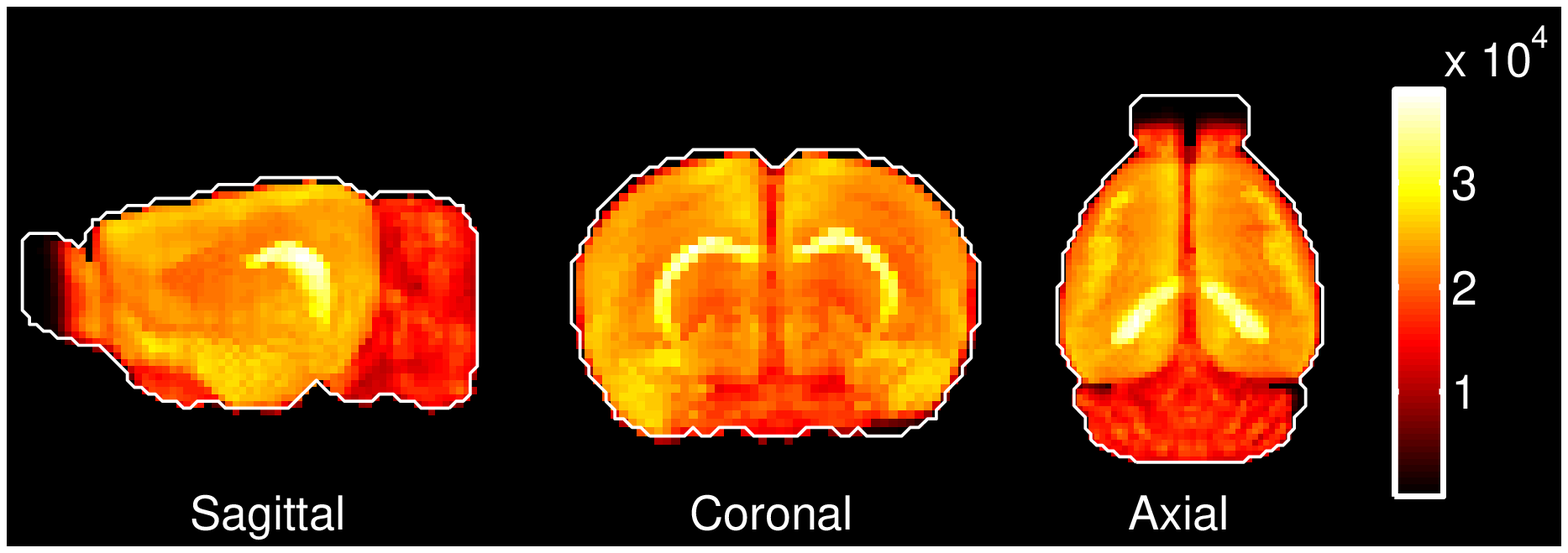}
\caption{Heat map of the maximal-intensity projections of
 the sum of the expression energies across all genes in the dataset. The cerebral cortex 
and the hippocampal region are clearly visible.}
\label{fig:sumOfAllGenes}
\end{figure}

\begin{figure}
\section{Appendix: Reference values of the localization scores for the coarsest atlas of the left hemisphere}
\begin{tabular}{|l|l|l|l|}
\hline
&\textbf{Uniform reference}&\textbf{Average reference}&\textbf{Region profile}\\\hline
\textbf{Cerebral cortex}&0.295&0.476 & \includegraphics[width=1.82in,keepaspectratio]{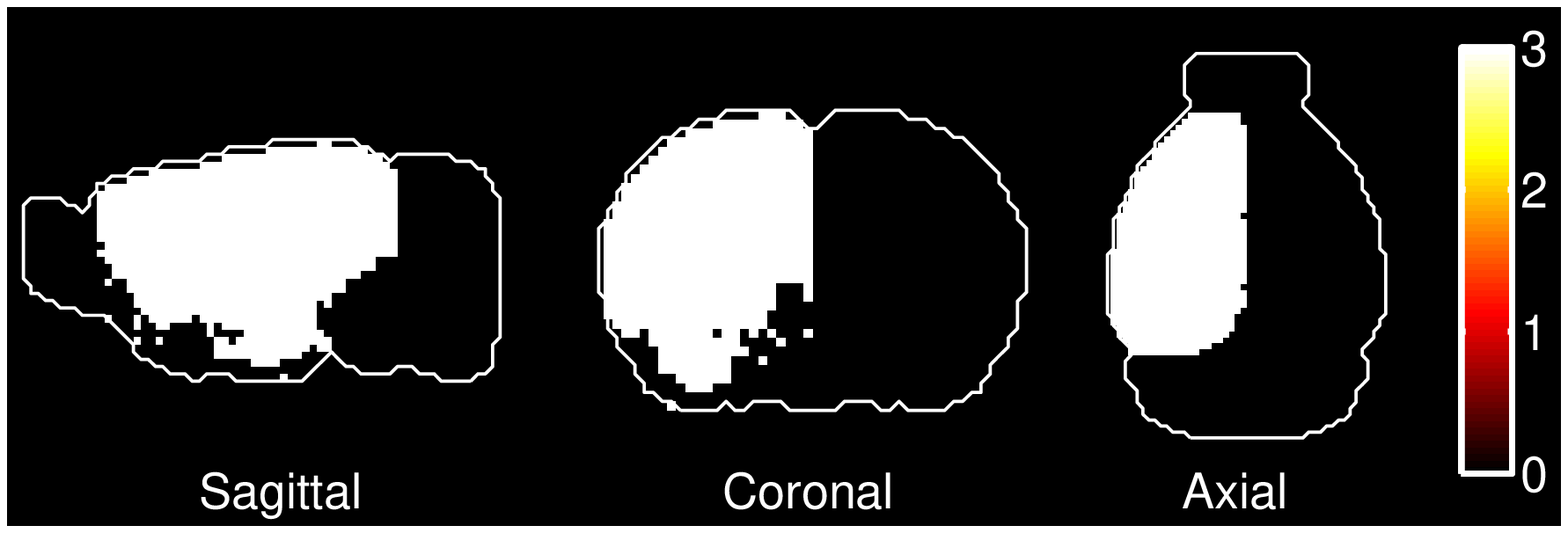}\\\hline
\textbf{Olfactory areas}&0.092&0.096& \includegraphics[width=1.82in,keepaspectratio]{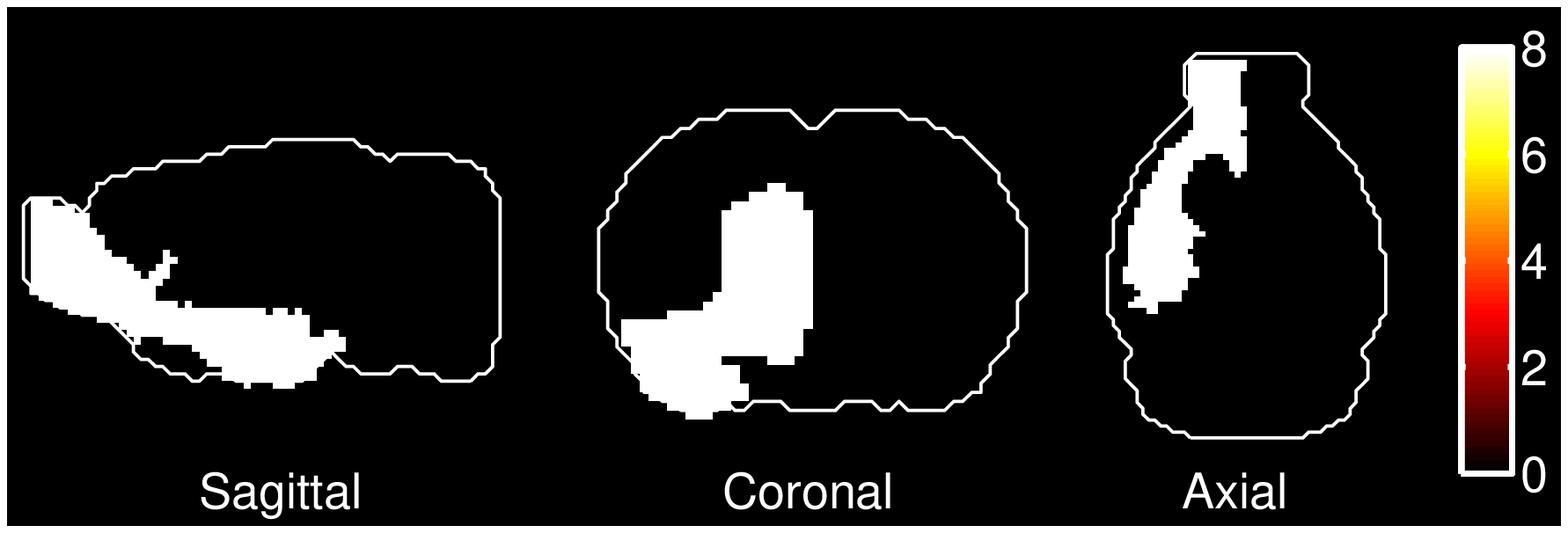}\\\hline
\textbf{Hippocampal region}&0.0426&0.0642& \includegraphics[width=1.82in,keepaspectratio]{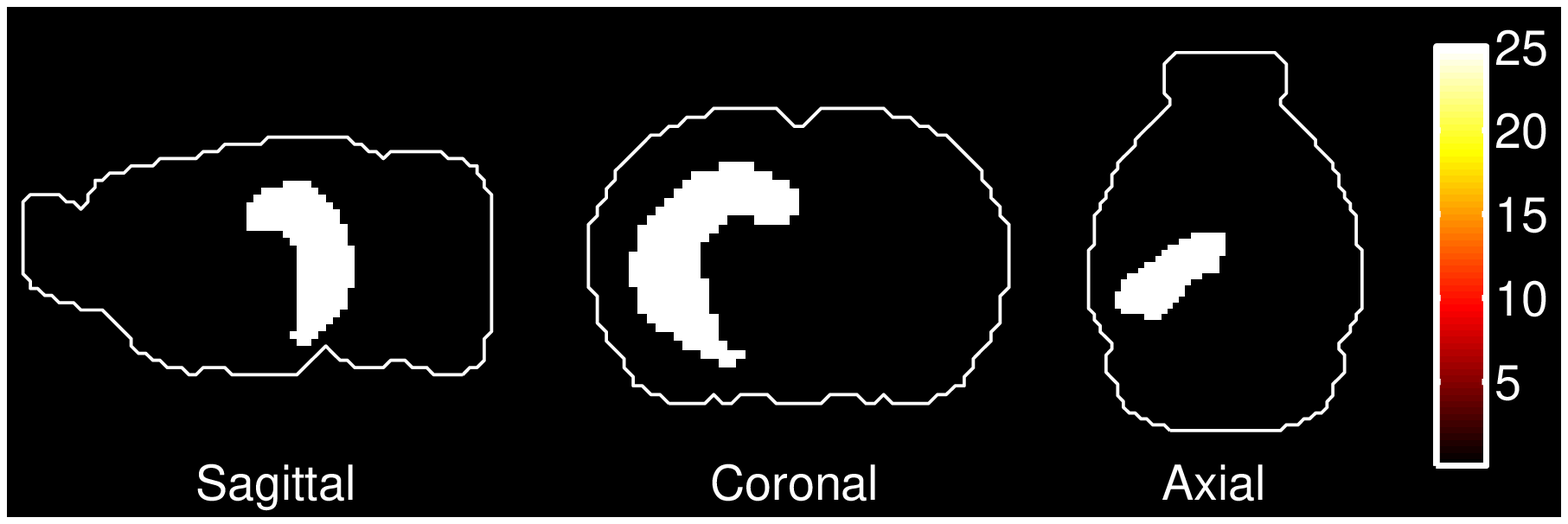}\\\hline
\textbf{Retrohippocampal region}&0.039&0.057& \includegraphics[width=1.82in,keepaspectratio]{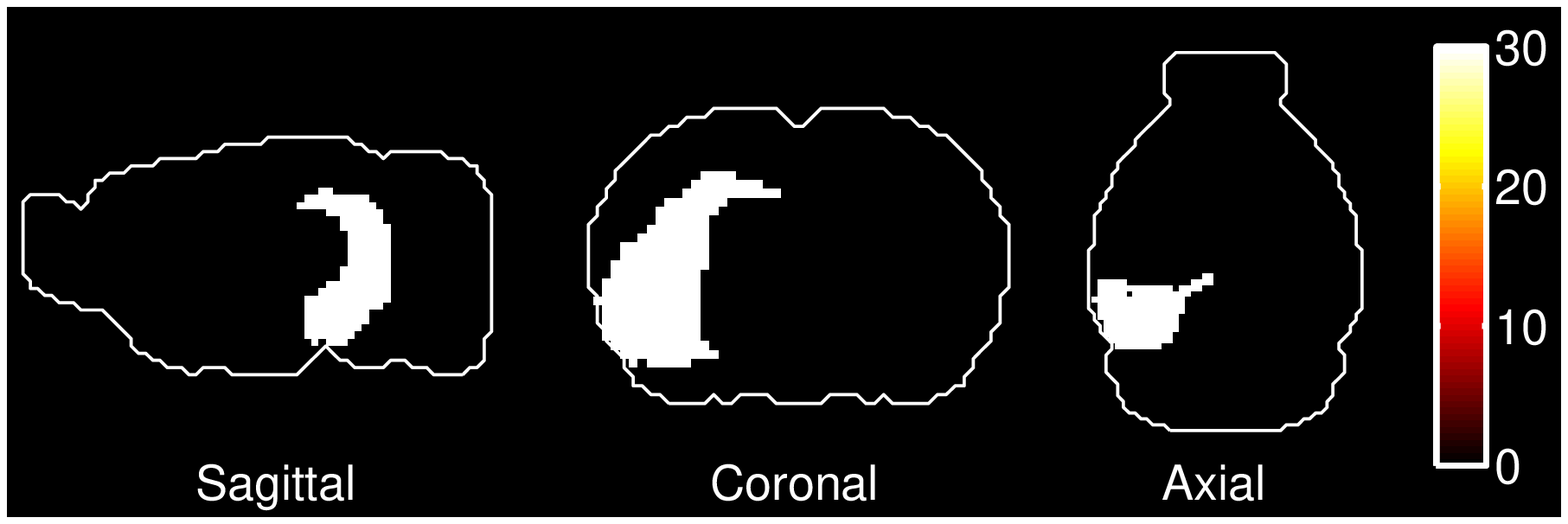}\\\hline
\textbf{Striatum}&0.086&0.0605& \includegraphics[width=1.82in,keepaspectratio]{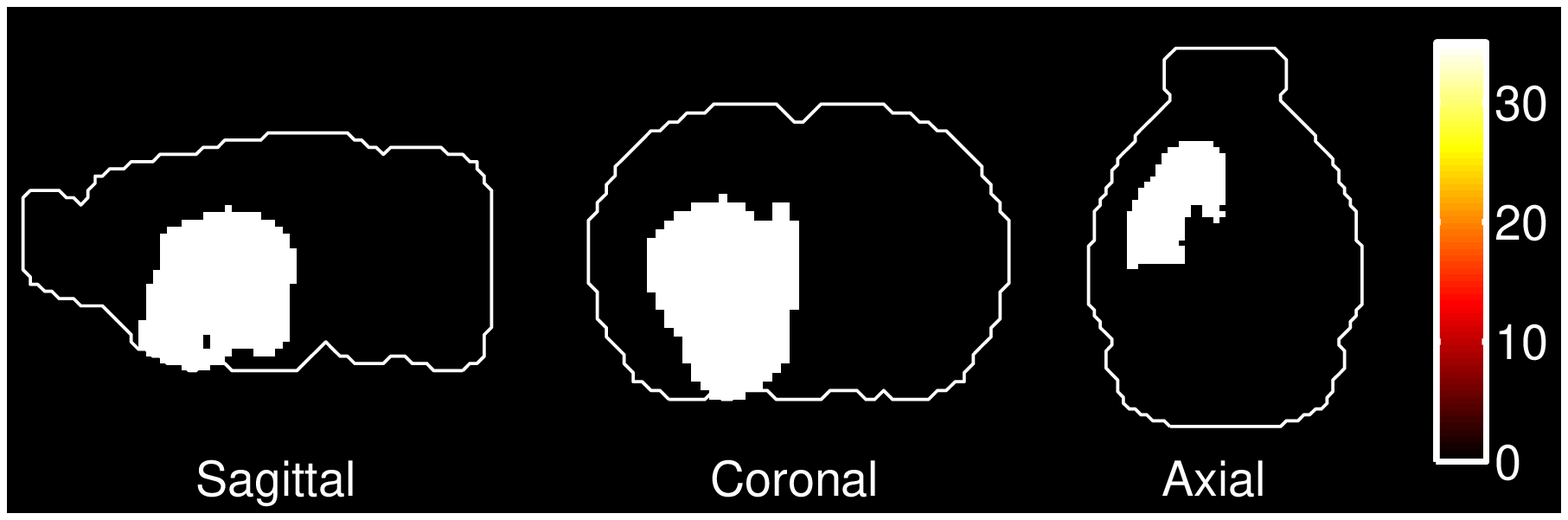}\\\hline
\textbf{Pallidum}&0.019&0.010& \includegraphics[width=1.82in,keepaspectratio]{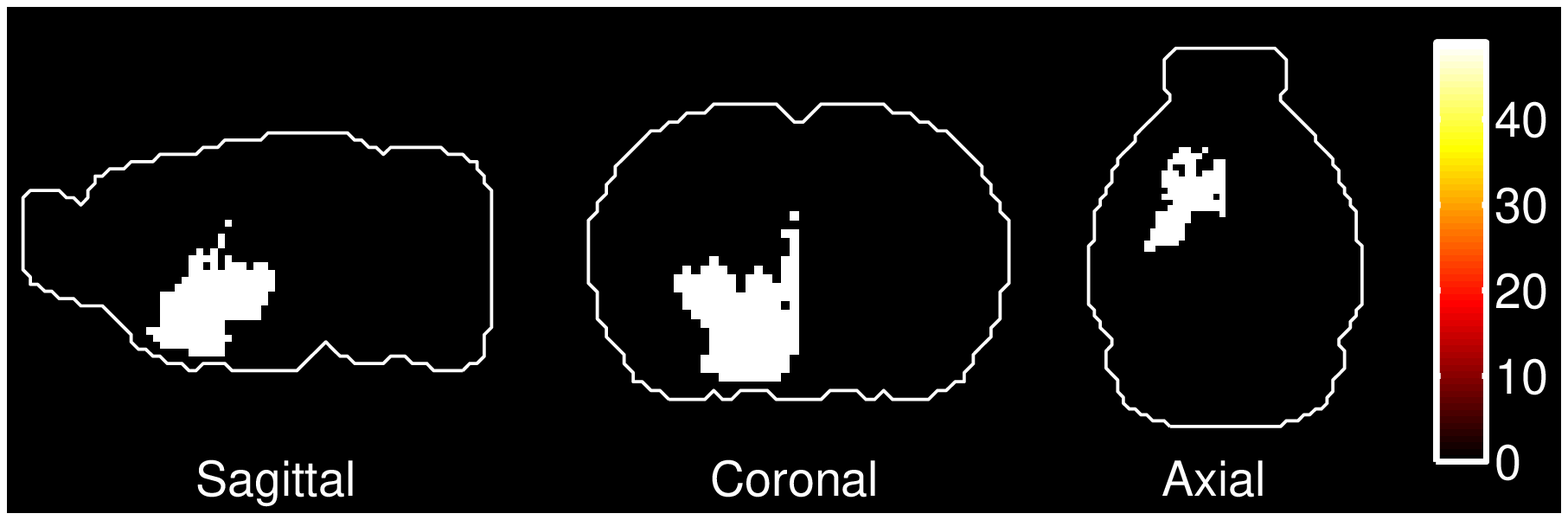}\\\hline
\textbf{Thalamus}&0.043&0.030& \includegraphics[width=1.82in,keepaspectratio]{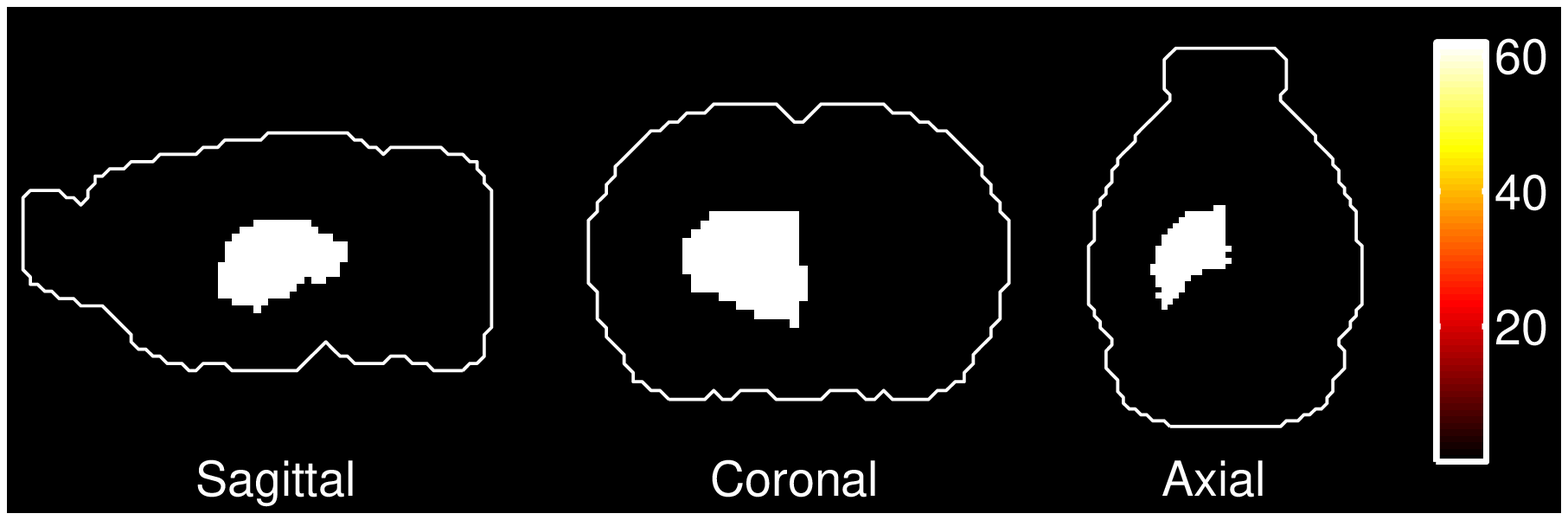}\\\hline
\textbf{Hypothalamus}&0.034&0.020&  \includegraphics[width=1.82in,keepaspectratio]{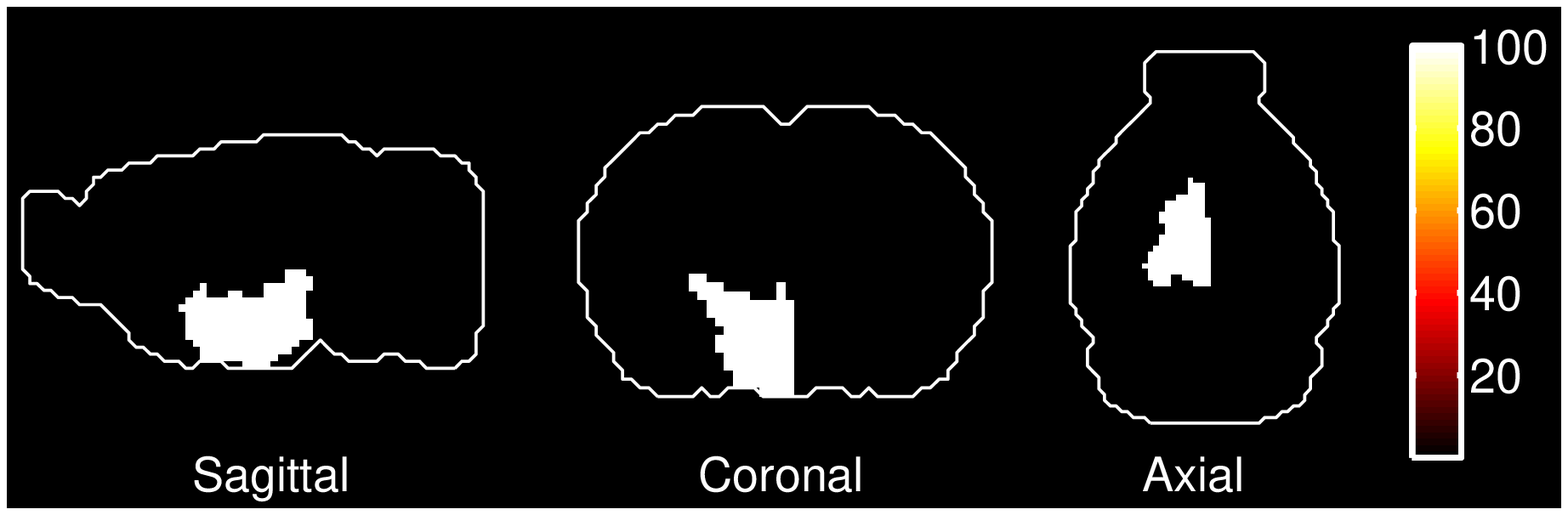}\\ \hline
\textbf{Midbrain}&0.078&0.041&  \includegraphics[width=1.82in,keepaspectratio]{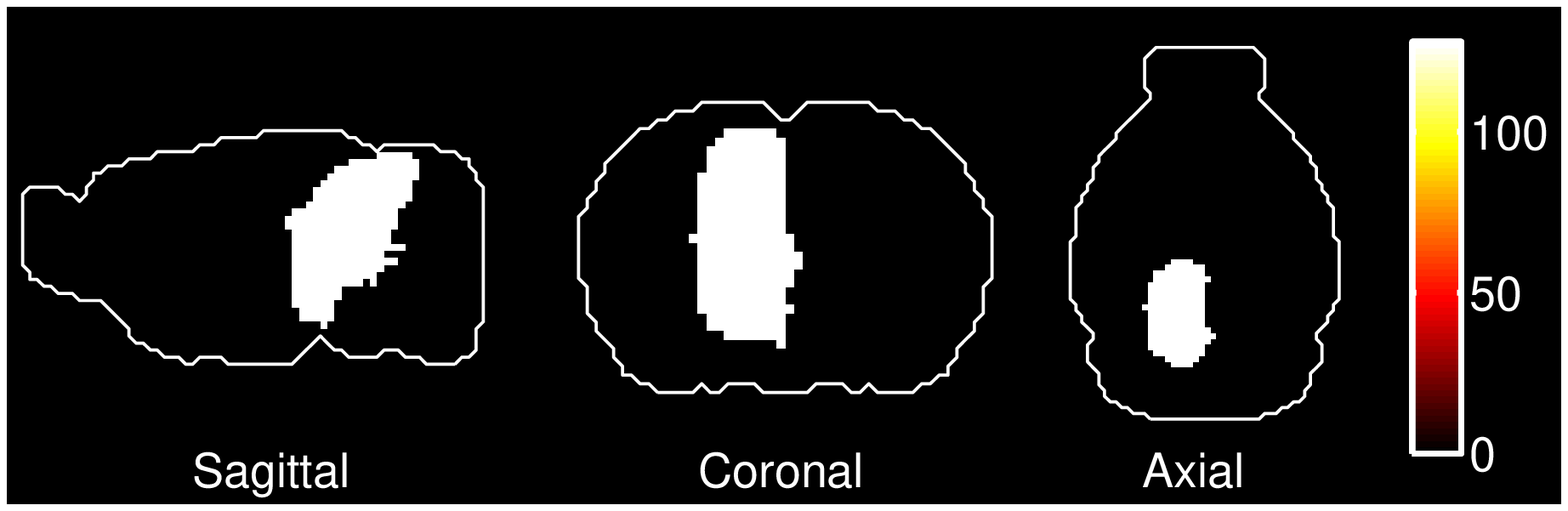}\\\hline
\textbf{Pons}&0.046&0.024&  \includegraphics[width=1.82in,keepaspectratio]{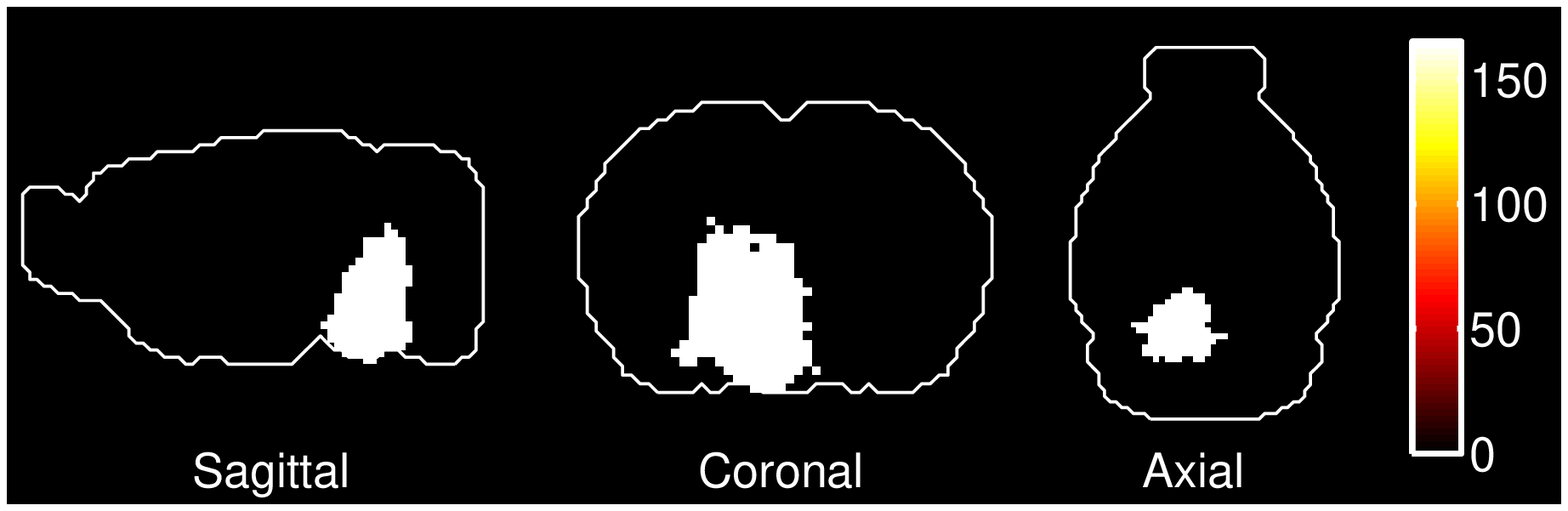}\\\hline
\textbf{Medulla}&0.061&0.042&  \includegraphics[width=1.82in,keepaspectratio]{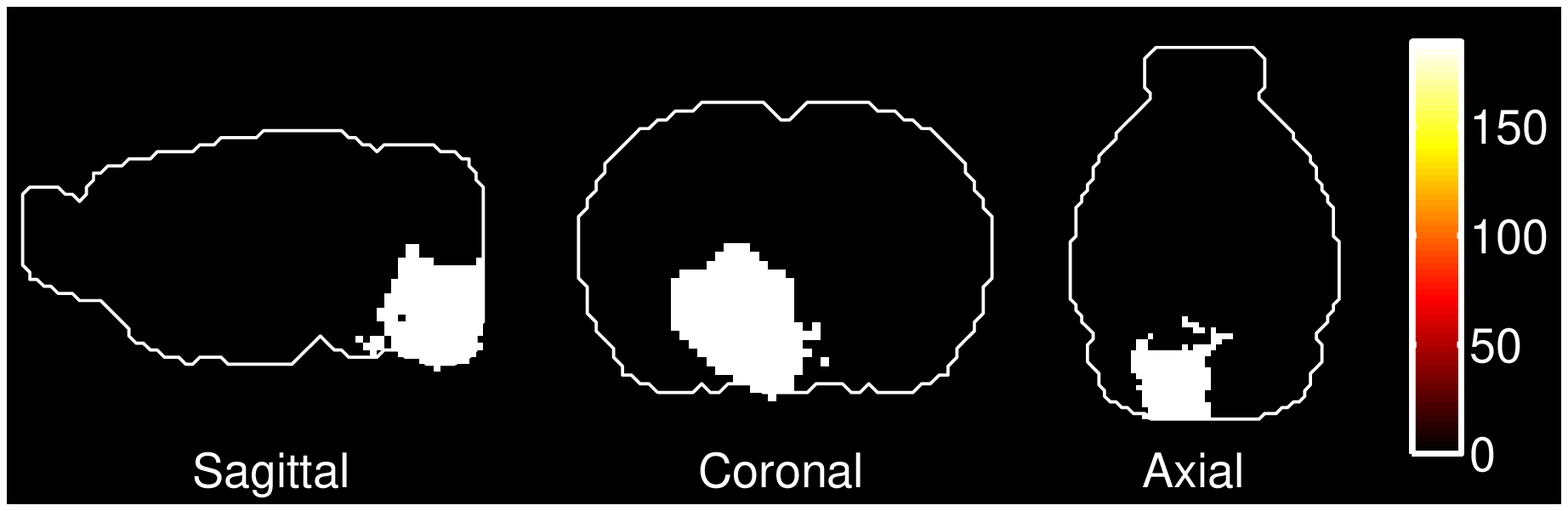} \\\hline
\textbf{Cerebellum}&0.114&0.055&  \includegraphics[width=1.82in,keepaspectratio]{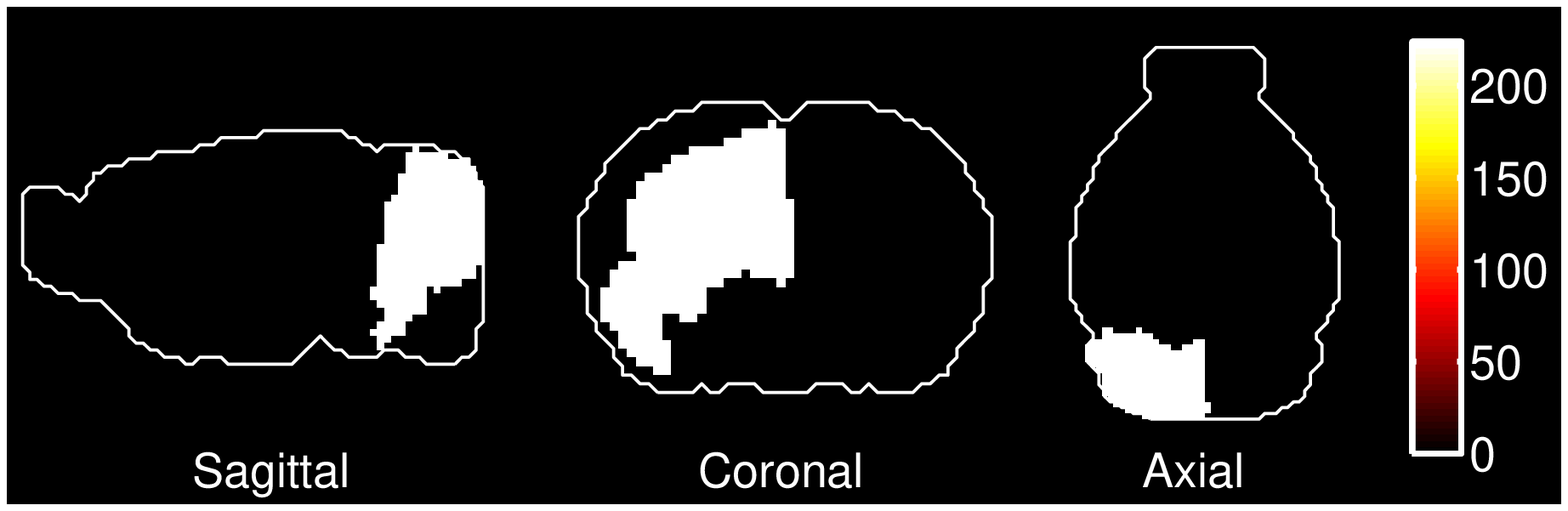}\\\hline
\end{tabular}
\caption{Values of the uniform and average reference scores for each of the 12 main regions of the left hemisphere.}
\label{fig:referenceTableBig12}

\end{figure}

For each region $\omega$, we can count the number of markers as the number of
genes whose localization score in the region is larger than the
fraction of the brain occupied by the region, as defined by
the uniform of average references. The results are illustrated on figure
(\ref{fig:tableOfRegionsbig12}).\\
\begin{figure}
\begin{tabular}{|l|l|l|}
\hline
\textbf{Region name}&\textbf{Nb of genes above average ref.}&\textbf{Nb of genes above volume ref.}\\\hline
Basic cell groups and regions&3041&3041\\\hline
Medulla&1418&920\\\hline
Pons&1315&600\\\hline
Cerebellum&1230&675\\\hline
Olfactory areas&1210&1241\\\hline
Thalamus&1144&614\\\hline
Midbrain&1126&381\\\hline
Hippocampal region&1055&1544\\\hline
Pallidum&1038&276\\\hline
Hypothalamus&1007&456\\\hline
Retrohippocampal region&998&1626\\\hline
Striatum&855&479\\\hline
Cerebral cortex&782&1791\\\hline
\end{tabular}

\caption{Ranking of regions by decreasing number of genes with localization 
score above average reference.}
\label{fig:tableOfRegionsbig12} 
\end{figure}
Moreover, the two  reference values defined above using either the volumes 
of the brain regions or the average of the expression of all genes
in the dataset show important distorsions, as can be seen from the table
(\ref{fig:referenceTableBig12}). In particular, the gene-expression
profiles are biased towards the cerebral cortex and the hippocampal 
region, as the value of $\lambda^{\mathrm{average}}_{\mathrm{cerebral cortex}}$ is higher than 
47 percent, while the value of $\lambda^{\mathrm{uniform}}_{\mathrm{cerebral cortex}}$ is lower than
30 percent. This distorsion is manifest of the maximal-intensity 
projection of the sum of all expression-energy profiles across 
the dataset, shown on figure (\ref{fig:sumOfAllGenes}).\\

\begin{figure}
\centering
\includegraphics[width=4in,keepaspectratio]{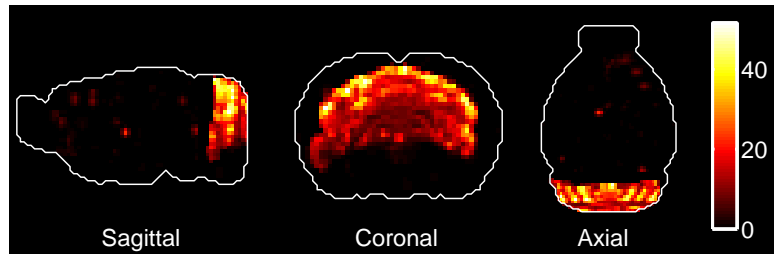}
\caption{ {\bf{Absolute best localization across all genes and regions.}} Heat map of the maximal-intensity projections of
 the expression energy of Gabra6.}
\label{fig:absoluteBestLoc}
\end{figure}

\begin{figure}
\centering
\includegraphics[width=4in,keepaspectratio]{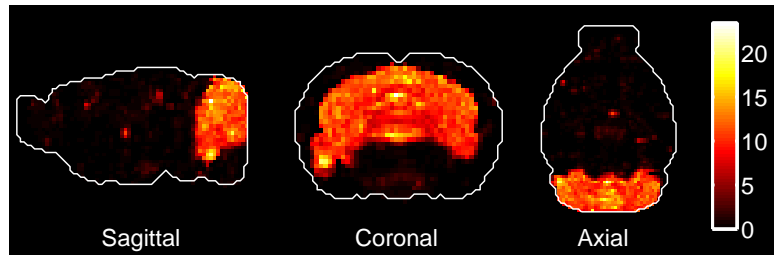}
\caption{ {\bf{Absolute best fitting across all genes and regions.}} Heat map of the maximal-intensity projections of
 the expression energy of 3110001A13Rik.}
\label{fig:absoluteBestFit}
\end{figure}

The highest scores across all genes and brain regions are for the cerebellum,
both by localization and by fitting. This comparison across all regions and genes 
makes more sense for fitting scores than for localization scores, as 
the fitting scores is not biased by the volume of the region. Both genes make good sense 
optically as markers of the cerebellum. The best-localized gene is Gabra6, 
at 98 percent localization score (see figure (\ref{fig:absoluteBestLoc})).
It is the 73rd best-fitted gene to the cerebellum. The best-fitted gene is 3110001A13Rik,
 at 89 percent fitting score (see figure (\ref{fig:absoluteBestFit})).
It is the 2nd best-fitted gene to the cerebellum. The distorsion is much larger for Gabra6 because its 
expression profile shows inhomogeneities inside the cerebellum. For 3110001A13Rik, the localization is optically extremely 
good.\\ 

\subsection{Sets of genes}
\begin{figure}
\begin{minipage}{0.48\textwidth} 
\includegraphics[width=3in,keepaspectratio]{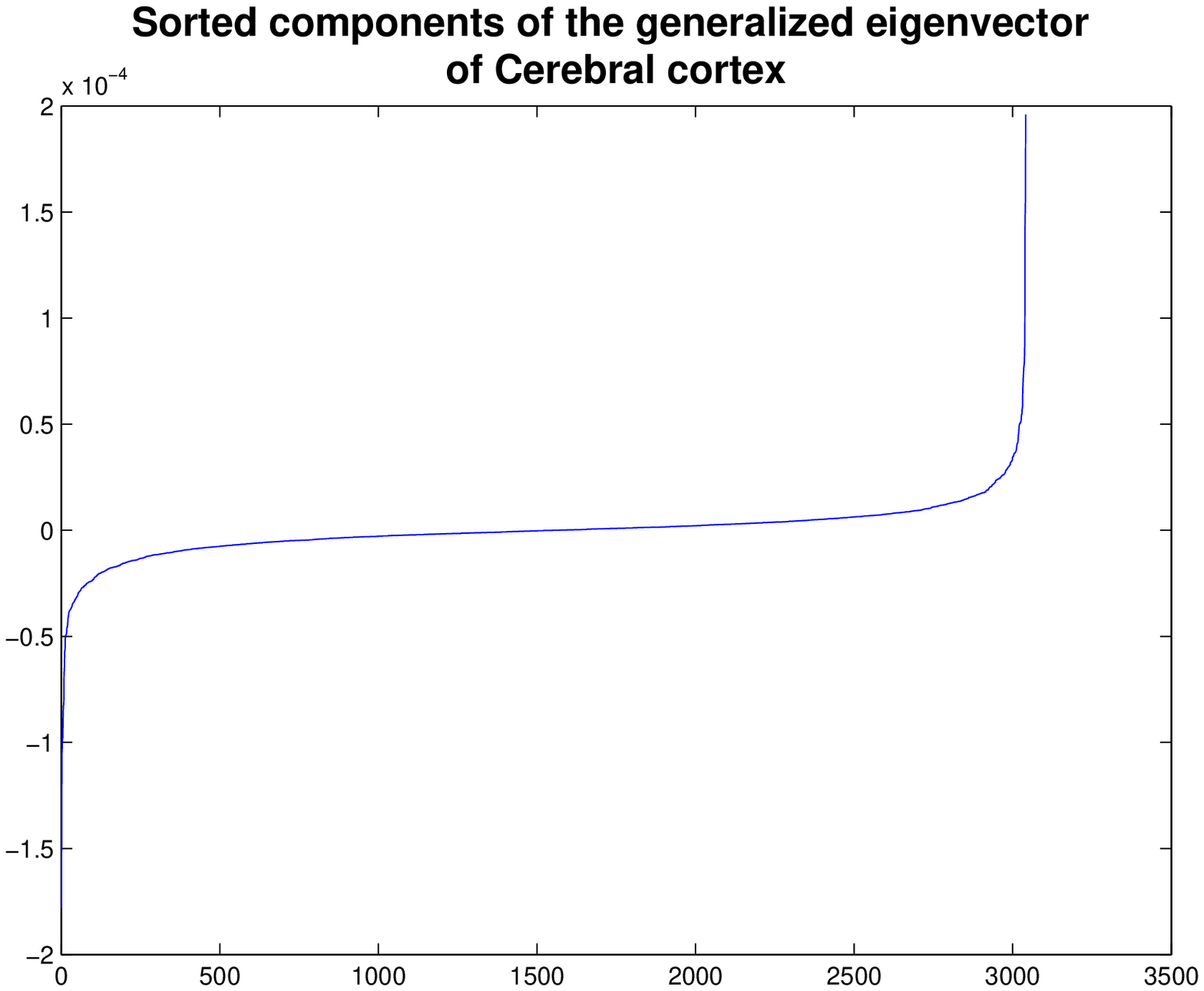}
\end{minipage}      
\begin{minipage}{0.48\textwidth}      
\includegraphics[width=2.5in,keepaspectratio]{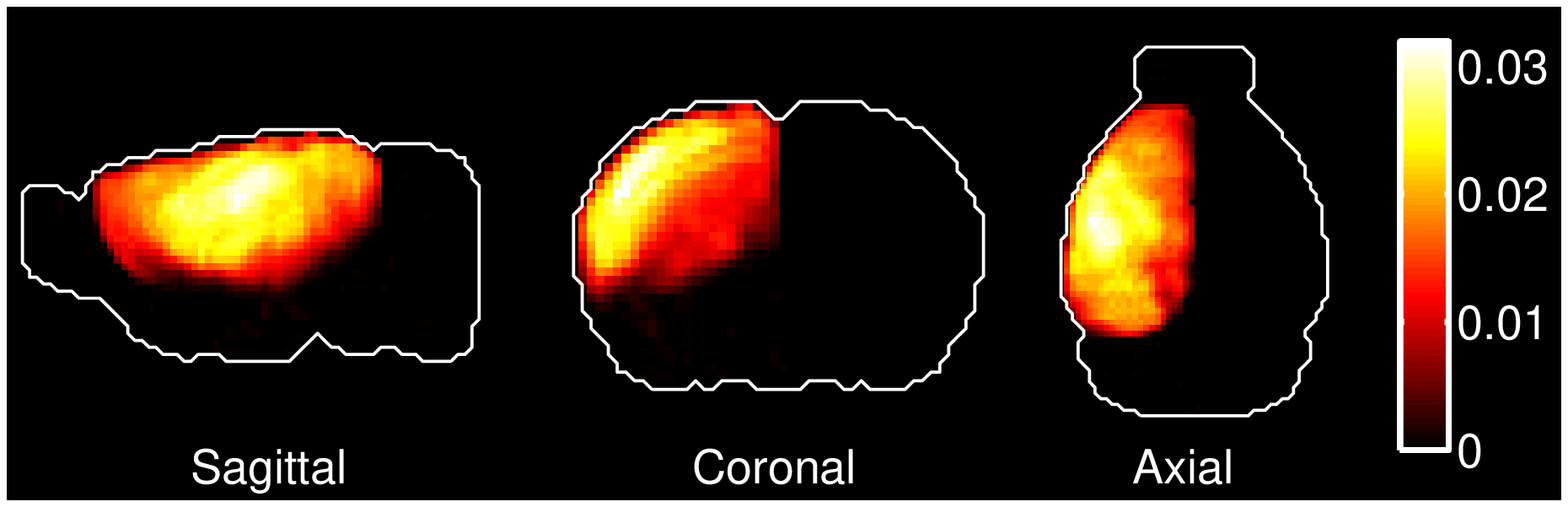}
\end{minipage} 
\caption{Optimal set of marker genes in the sense of generalized localization scores. Sorted components
 (left), heat map of the maximal-intensity projections of
 the expression energies of the genes weighted by these coefficients (right).}
\label{fig:genCortex}
\end{figure}

Sorting the coefficients of the generalized eigenvector associated to
the largest generalized eigenvalue for the cerebral cortex yields a
profile (illustrated in figure (\ref{fig:genCortex}) ), which has
coefficients of both signs (as above $\omega$ is chosen to be the
cerebral cortex), with a localization score of $0.979$ (higher than
the optimum for single genes, as it should be).\\
 It is interesting to note that in some regions the gene-expression profile correponding to the 
generalized eigenvector looks much more coherent as a marker than the best-localized gene.
More examples can be found in figure (\ref{fig:generalizedTableBig12}). Taking combinations 
of genes therefore enables one to get closer to the characteirstic function of the regions.
It is therefore tempting to go back to the fitting scores and to adapt it to sets of genes,
with positivity constraints that would produce sparser sets of genes.  
 \begin{figure}
\centering
\begin{tabular}{|l|l|l|}
\hline
&\textbf{Fitting rank}&\textbf{Fitting scores, score of set 0.89656}\\\hline
\textbf{Satb2}&1&0.89\\\hline
\textbf{Ephb6}&3&0.84\\\hline
\textbf{Igfbp6}&18&0.81\\\hline
\textbf{Map3k5}&38&0.79\\\hline
\end{tabular}
\\
\centering
\includegraphics[width=4in,keepaspectratio]{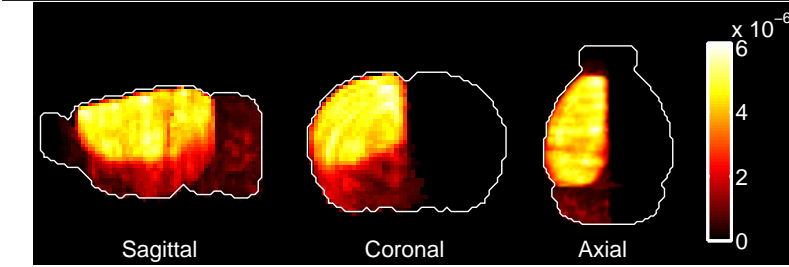} 
\caption{Optimal set of marker genes in the sense of generalized fitting scores. Heat map of the maximal-intensity projections 
of the (normalized) sum of the expression energies of the genes.}
\label{fig:genFitCortex}
\end{figure}     
Midbrain is a brain region for which the generalized eigenvector gives a much better visual
impression of the whole structure than the best-localized gene. The generalized fitting score, in the case 
of midbrain, gets rid of much of the signal outside the region but does not achieve much homogeneity inside.
 Generally speaking, and not too
surprisingly, the sparse sets of genes are much less impressive markers than the generalized eigenvectors,
as they rely on much fewer degrees of freedom for optimization. On the other hand, 
the best individual marker returned by the global fitting criterion is much more convincing than the one returned 
by the localization score.

\subsection{Good separators}
\begin{figure}
\section{Appendix: Best separators} 
\includegraphics[width=2in,keepaspectratio]{regionProfile6.eps}
\caption{Characteristic function of the reunion of the striatum and the cerebellum.}
\begin{tabular}{|l|l|l|l|}
\hline
&\textbf{Fitting rank}&\textbf{Localization rank}&\textbf{Heat map}\\\hline
\textbf{Slc32a1}&2327&218&\includegraphics[width=2.25in,keepaspectratio]{separatorProfileStriatum1.eps}\\\hline
\textbf{Abat}&979&2997&\includegraphics[width=2.25in,keepaspectratio]{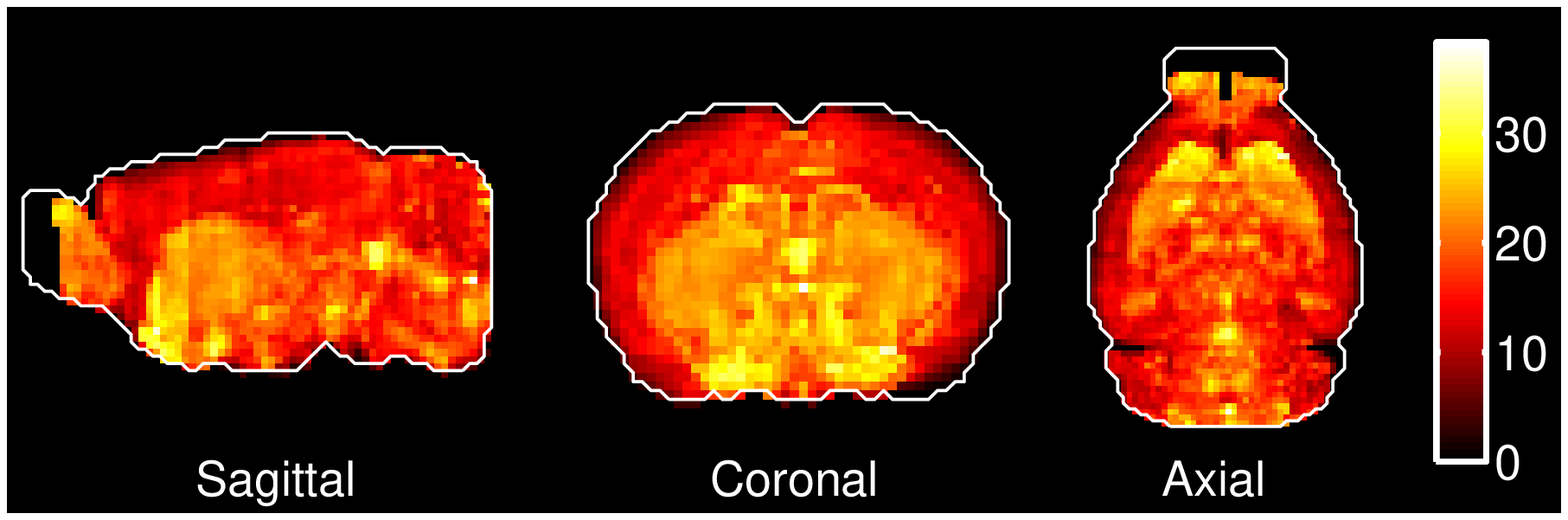}\\\hline
\textbf{Ptpn5}&634&2056&\includegraphics[width=2.25in,keepaspectratio]{separatorProfileStriatum3.eps}\\\hline
\textbf{Serpine2}&107&807&\includegraphics[width=2.25in,keepaspectratio]{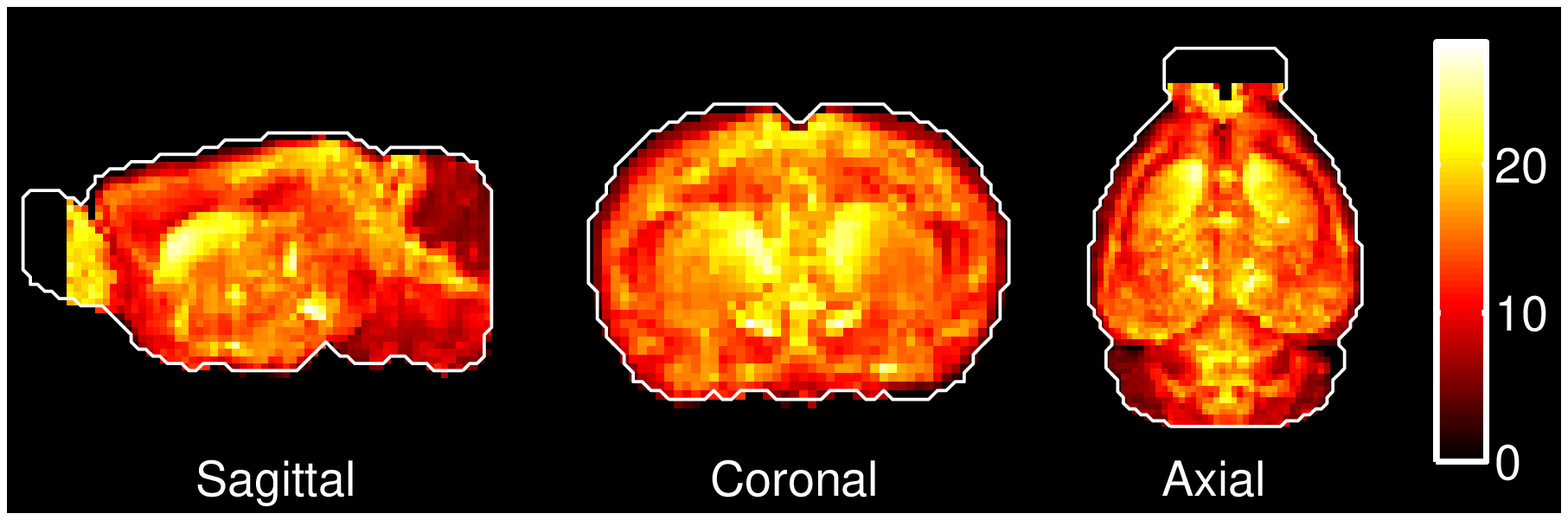}\\\hline
\textbf{Gad2}&898&2267&\includegraphics[width=2.25in,keepaspectratio]{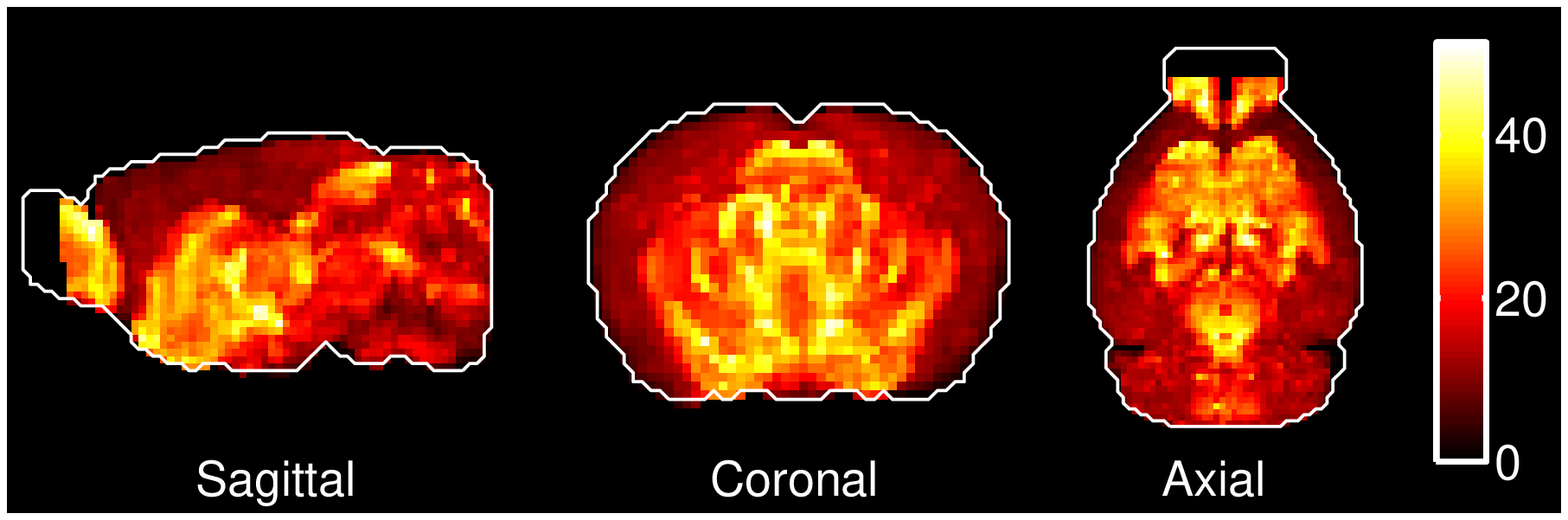}\\\hline
\textbf{Alcam}&651&424&\includegraphics[width=2.25in,keepaspectratio]{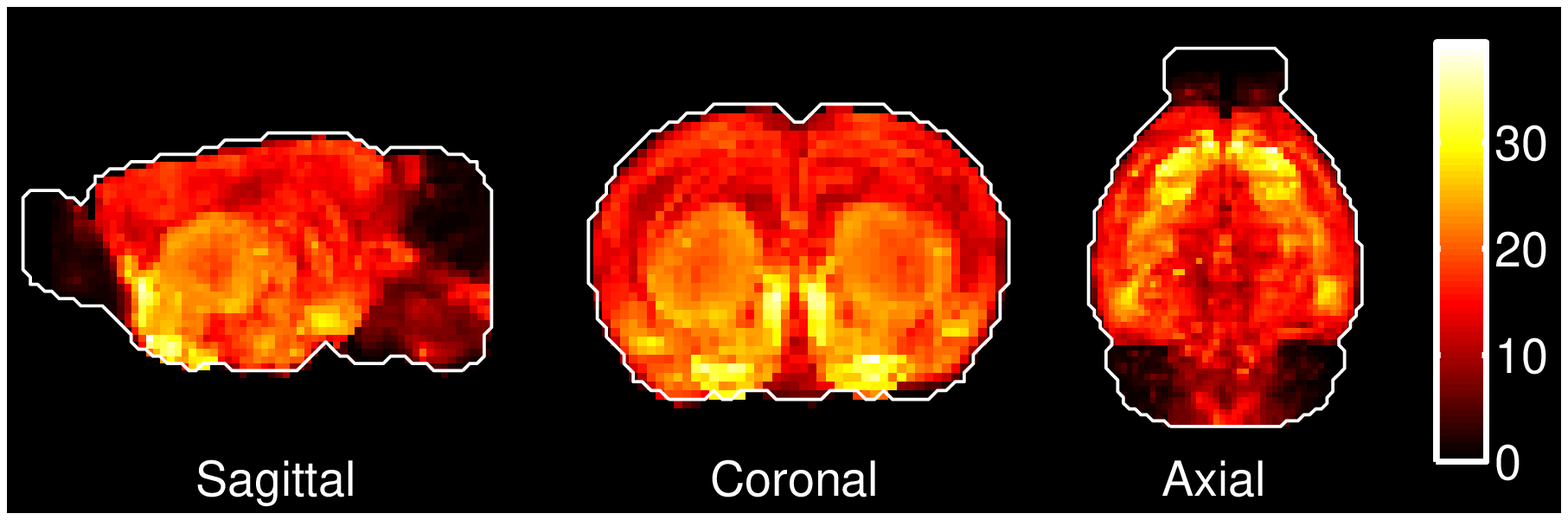}\\\hline
\textbf{Pcp4l1}&2839&968&\includegraphics[width=2.25in,keepaspectratio]{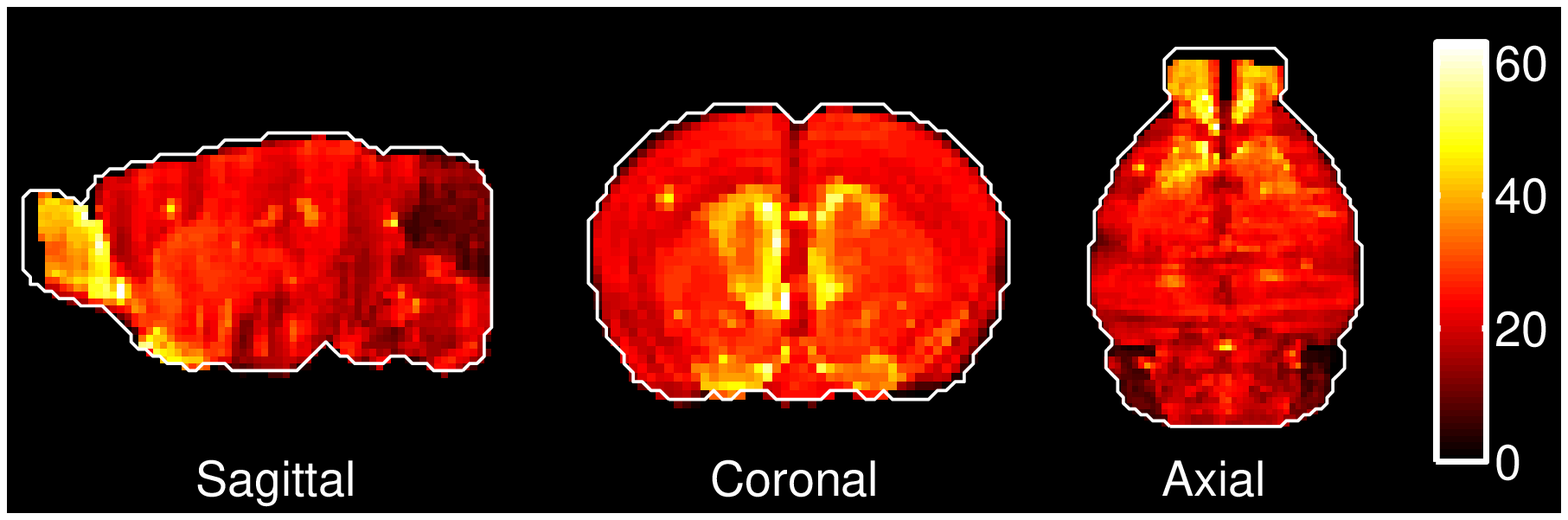}\\\hline
\textbf{Grik2}&70&629&\includegraphics[width=2.25in,keepaspectratio]{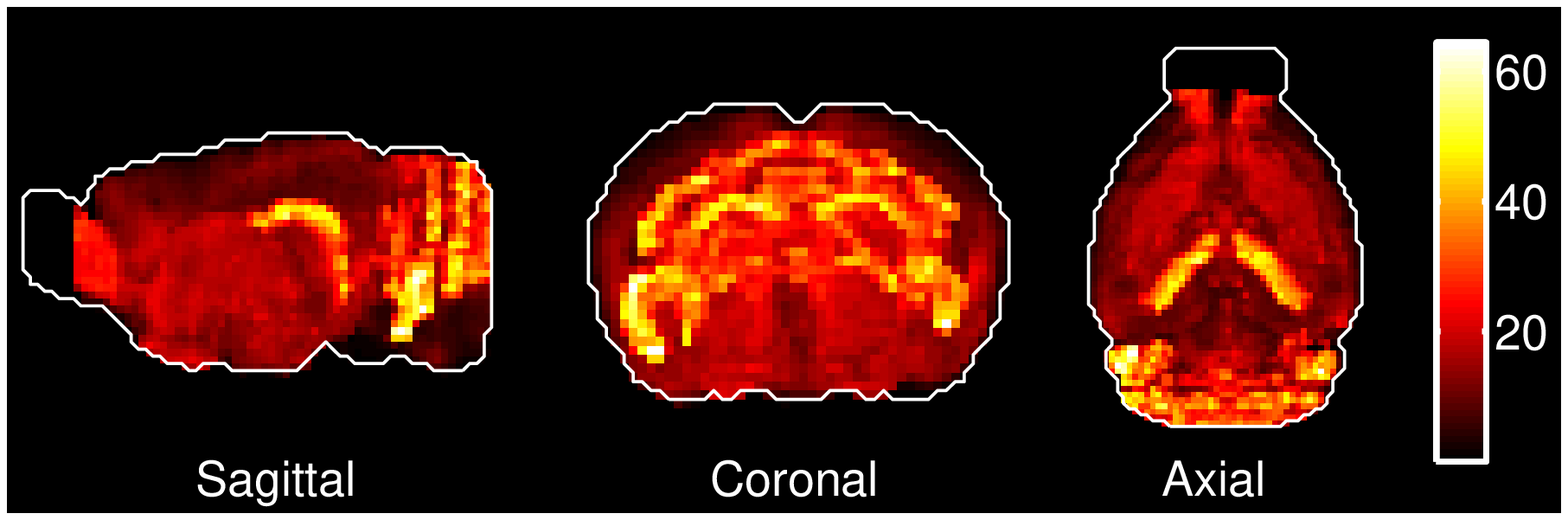}\\\hline
\textbf{Ckb}&1733&2794&\includegraphics[width=2.25in,keepaspectratio]{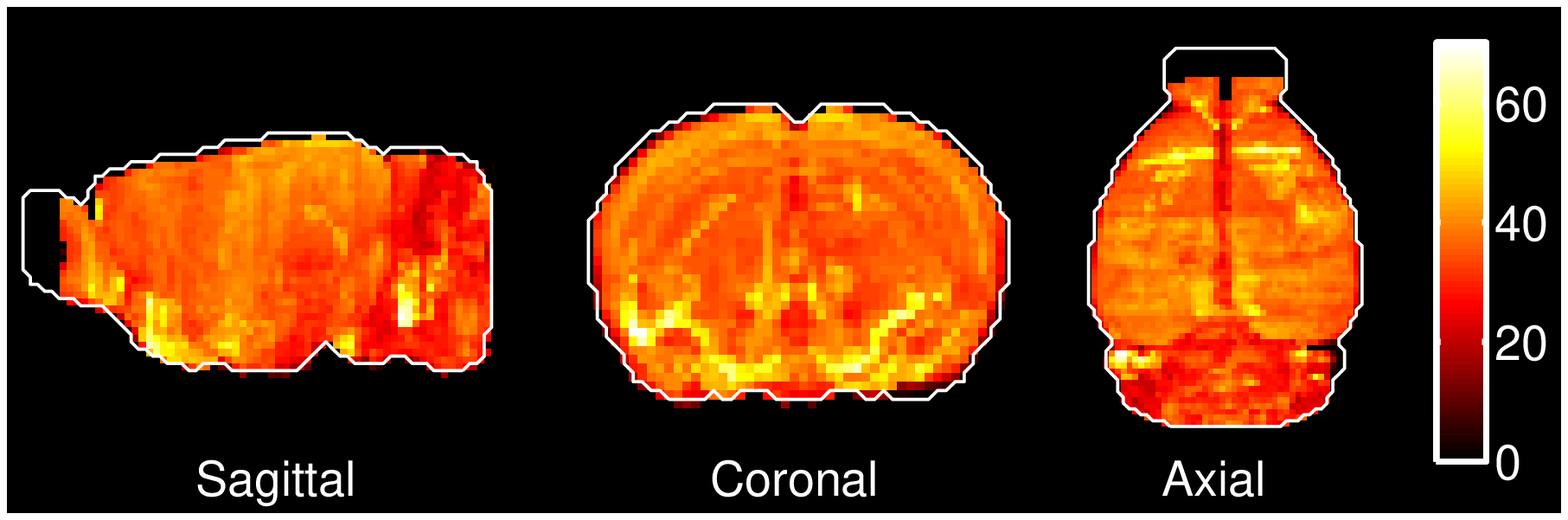}\\\hline
\textbf{Gad1}&861&722&\includegraphics[width=2.25in,keepaspectratio]{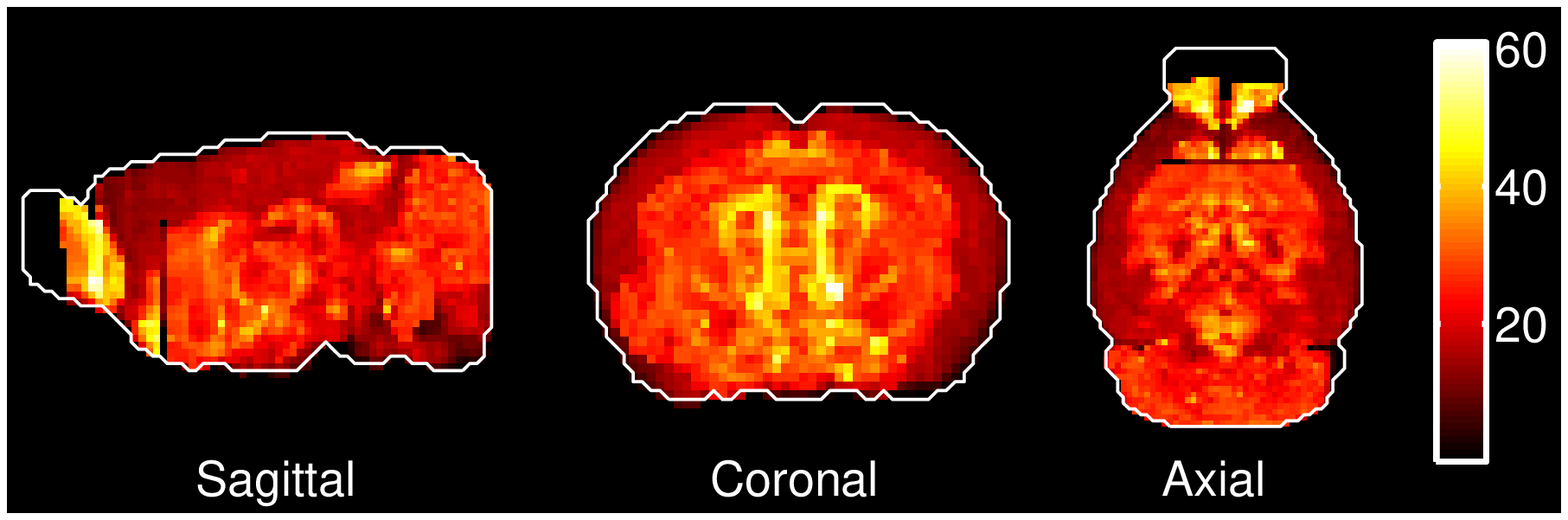}\\\hline
\end{tabular}

\caption{The best few separators of the striatum, with their rankings according to the
global fitting and localization criteria. It turns out that the genes were rescued from low ranks
by the local algorithms, and most of them clearly show the profile of the striatum.}
\label{fig:separatorStriatum}
\end{figure}

For every region in the coarsest annotation of the left hemisphere,
we ran an algorithm with a range of values of the internal an 
external paramater for the model function and for the
local mask. As the boundary of the striatum does not have too much overlap
with the boundary of the brain, it is easier to visualize than the cerebral cortex in a
maximum-intensity  projection and we chose it for illustration (see figure (\ref{fig:separatorStriatum}) 
for the first 10 genes returned by the algorithm, none of which has better rank than 218 for localization and 70).\\

\begin{figure}
\centering
\includegraphics[width=5in,keepaspectratio]{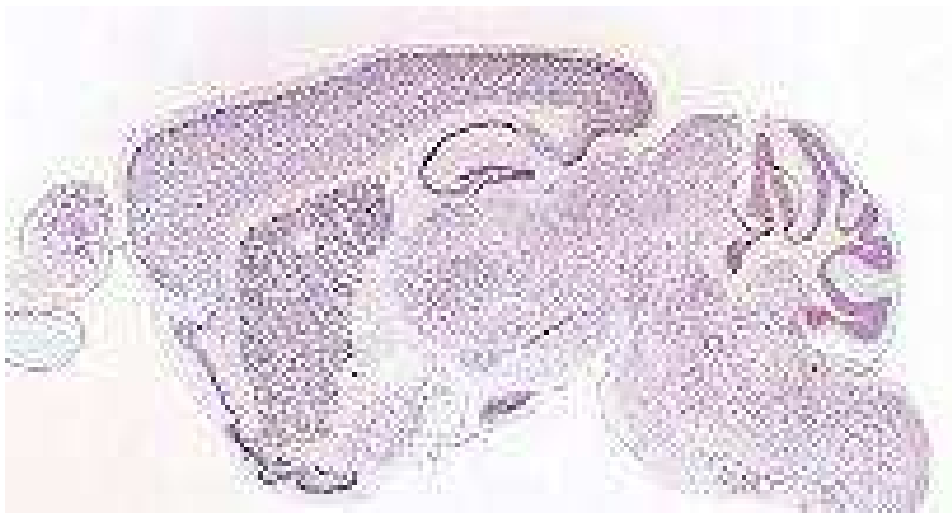}
\caption{{\bf{Ptpn5: ISH on a sagittal section for Ptpn5.}} Ptpn5 is detected as the third best local marker of striatum.
Digitized expression energies suggested that this gene is also highly expressed in the cerebral cortex. The two structures
are indeed visible and separable.}
\label{fig:Ptpn5Sag}
\end{figure}

Since the maximum-intensity projection can hide some well-sepearated regions, it is
not as reliable as sections to evaluate separation properties,
but still it is instructive to see how this local criterion can return genes that score 
 low for localization and/or fitting but still have a distinct pattern around striatum.
 Slc32a1 has a clear but inhomogeneous pattern in striatum, and a high expression in the main olfactory bulb. 
Both features penalize the global fitting score, but only the second one penalizes the global localization score, which
is consistent with the fitting rank being much lower than the localization rank.
 Ptpn5 exhibits a less contrasted but more homogeneous pattern in the striatum (it rather follows the caudoputamen
than the striatum), but the expression in also quite high in the cerebral cortex. Caudoputamen is still
striking and the gene was rescued by the local algorithm, even though the expression 
in the cerebral cortex severely penalizes Ptpn5 both for glocal localization (according to which it is ranked 2056) and global fitting (according to which it is ranked 634). It would be interesting, when repeating the experiment 
for the same gene, either in the same species or in different species, to see if 
the local separation property is as the global properties measured by the localization and fitting scores. 
\begin{figure}
\centering
\includegraphics[width=4in,keepaspectratio]{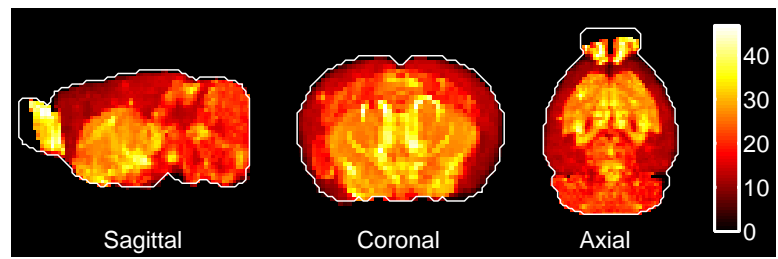}
\caption{{\bf{Best separator of striatum.}} Heat map of the maximal-intensity projection of Slc32a1.}
\label{fig:coMarkerStriCer1}
\end{figure}
\begin{figure}
\centering
\includegraphics[width=4in,keepaspectratio]{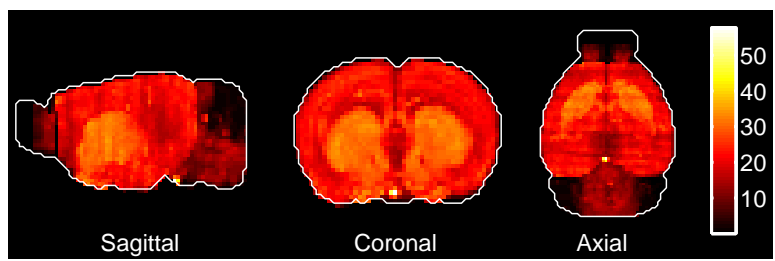}
\caption{{\bf{Third best separator of striatum.}} Heat map of the maximal-intensity projection of Ptpn5.}
\label{fig:coMarkerStriCer2}
\end{figure}   
 A sagittal section drawn from the ISH data (figure (\ref{fig:Ptpn5Sag})) confirms that 
Ptpn5 is highly expressed in the striatum, but also in the cortex, albeit to a lesser extent. The
separation between cerebral cortex and striatum is clearly visible on the section. This is 
the separation property that our local criterion is supposed to detect.  

\subsection{Good co-markers}

\begin{figure}
\section{Appendix: Best co-markers} 
\includegraphics[width=2in,keepaspectratio]{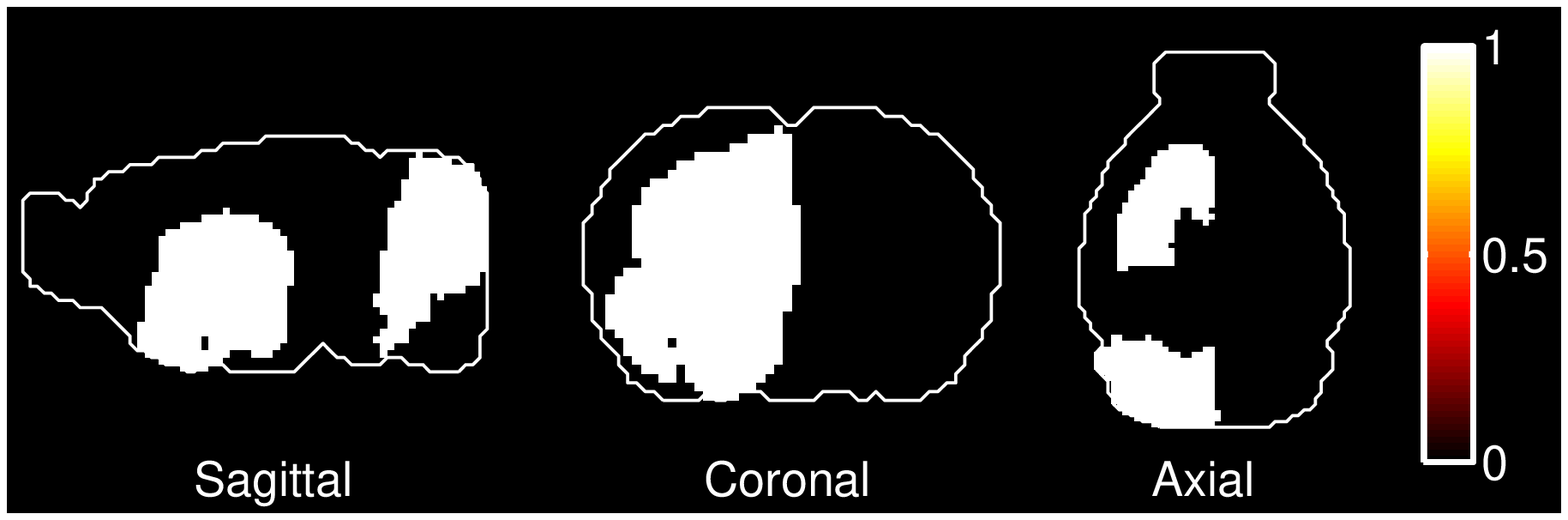}
\caption{Characteristic function of the reunion of the striatum and the cerebellum.}
\begin{tabular}{|l|l|l|l|}
\hline
&\textbf{Score for pair}&\textbf{Tangent coefficient}&\textbf{Heat map}\\\hline
\textbf{Id4}&0.54781&1.0085&\includegraphics[width=2.25in,keepaspectratio]{geneProfileForPair1big12613.eps}\\\hline
\textbf{D330017J20Rik}&0.52928&0.85764&\includegraphics[width=2.25in,keepaspectratio]{geneProfileForPair2big12613.eps}\\\hline
\textbf{Grik2}&0.45338&0.81055&\includegraphics[width=2.25in,keepaspectratio]{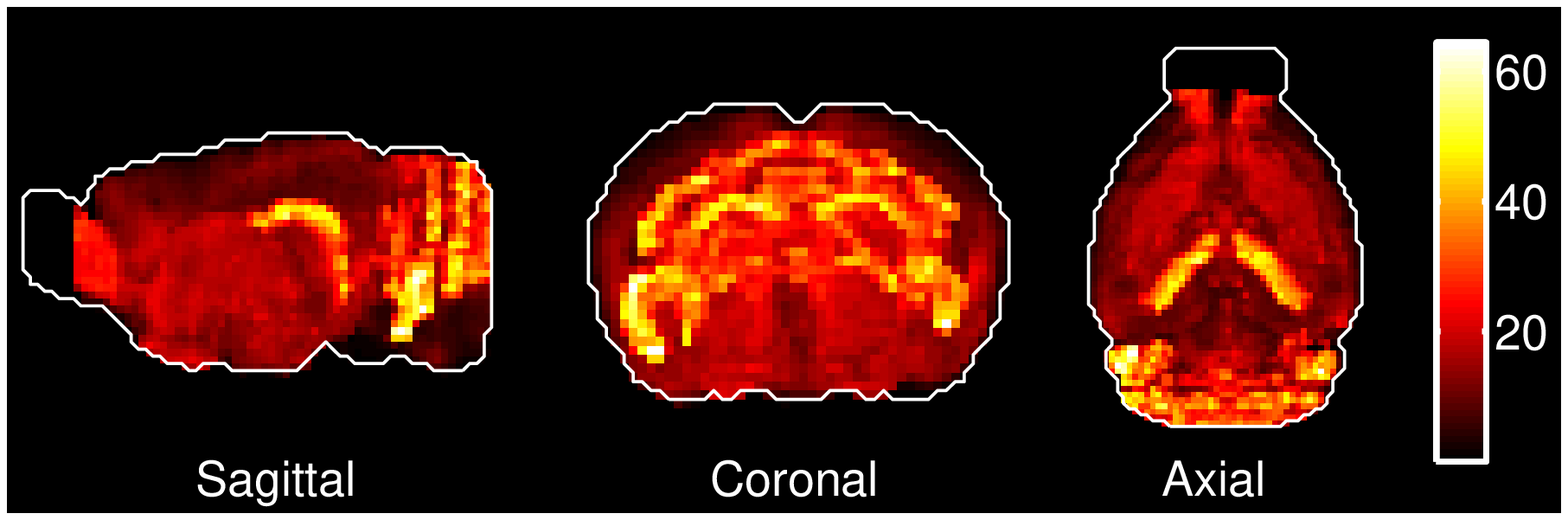}\\\hline
\textbf{Vldlr}&0.42963&0.80894&\includegraphics[width=2.25in,keepaspectratio]{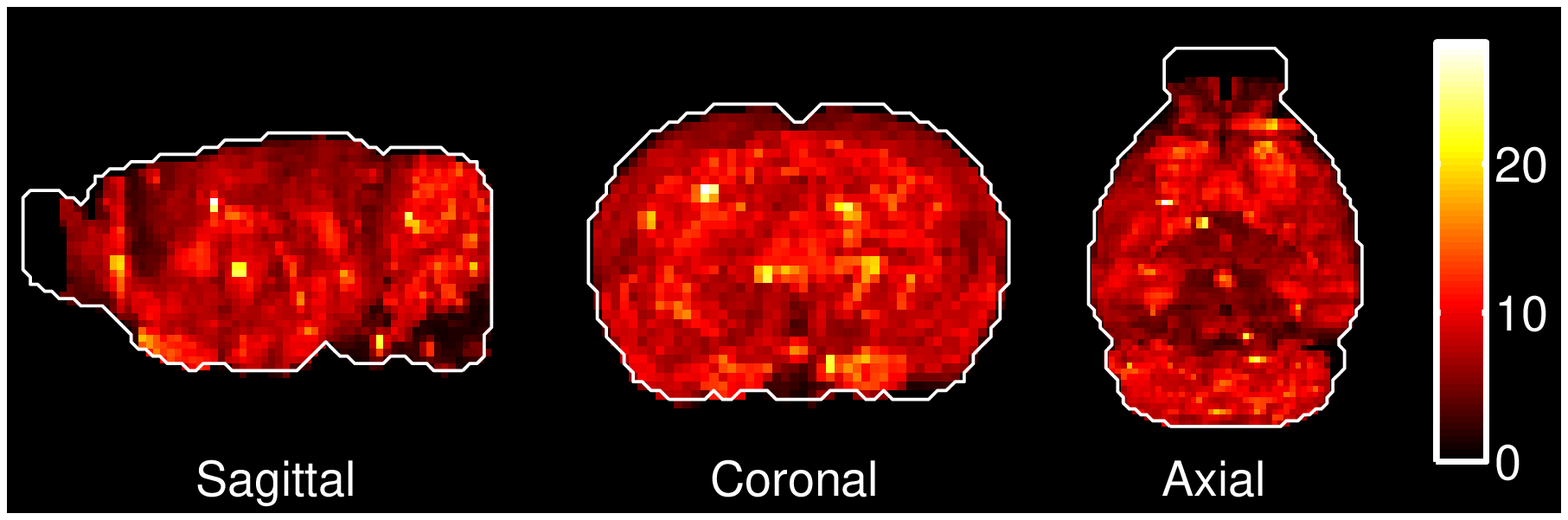}\\\hline
\textbf{8430415E04Rik}&0.42156&0.80471&\includegraphics[width=2.25in,keepaspectratio]{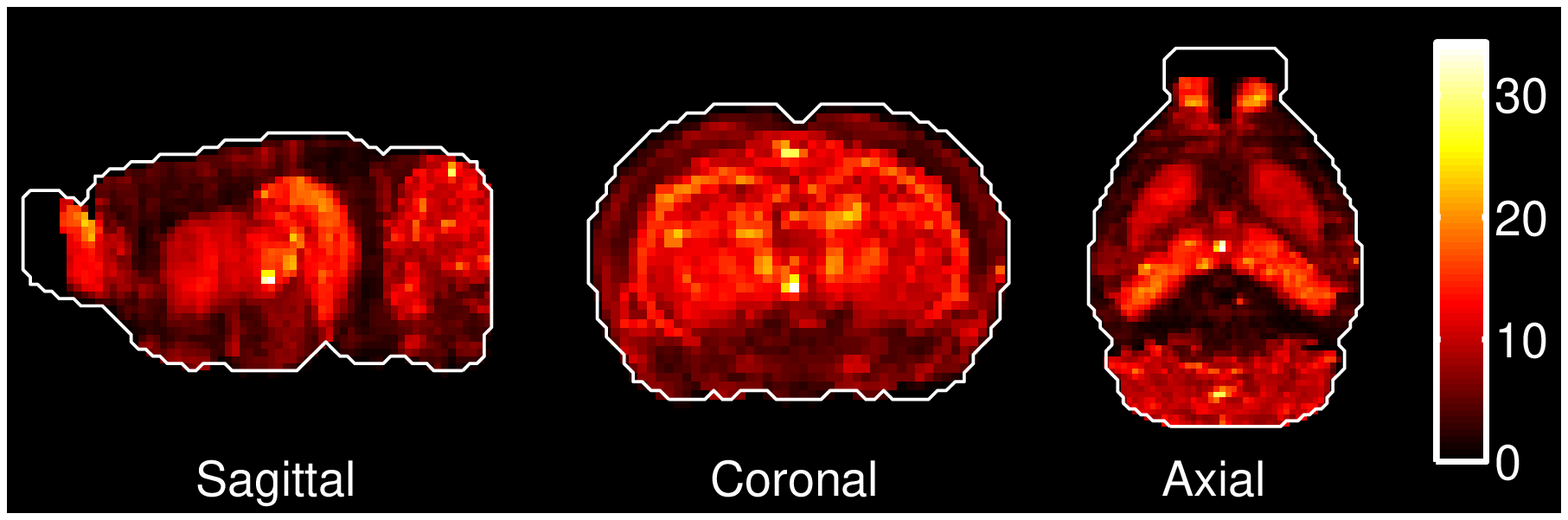}\\\hline
\textbf{Gad1}&0.41918&1.0063&\includegraphics[width=2.25in,keepaspectratio]{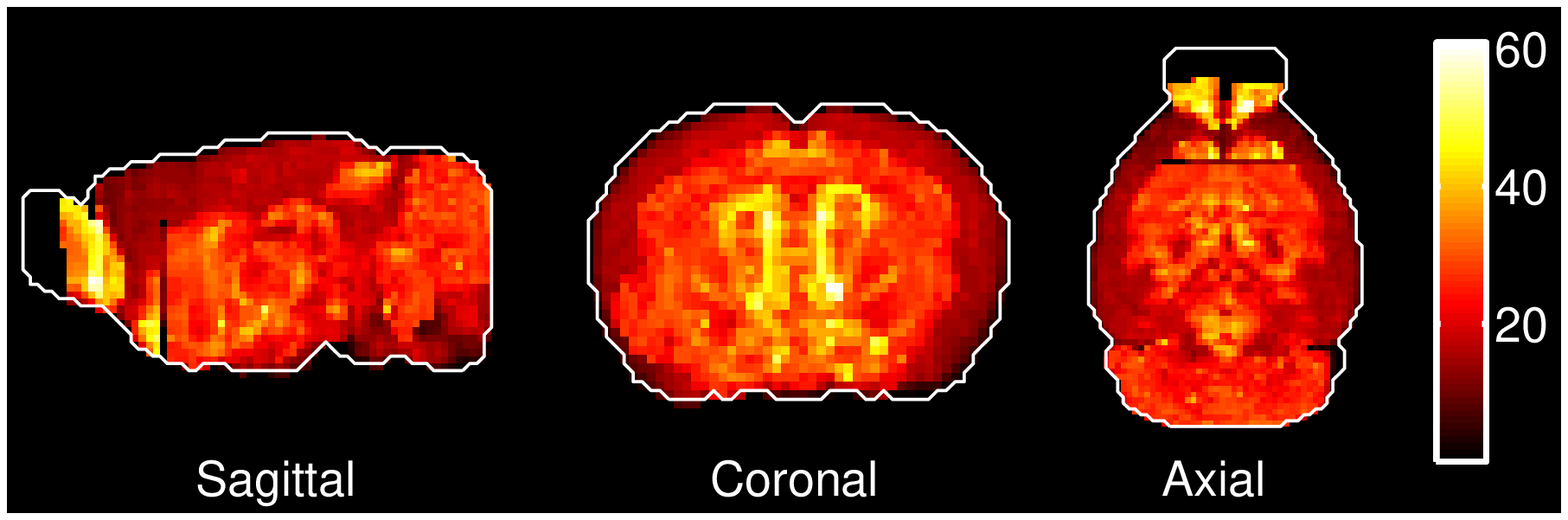}\\\hline
\textbf{Dpy19l3}&0.39955&1.1353&\includegraphics[width=2.25in,keepaspectratio]{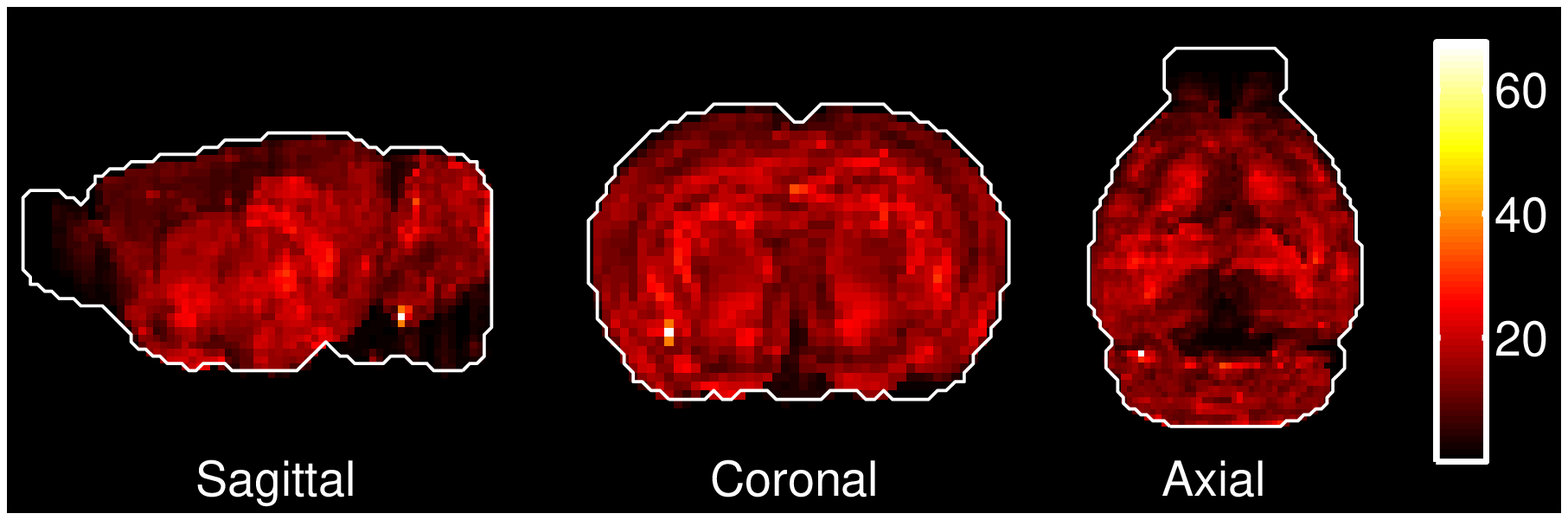}\\\hline
\textbf{D13Bwg1146e}&0.39028&1.1011&\includegraphics[width=2.25in,keepaspectratio]{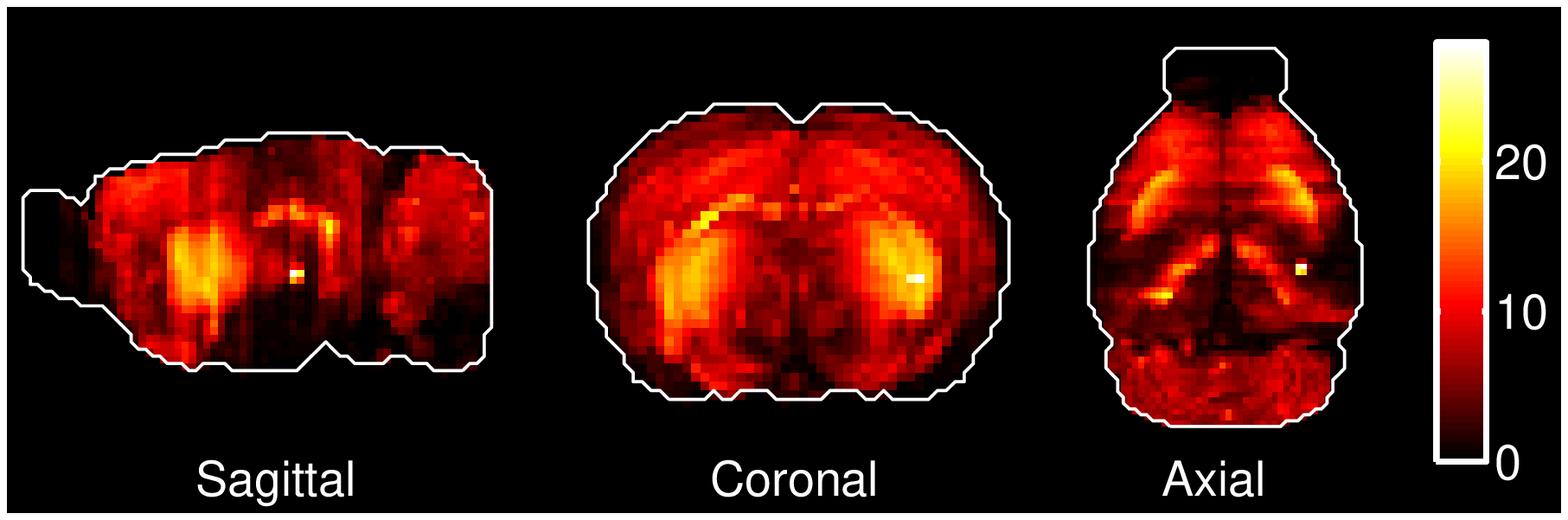}\\\hline
\textbf{Ppp2r5d}&0.37326&1.0788&\includegraphics[width=2.25in,keepaspectratio]{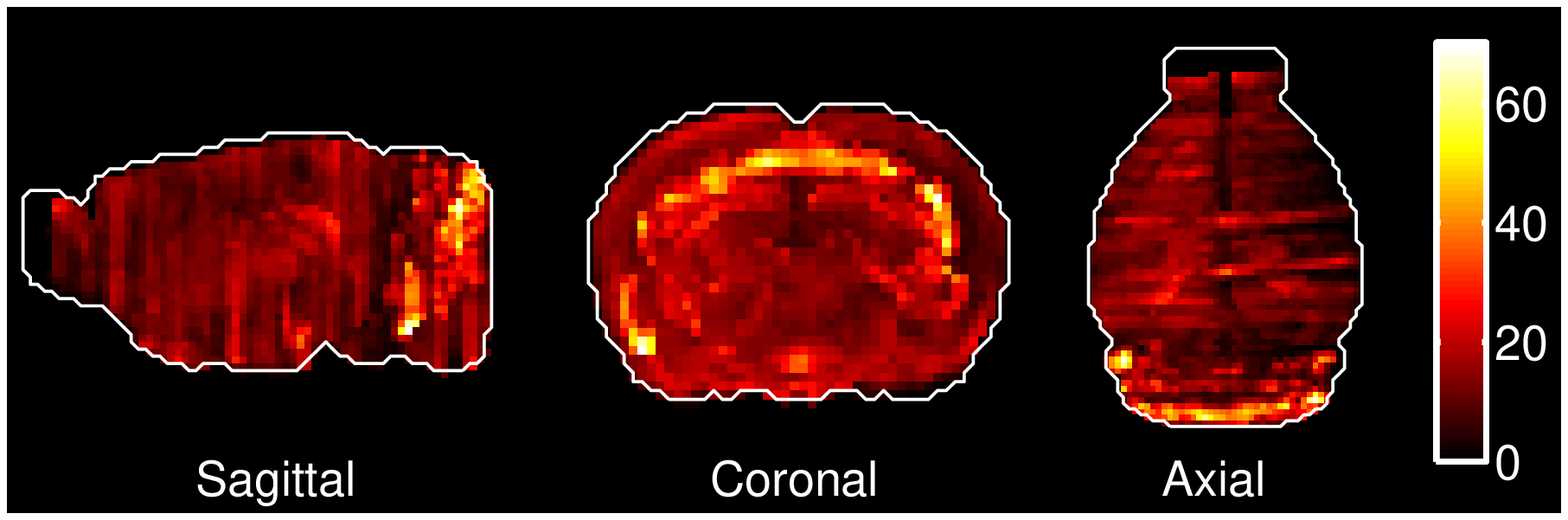}\\\hline
\textbf{Itpr1}&0.36654&0.96994&\includegraphics[width=2.25in,keepaspectratio]{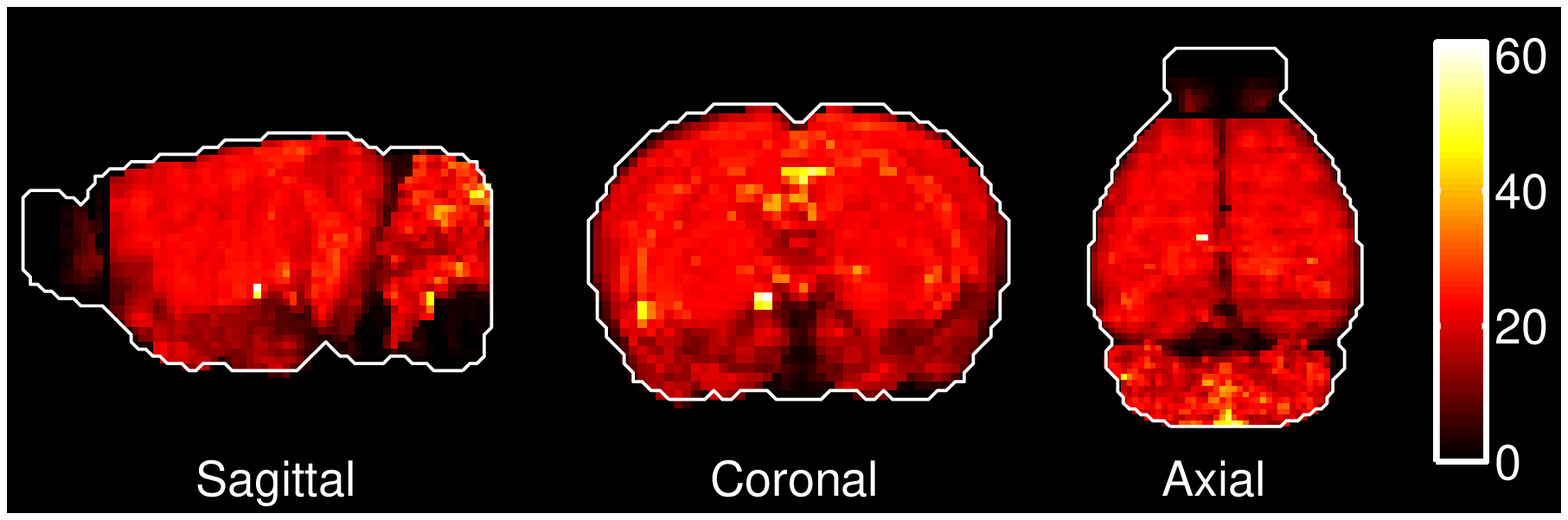}\\\hline
\end{tabular}

\caption{The best few co-markers of striatum and cerebellum.}
\label{fig:coMarkersStriCer}
\end{figure}

A table of the best few co-markers for striatum and cerebellum (see figure(\ref{fig:coMarkersStriCer})) can be found in an 
appendix. It was obtained at a value $\tau=0.5$ of the $\tau$-balance constraint defined above. This 
value is somewhat arbitrary and and by the look of the tangent coefficients
and fitting scores for the genes, there is no monitonicity of
the value of the tangent wrt the value of the score. The user of the software can input a higher value of $\tau$
in order to explore genes with a higher tolerance on the relative fittings. We 
do not have a natural optimization criterion to propose to choose $\tau$ 
and therefore leave it as a parameter. It can be noted, however, than the $\tau$-balance constraint
at $\tau = 0.5$ for pallidum and cerebellum returns an empty set of markers.
Thus, fixing the level of the balance constraint and counting the number of marker genes
returned by the algorithm suggests an indication on the degree of solidarity between pairs of regions.\\
 
The best two co-markers of the cerebellum and the striatum are Id4 (see figure (\ref{fig:coMarkerStriCer1})) and D330017J20Rik 
(see figure (\ref{fig:coMarkerStriCer2})). The first one
has a value of tangent very close to one, the second has  a value of tangent close t0 0.86. Id4 has indeed a more homogeneous
expression across striatum and cerebellum, and D330017J20Rik has a pattern of higher expression inside cerebellum, hence a tangent
more remote from one, but the two genes show a clear pattern for the pair striatum-cerebellum.
\begin{figure}
\centering
\includegraphics[width=4in,keepaspectratio]{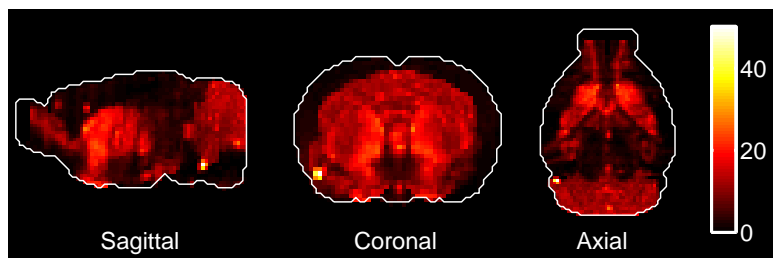}
\caption{{\bf{Best co-marker of striatum and cerebellum.}} Heat map of the maximal-intensity projection of Id4.}
\label{fig:coMarkerStriCer1}
\end{figure}
\begin{figure}
\centering
\includegraphics[width=4in,keepaspectratio]{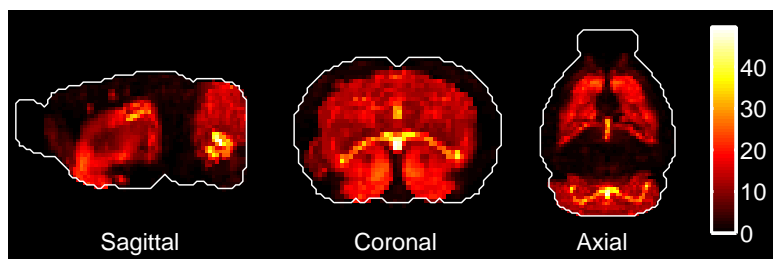}
\caption{{\bf{Second Best co-marker of striatum and cerebellum.}} Heat map of the maximal-intensity projection of D330017J20Rik.}
\label{fig:coMarkerStriCer2}
\end{figure}



\section*{Acknowledgments}
 We thank
Claudio Mello and Peter Lovell (Oregon Health and Science University)
for discussions and correspondence. This research is supported by the NIDA Grant
1R21DA027644-01, {\it{Co-expression networks of addiction-related genes in the mouse and human brain}}.

\newpage
\newpage
\begin{figure}
\section{Appendix: Best-localized genes and characteristic functions for the 12 main regions of the left hemisphere} 
\begin{tabular}{|l|l|l|l|}
\hline
{\bf{Brain Region}}&{\bf{Best marker}}&{\bf{Best-localized gene}}&{\bf{Region profile}}\\\hline
Cerebral cortex&Pak7&\includegraphics[width=2in,keepaspectratio]{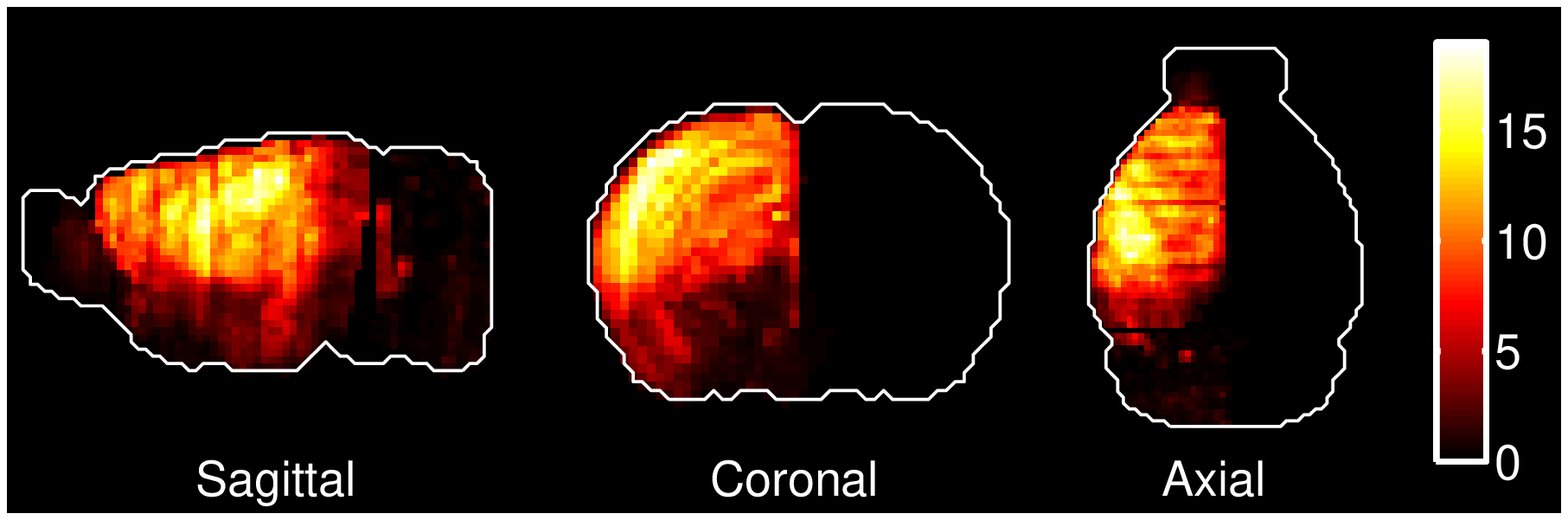}&\includegraphics[width=2in,keepaspectratio]{regionProfile1.eps}\\\hline
Olfactory areas&C230040D10Rik&\includegraphics[width=2in,keepaspectratio]{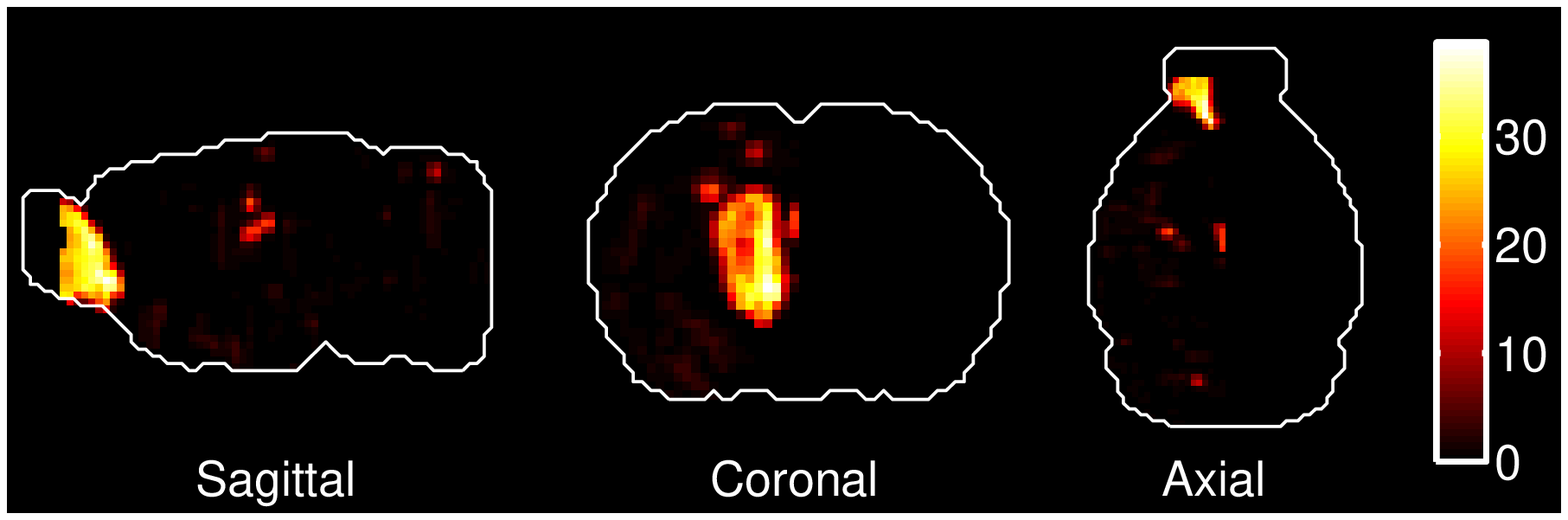}&\includegraphics[width=2in,keepaspectratio]{regionProfile3.eps}\\\hline
Hippocampal region&Prox1&\includegraphics[width=2in,keepaspectratio]{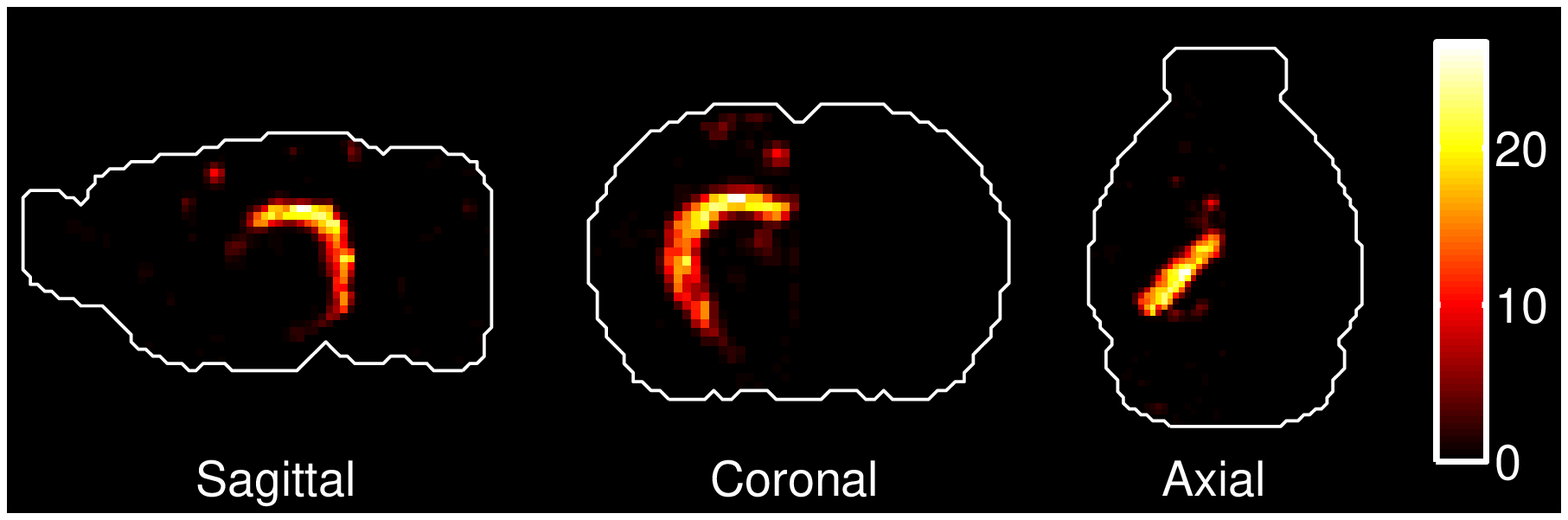}&\includegraphics[width=2in,keepaspectratio]{regionProfile4.eps}\\\hline
Retrohippocampal region&Rxfp1&\includegraphics[width=2in,keepaspectratio]{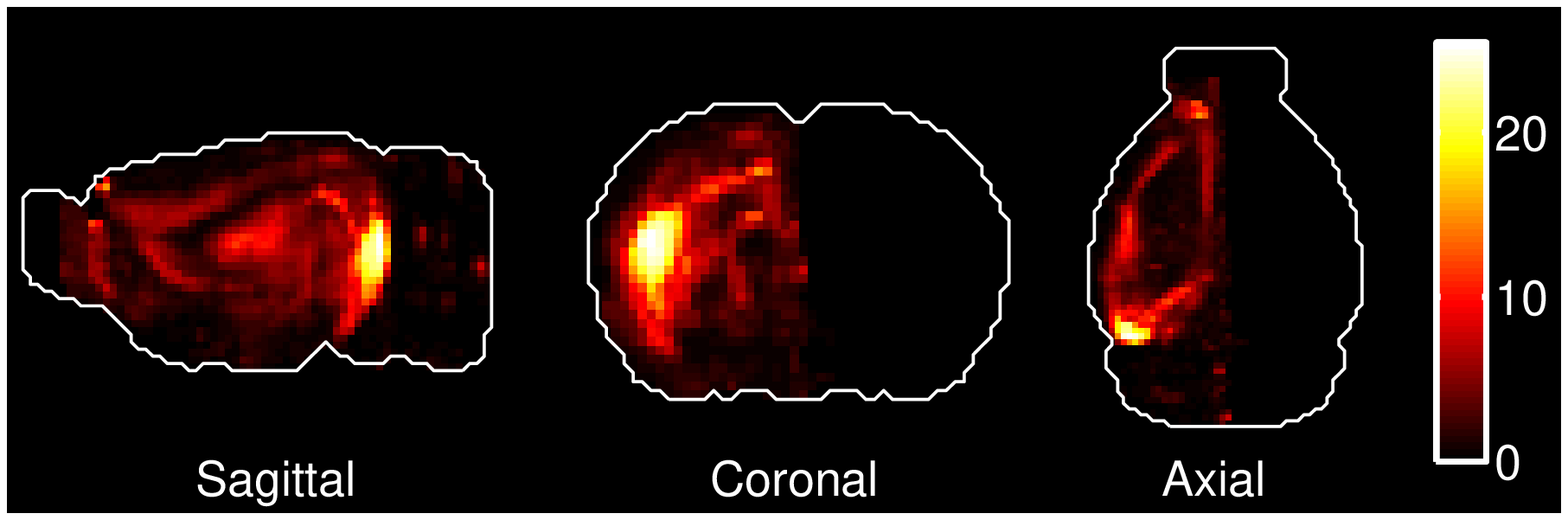}&\includegraphics[width=2in,keepaspectratio]{regionProfile5.eps}\\\hline
Striatum&Ric8b&\includegraphics[width=2in,keepaspectratio]{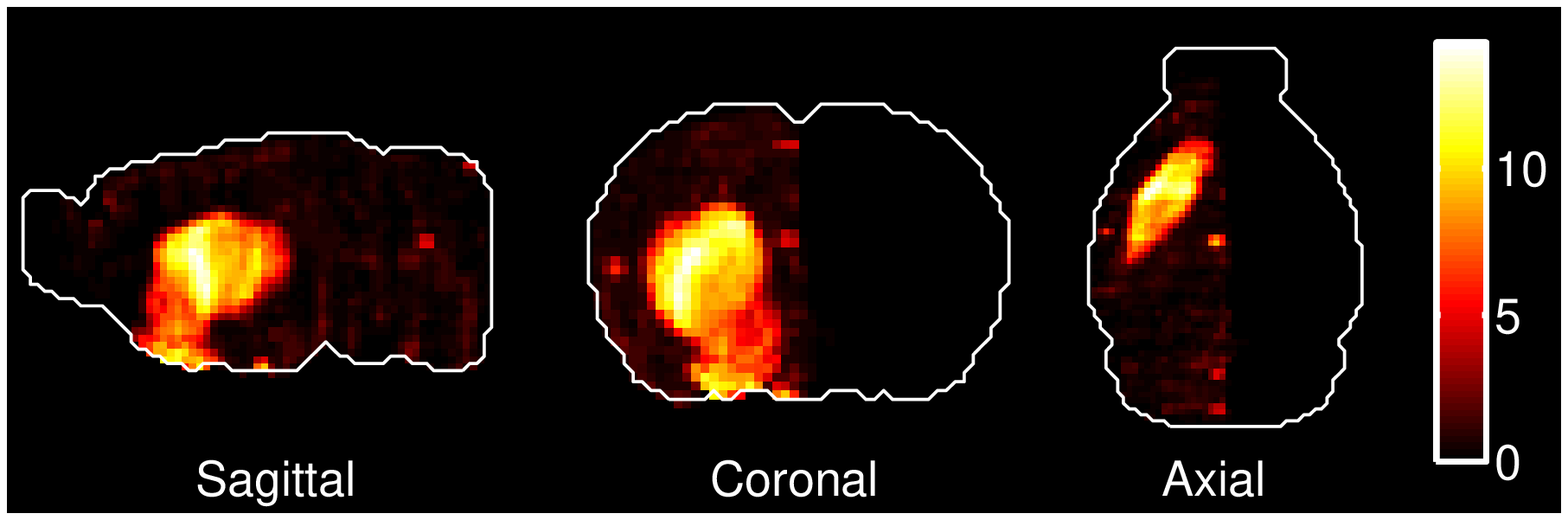}&\includegraphics[width=2in,keepaspectratio]{regionProfile6.eps}\\\hline
Pallidum&Ebf4&\includegraphics[width=2in,keepaspectratio]{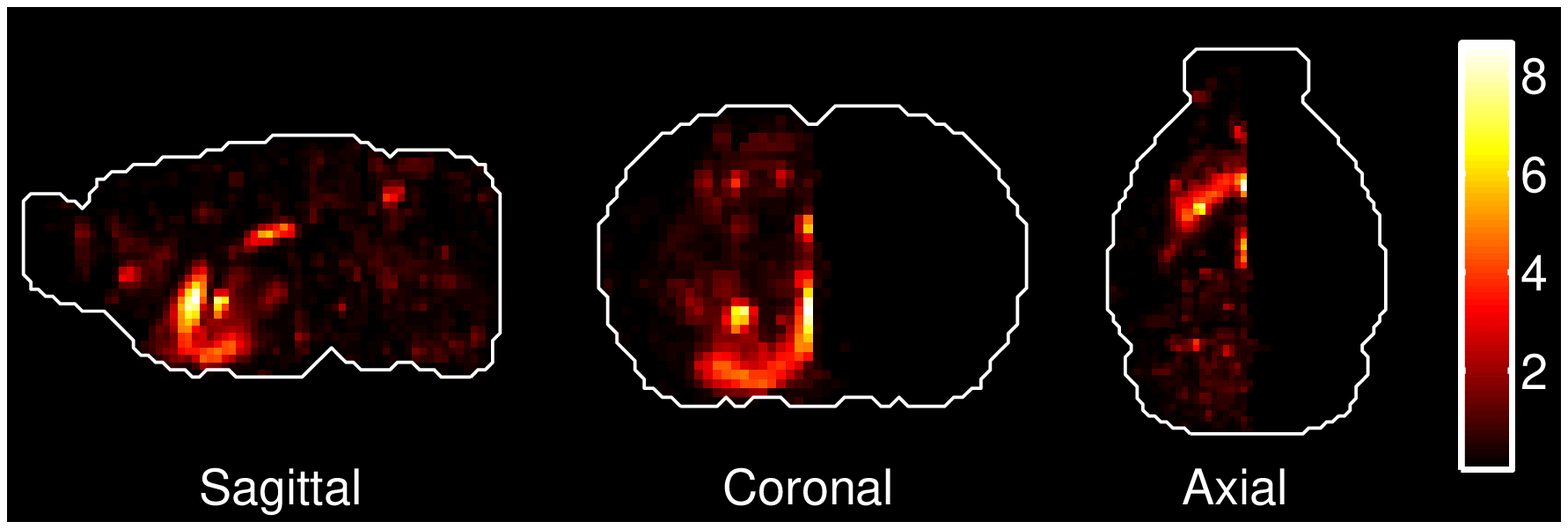}&\includegraphics[width=2in,keepaspectratio]{regionProfile7.eps}\\\hline
Thalamus&Lef1&\includegraphics[width=2in,keepaspectratio]{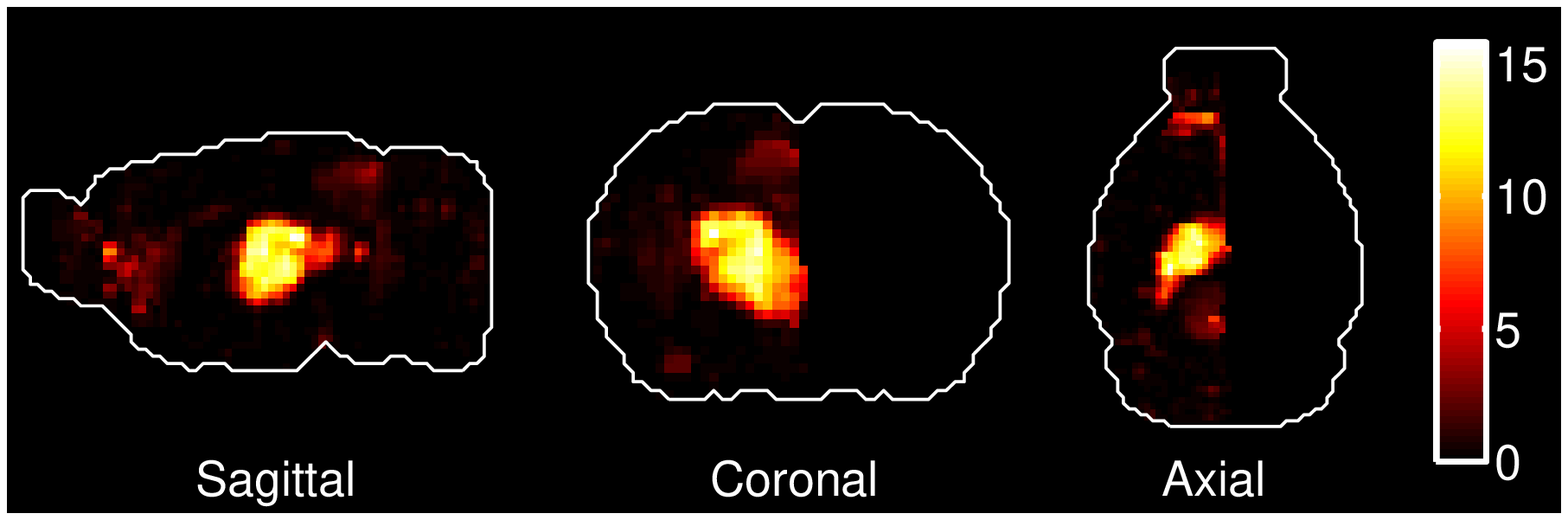}&\includegraphics[width=2in,keepaspectratio]{regionProfile8.eps}\\\hline
Hypothalamus&Nr5a1&\includegraphics[width=2in,keepaspectratio]{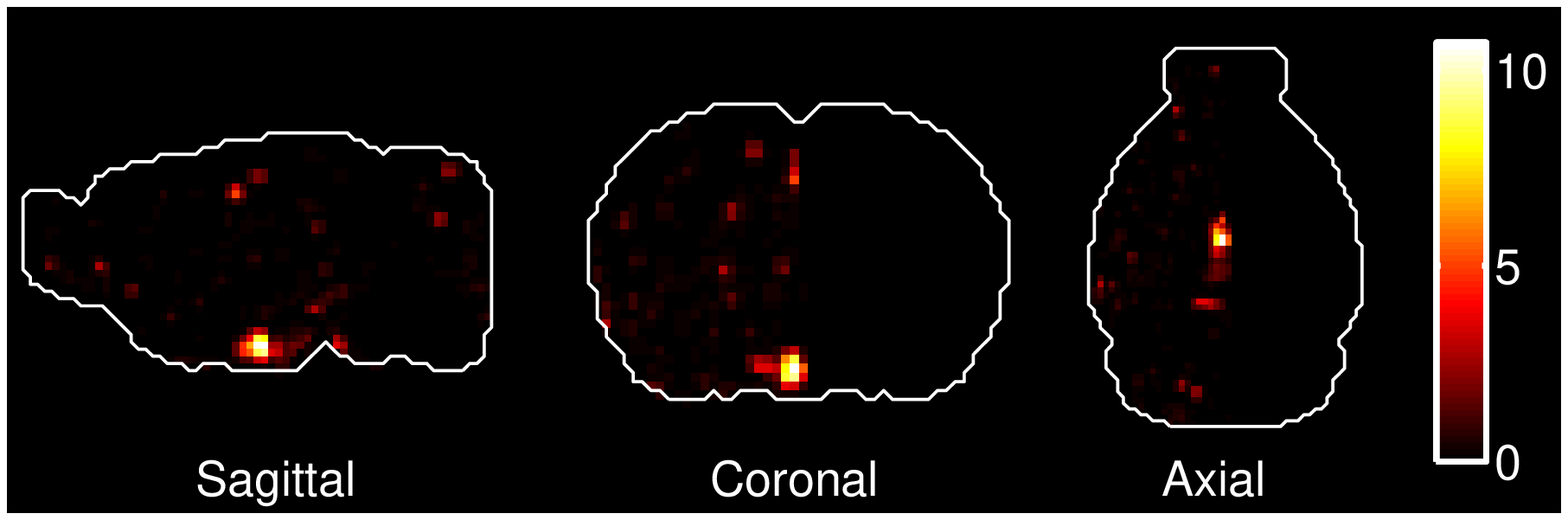}&\includegraphics[width=2in,keepaspectratio]{regionProfile9.eps}\\\hline
Midbrain&Ntsr1&\includegraphics[width=2in,keepaspectratio]{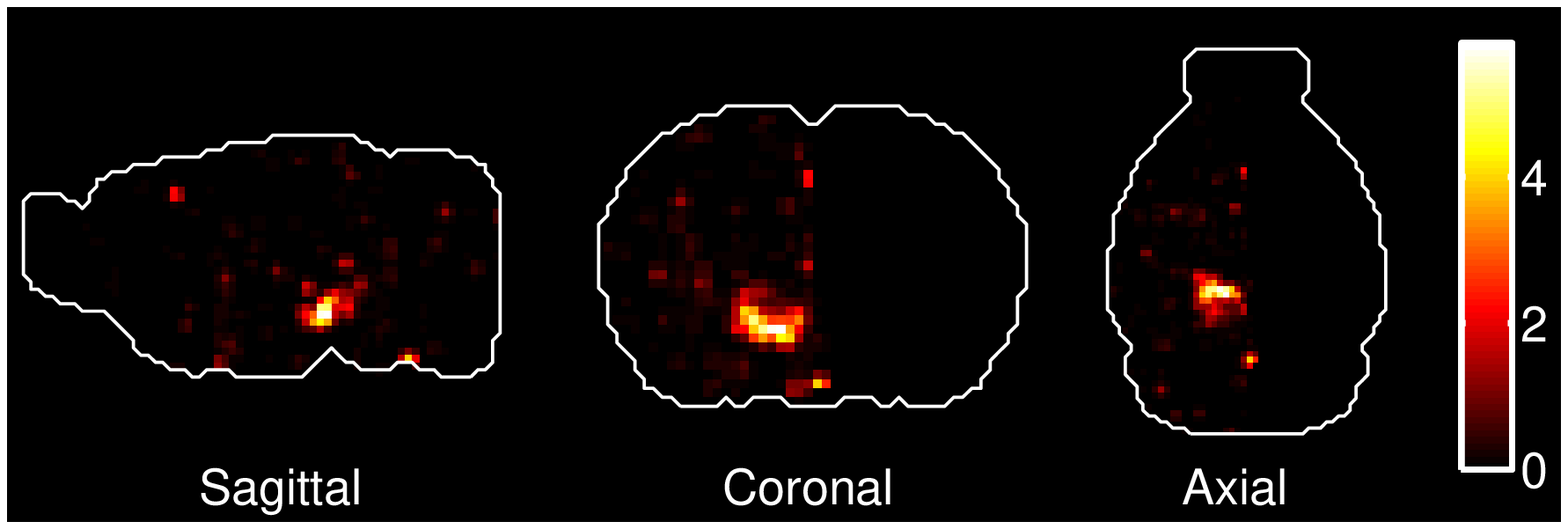}&\includegraphics[width=2in,keepaspectratio]{regionProfile10.eps}\\\hline
Pons&Dbh&\includegraphics[width=2in,keepaspectratio]{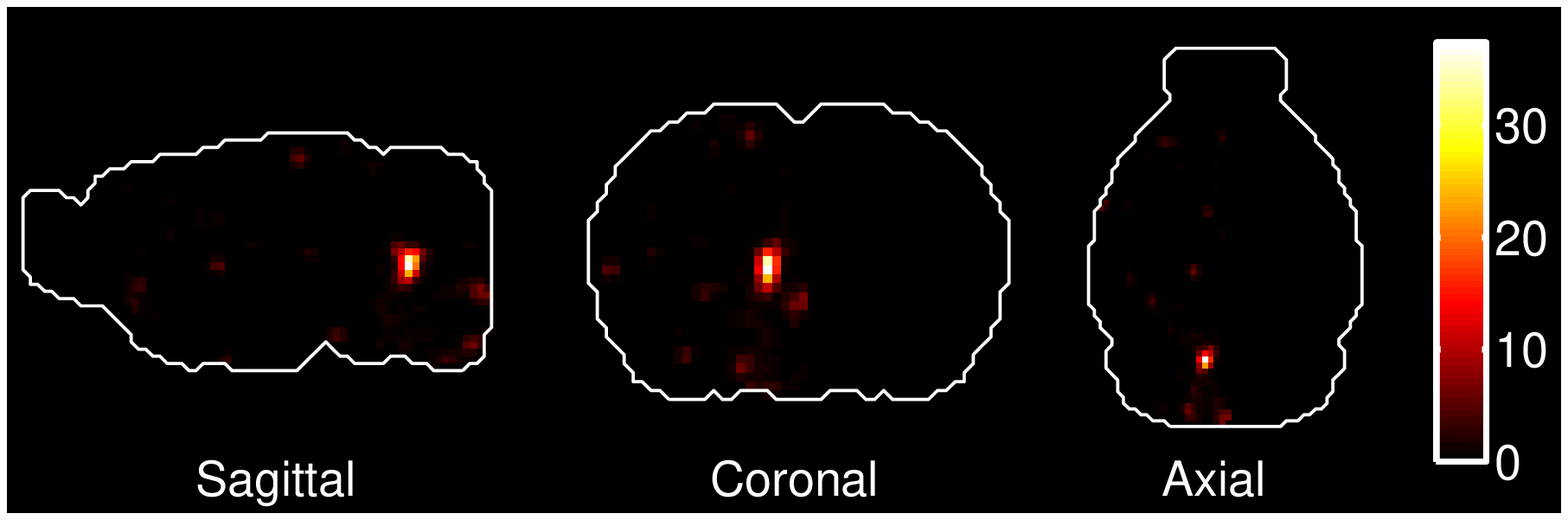}&\includegraphics[width=2in,keepaspectratio]{regionProfile11.eps}\\\hline
Medulla&Vil2&\includegraphics[width=2in,keepaspectratio]{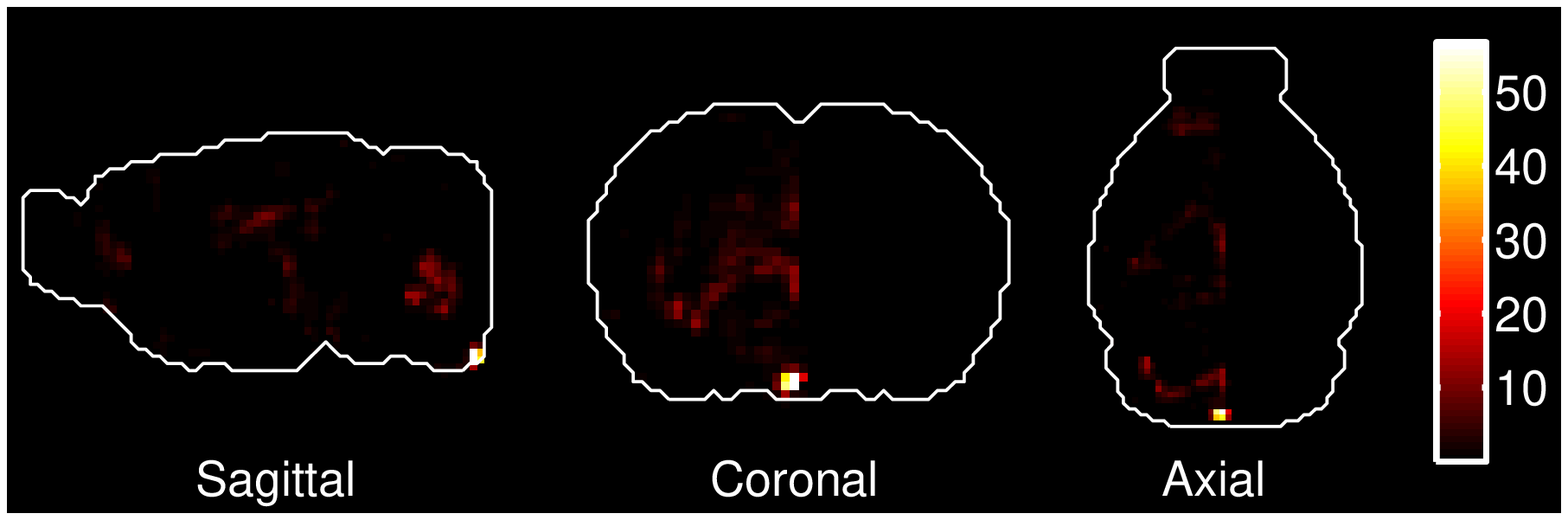}&\includegraphics[width=2in,keepaspectratio]{regionProfile12.eps}\\\hline
Cerebellum&Gabra6&\includegraphics[width=2in,keepaspectratio]{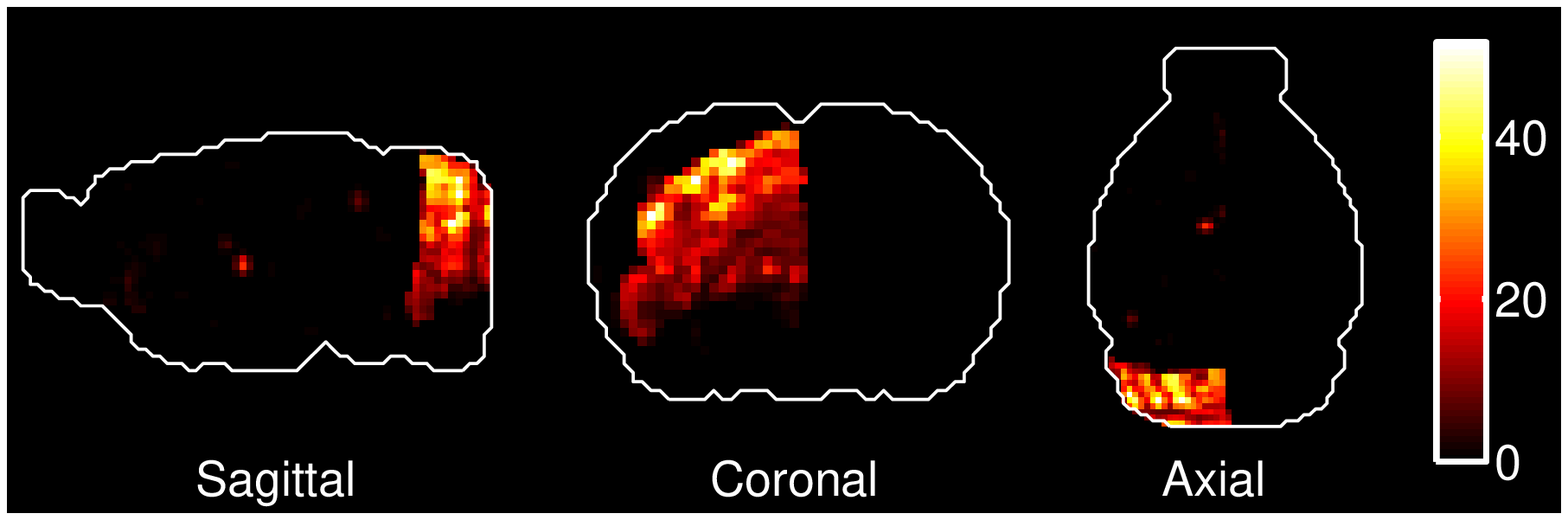}&\includegraphics[width=2in,keepaspectratio]{regionProfile13.eps}\\
\hline
\end{tabular}
\caption{The best few genes by localization for each of the 12 main regions of the left hemisphere.}
\label{fig:localizedTableBig12}
\end{figure}

\newpage
\begin{figure}
\section{Appendix: Best-fitted genes and characteristic functions for the 12 main regions of the left hemisphere} 
\begin{tabular}{|l|l|l|l|}
\hline
\textbf{Brain region}&\textbf{Best marker}&\textbf{Heat map of best marker}&\textbf{Region profile}\\\hline
Cerebral cortex&Satb2&\includegraphics[width=2in,keepaspectratio]{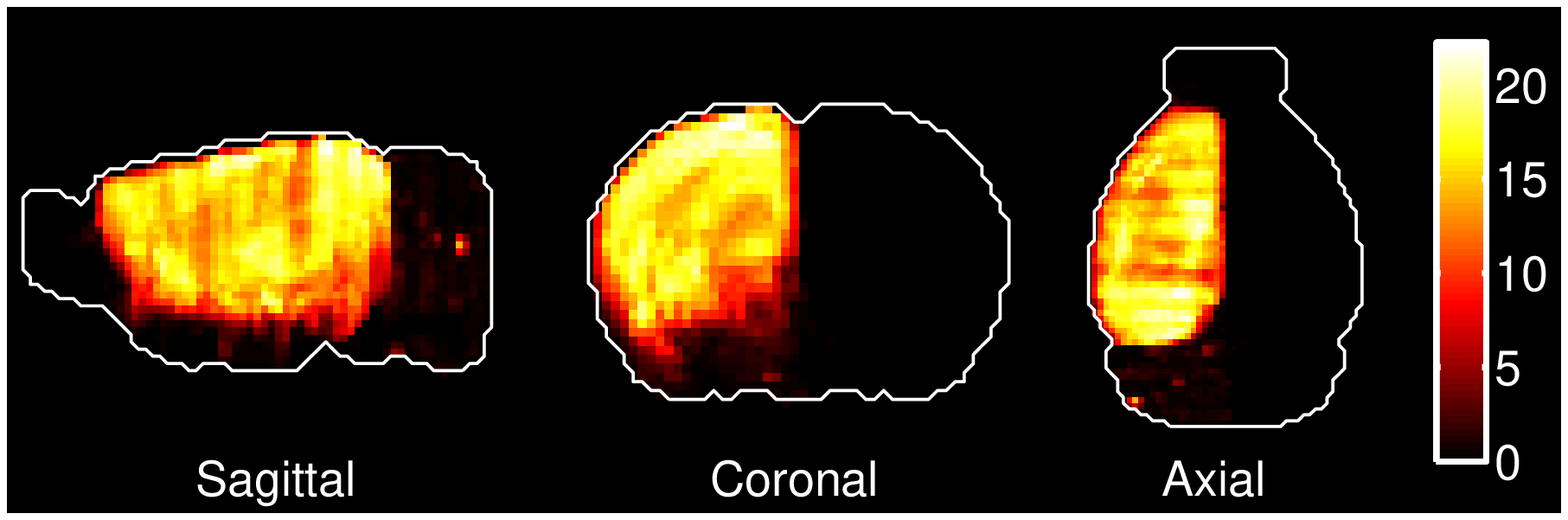}&\includegraphics[width=2in,keepaspectratio]{regionProfile2.eps}\\\hline
Olfactory areas&Ppfibp1&\includegraphics[width=2in,keepaspectratio]{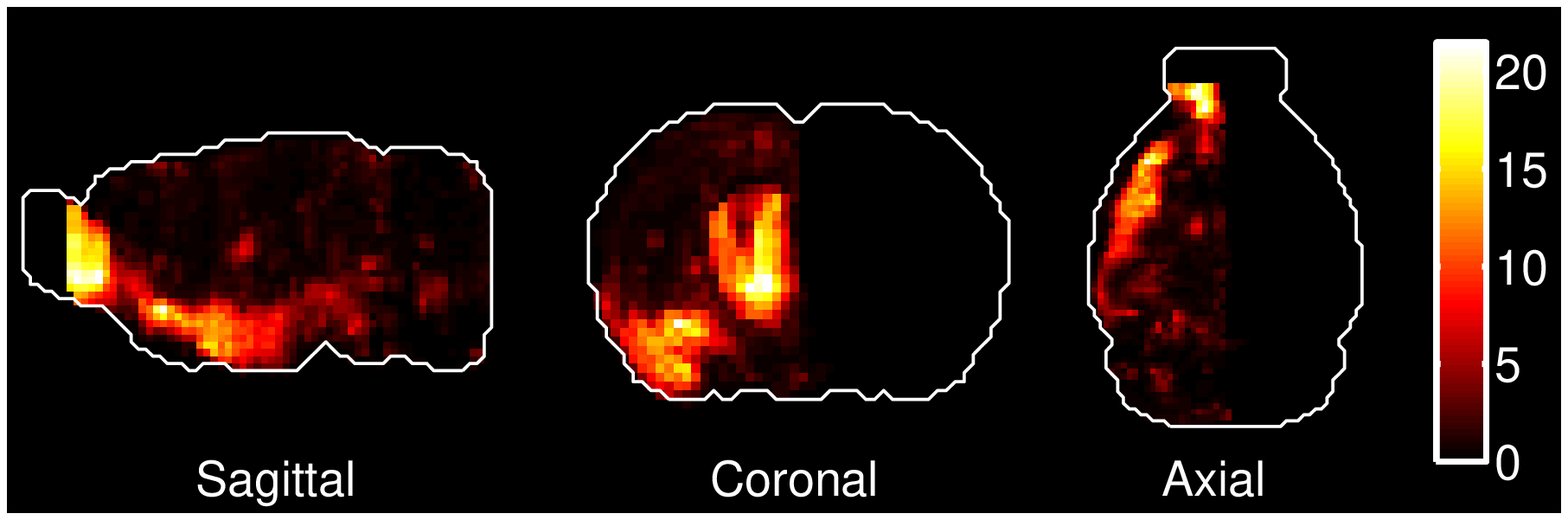}&\includegraphics[width=2in,keepaspectratio]{regionProfile3.eps}\\\hline
Hippocampal region&TC1412430&\includegraphics[width=2in,keepaspectratio]{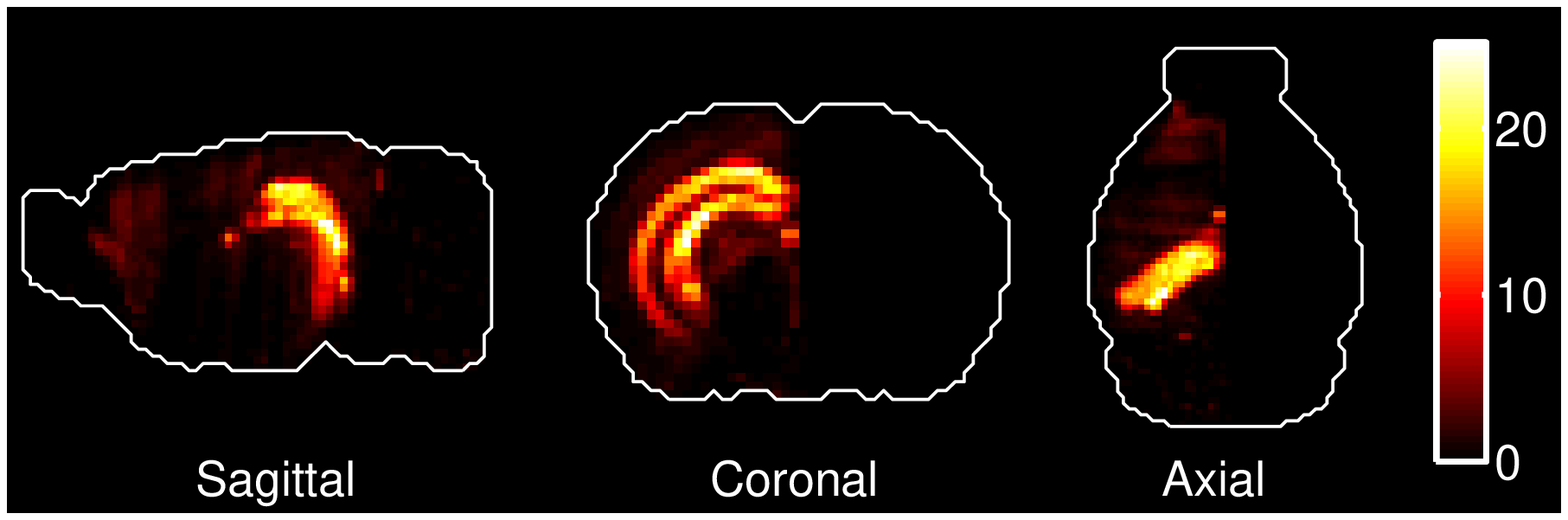}&\includegraphics[width=2in,keepaspectratio]{regionProfile4.eps}\\\hline
Retrohippocampal region&Rspo2&\includegraphics[width=2in,keepaspectratio]{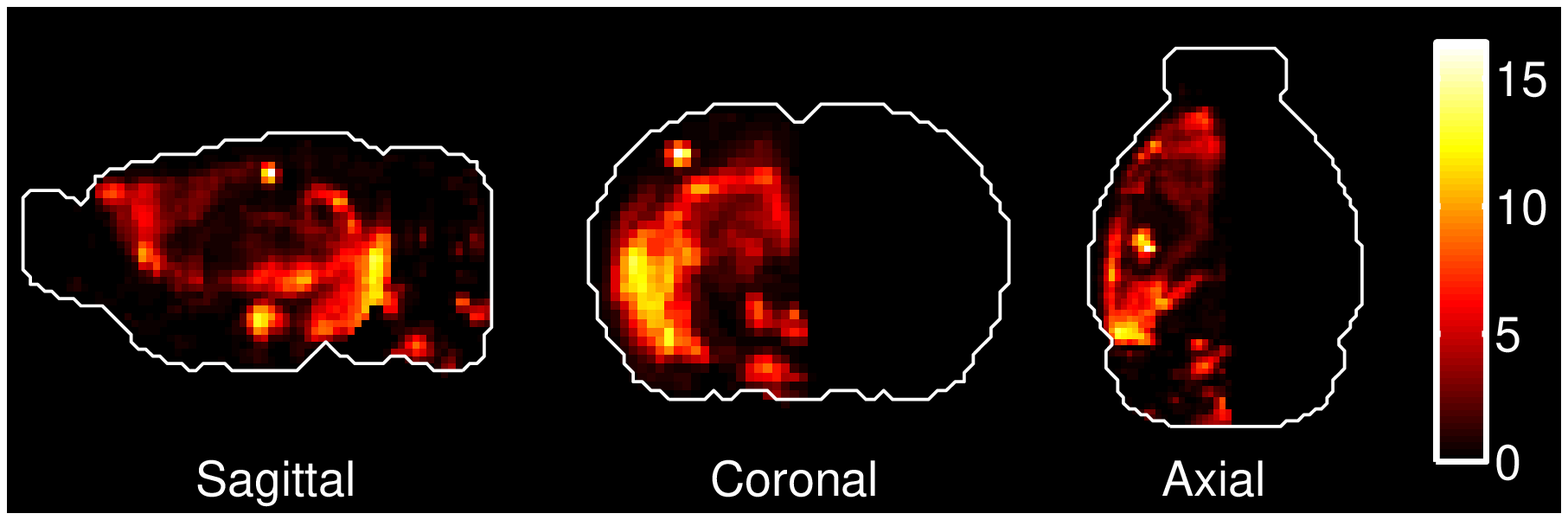}&\includegraphics[width=2in,keepaspectratio]{regionProfile5.eps}\\\hline
Striatum&Rgs9&\includegraphics[width=2in,keepaspectratio]{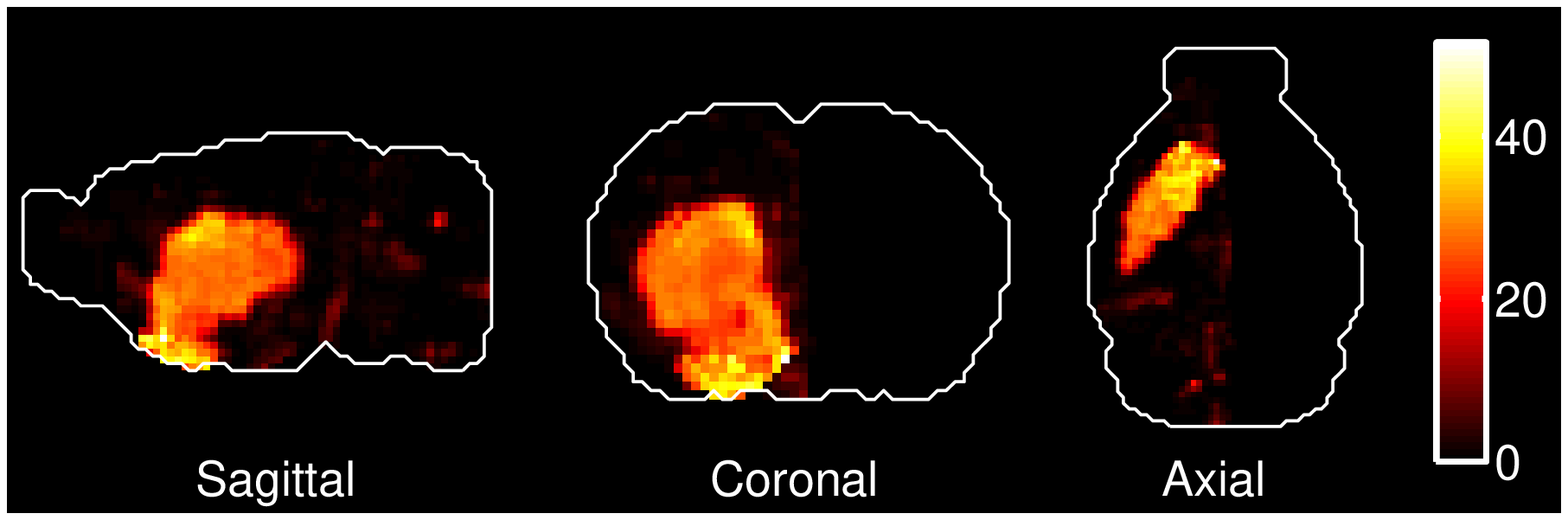}&\includegraphics[width=2in,keepaspectratio]{regionProfile6.eps}\\\hline
Pallidum&Ebf4&\includegraphics[width=2in,keepaspectratio]{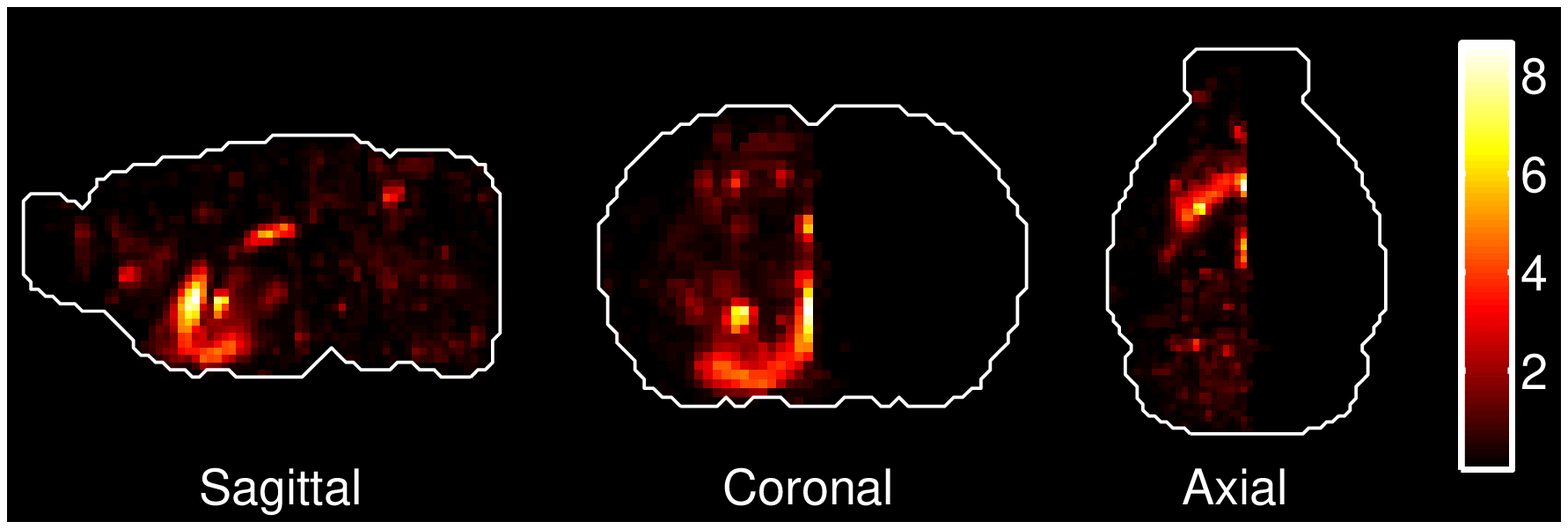}&\includegraphics[width=2in,keepaspectratio]{regionProfile7.eps}\\\hline
Thalamus&Lef1&\includegraphics[width=2in,keepaspectratio]{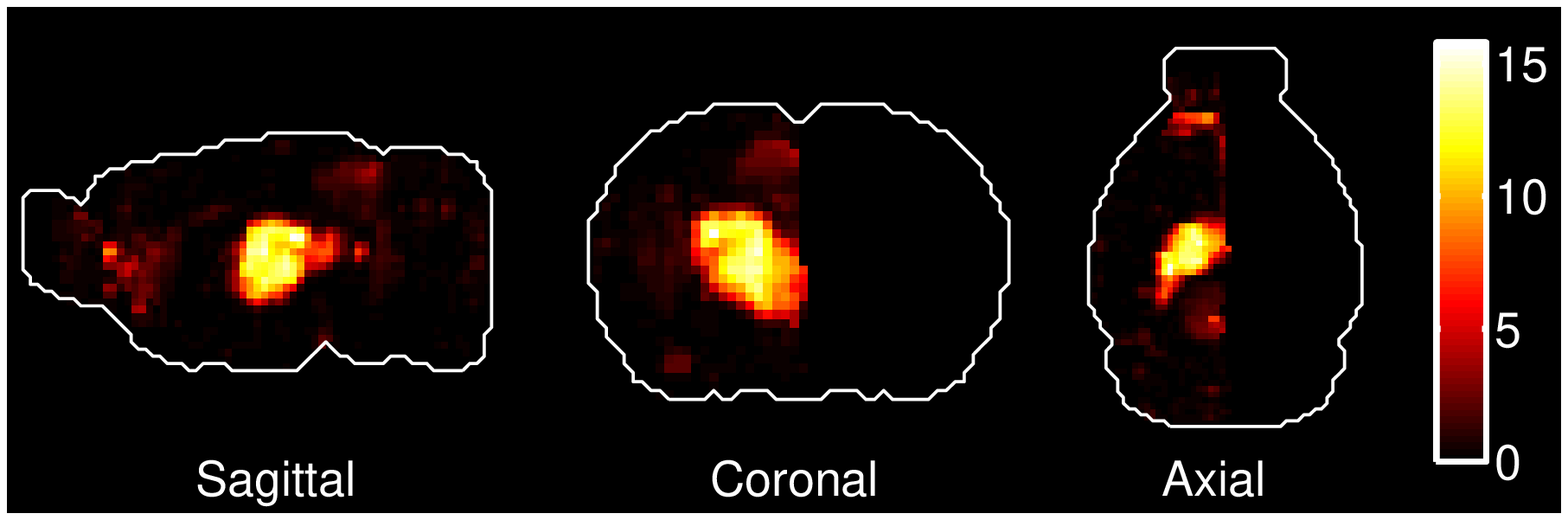}&\includegraphics[width=2in,keepaspectratio]{regionProfile8.eps}\\\hline
Hypothalamus&Gpr165&\includegraphics[width=2in,keepaspectratio]{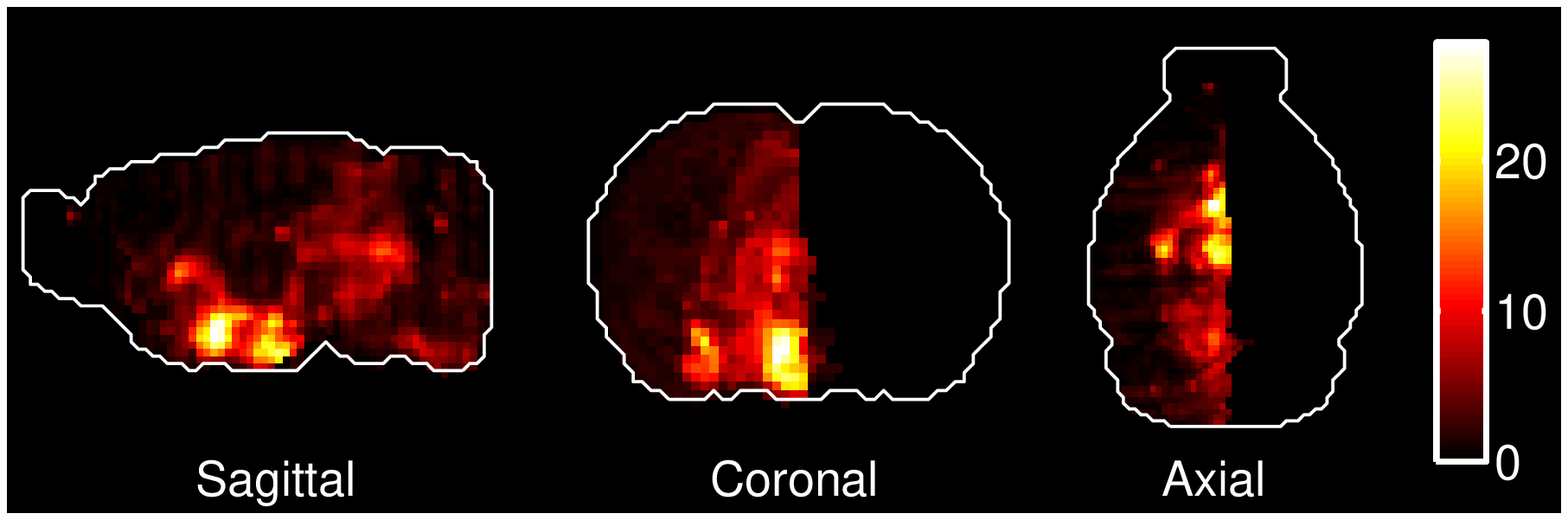}&\includegraphics[width=2in,keepaspectratio]{regionProfile9.eps}\\\hline
Midbrain&Slc17a6&\includegraphics[width=2in,keepaspectratio]{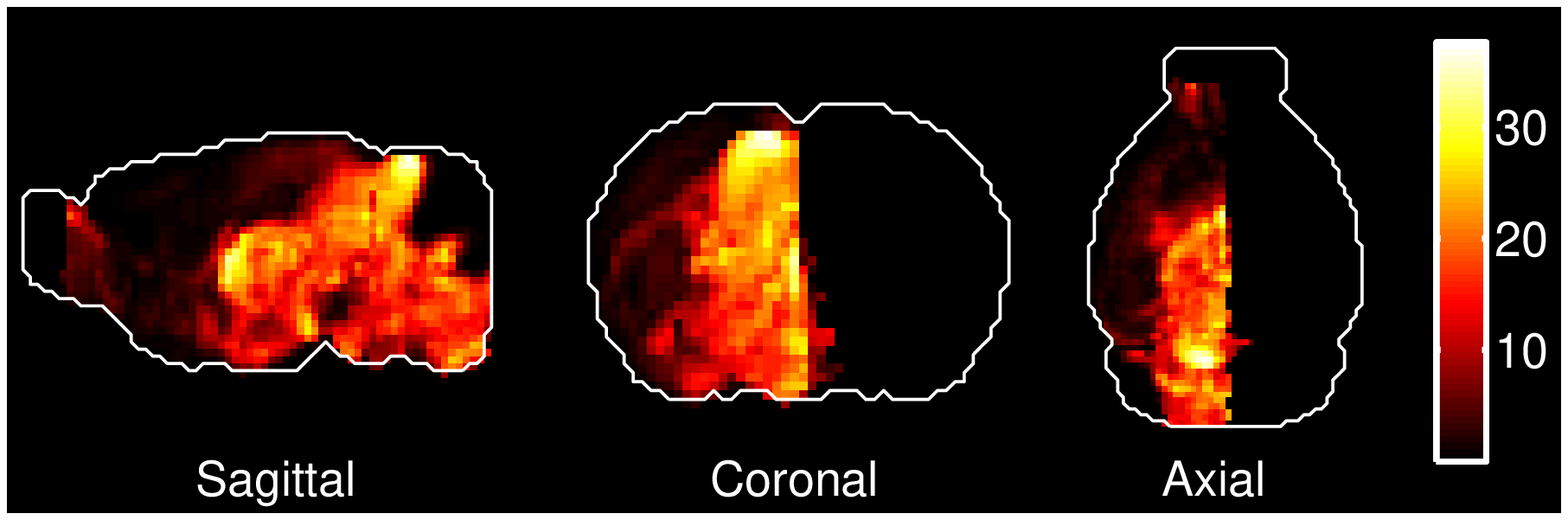}&\includegraphics[width=2in,keepaspectratio]{regionProfile10.eps}\\\hline
Pons&Klk6&\includegraphics[width=2in,keepaspectratio]{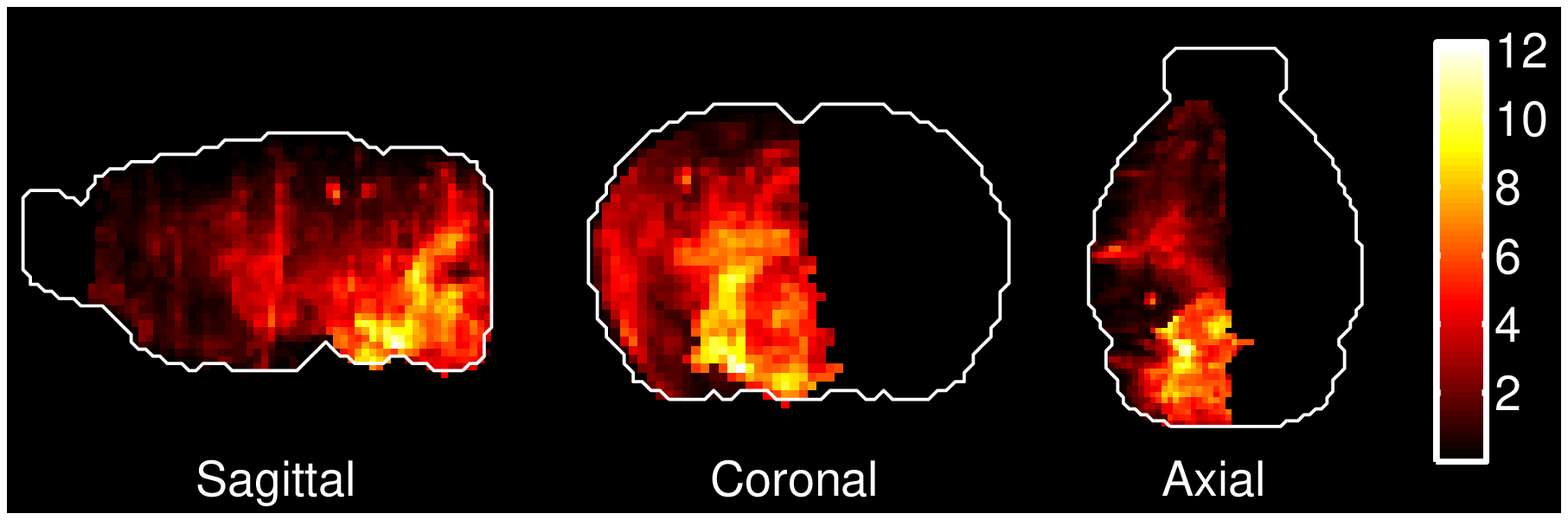}&\includegraphics[width=2in,keepaspectratio]{regionProfile11.eps}\\\hline
Medulla&Glra1&\includegraphics[width=2in,keepaspectratio]{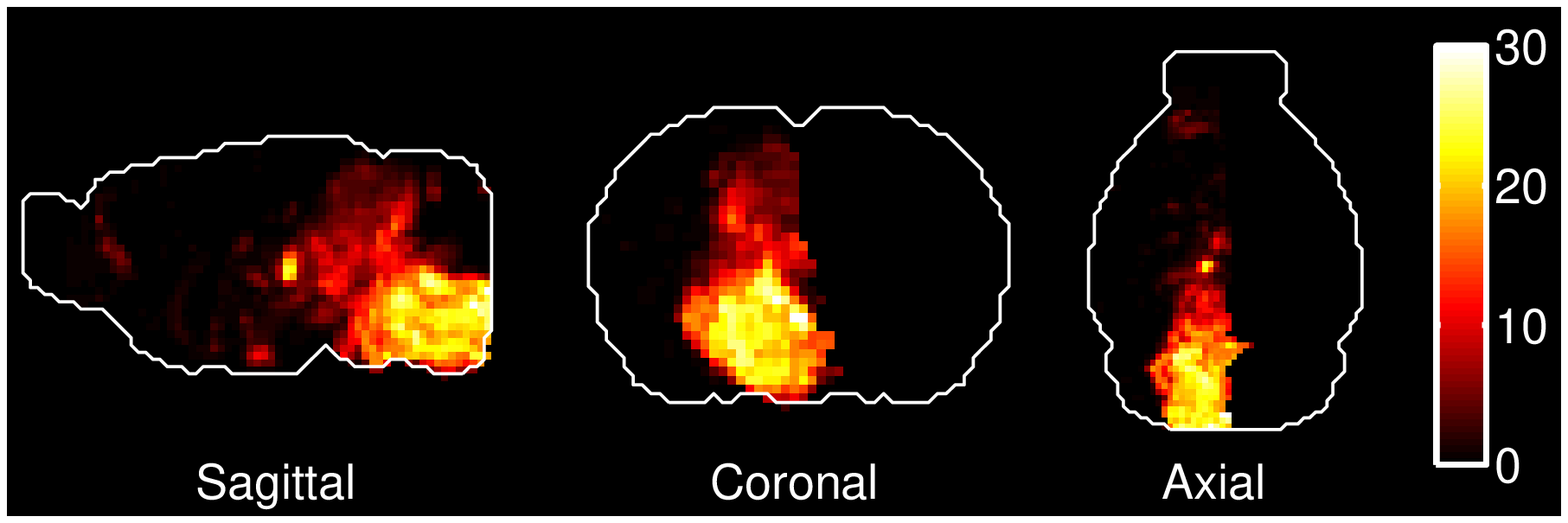}&\includegraphics[width=2in,keepaspectratio]{regionProfile12.eps}\\\hline
Cerebellum&3110001A13Rik&\includegraphics[width=2in,keepaspectratio]{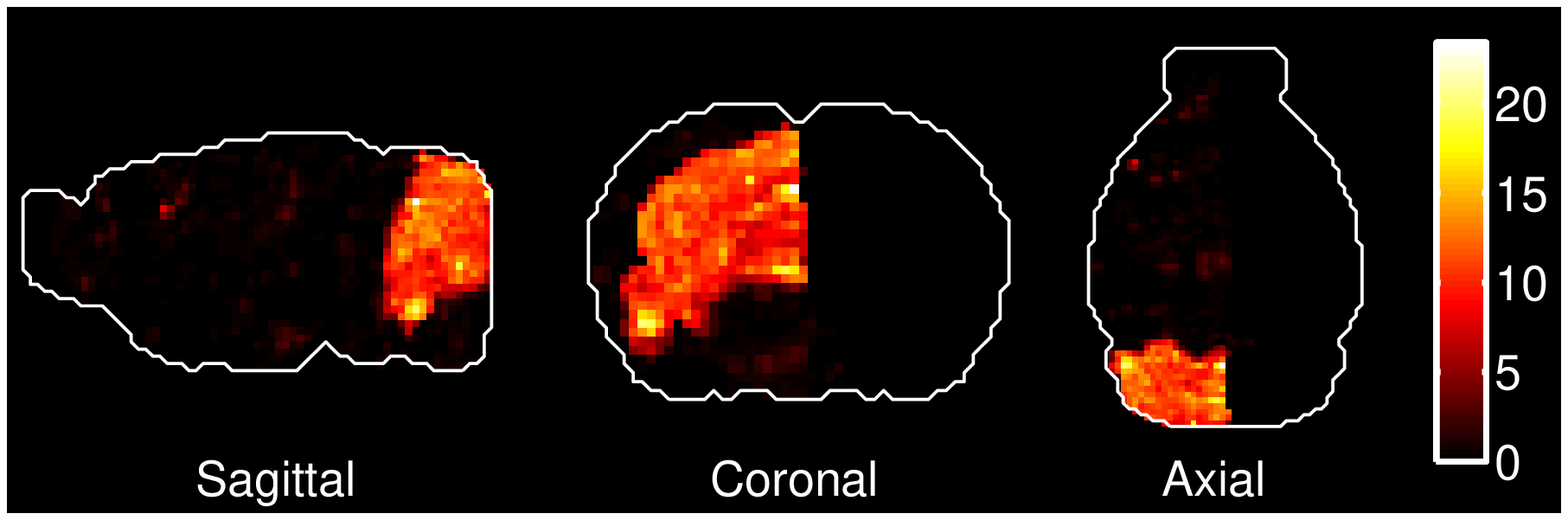}&\includegraphics[width=2in,keepaspectratio]{regionProfile13.eps}\\\hline
\end{tabular}

\caption{The best few genes by fitting score
for each of the 12 main regions of the left hemisphere.}
\label{fig:generalizedTableBig12}
\end{figure}

\newpage
\begin{figure}
\section{Appendix: Best sets of genes for localization in the twelve main regions of the left hemisphere} 
\begin{tabular}{|l|l|l|}
\hline
\textbf{Brain region}&\textbf{Heat map of best generalized marker}&\textbf{Region profile}\\\hline
Cerebral cortex&\includegraphics[width=2in,keepaspectratio]{genProfile2.eps}&\includegraphics[width=2in,keepaspectratio]{regionProfile2.eps}\\\hline
Olfactory areas&\includegraphics[width=2in,keepaspectratio]{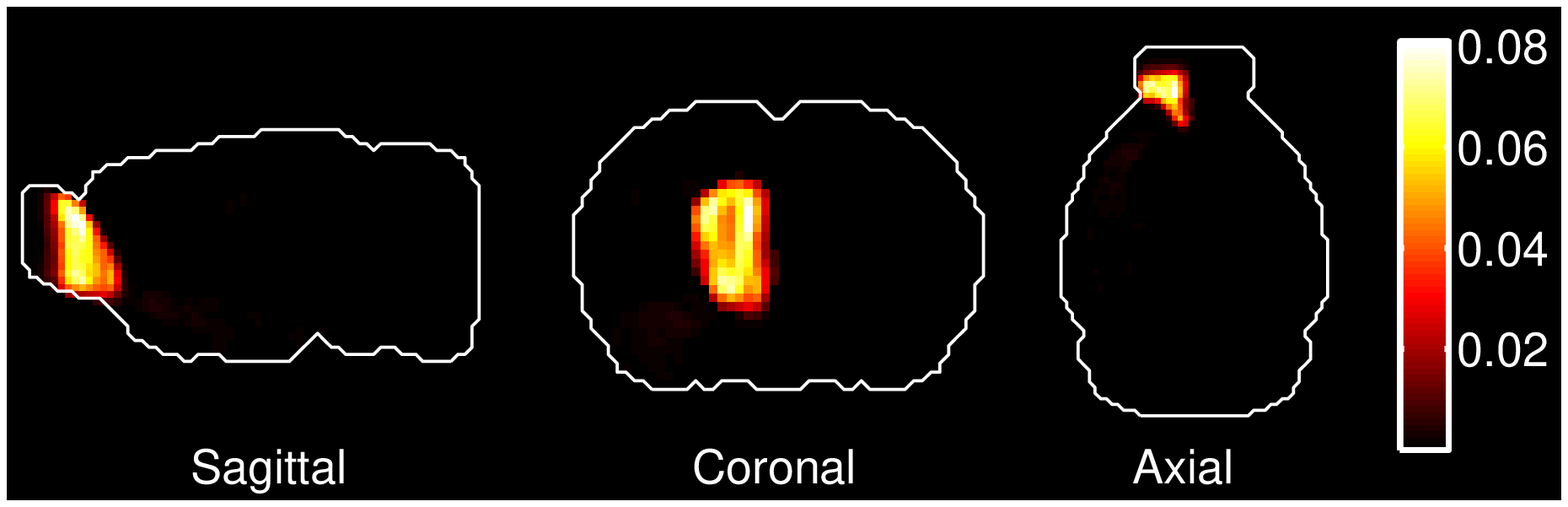}&\includegraphics[width=2in,keepaspectratio]{regionProfile3.eps}\\\hline
Hippocampal region&\includegraphics[width=2in,keepaspectratio]{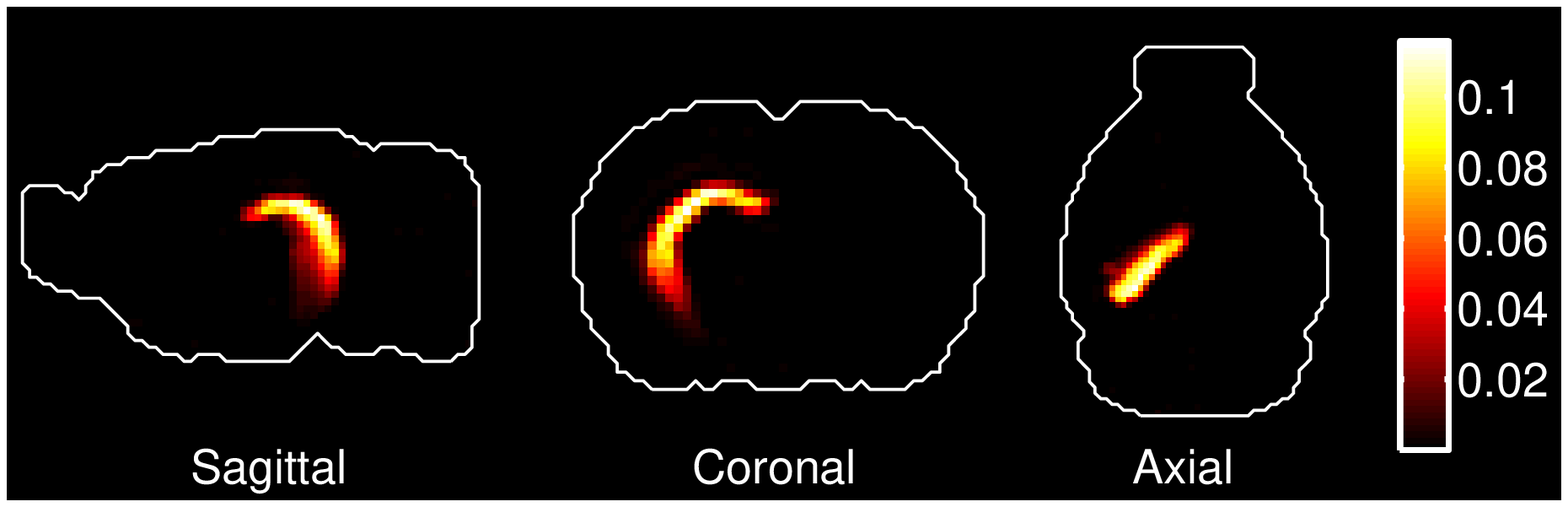}&\includegraphics[width=2in,keepaspectratio]{regionProfile4.eps}\\\hline
Retrohippocampal region&\includegraphics[width=2in,keepaspectratio]{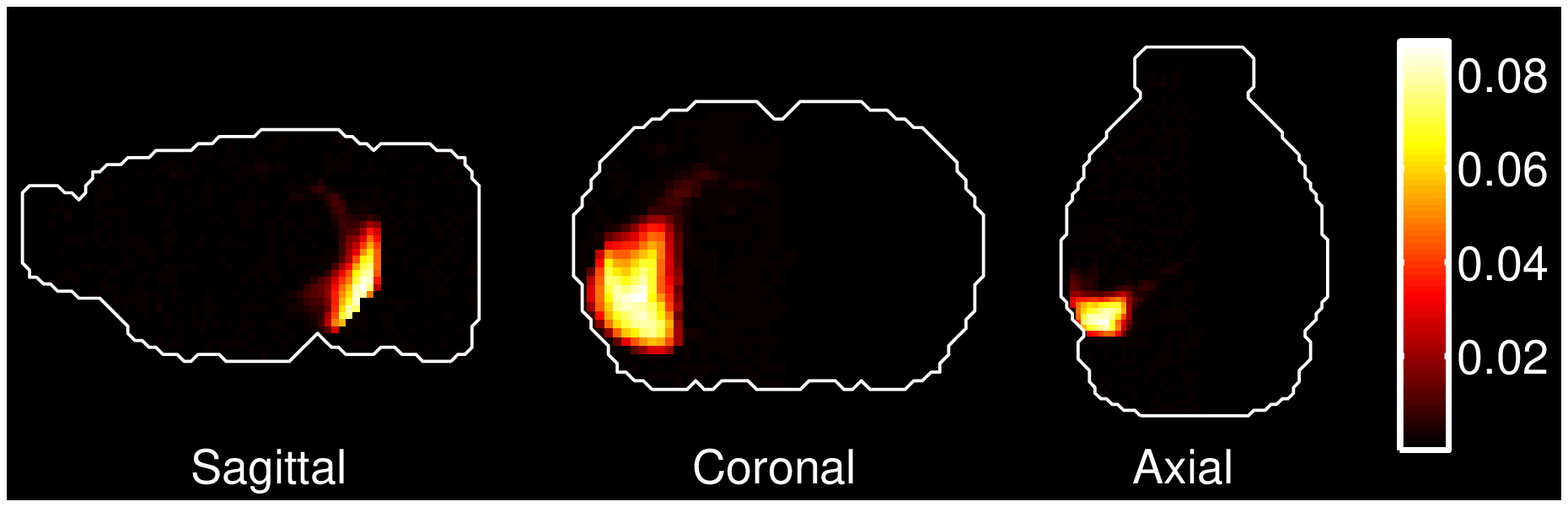}&\includegraphics[width=2in,keepaspectratio]{regionProfile5.eps}\\\hline
Striatum&\includegraphics[width=2in,keepaspectratio]{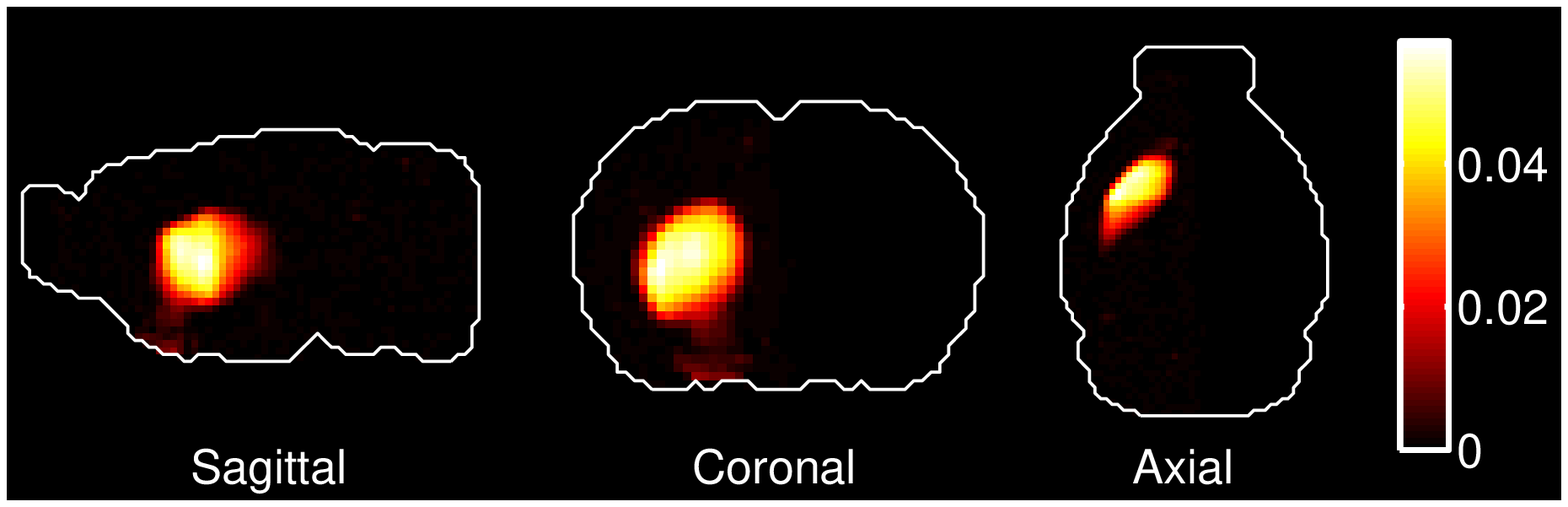}&\includegraphics[width=2in,keepaspectratio]{regionProfile6.eps}\\\hline
Pallidum&\includegraphics[width=2in,keepaspectratio]{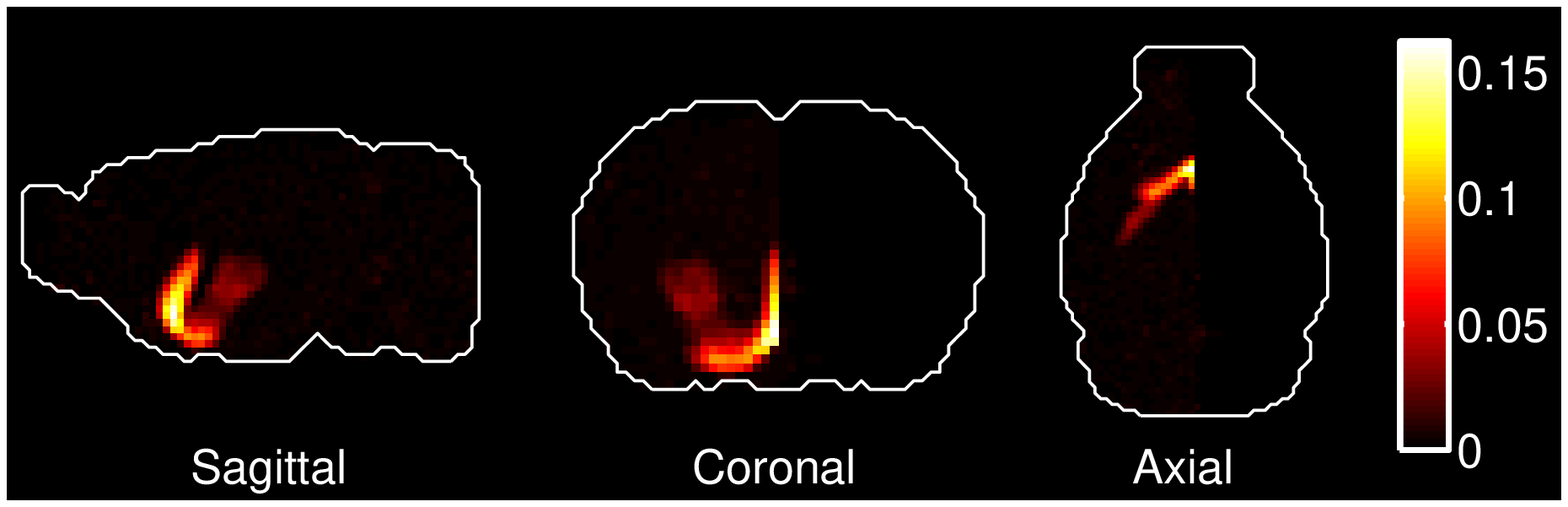}&\includegraphics[width=2in,keepaspectratio]{regionProfile7.eps}\\\hline
Thalamus&\includegraphics[width=2in,keepaspectratio]{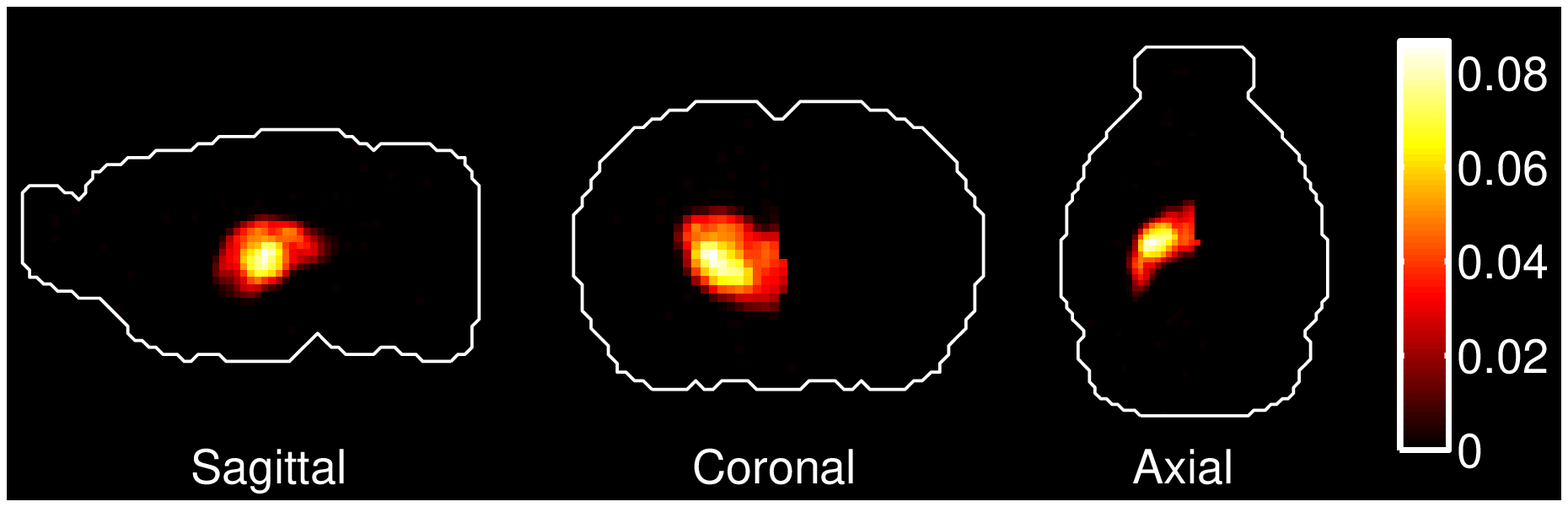}&\includegraphics[width=2in,keepaspectratio]{regionProfile8.eps}\\\hline
Hypothalamus&\includegraphics[width=2in,keepaspectratio]{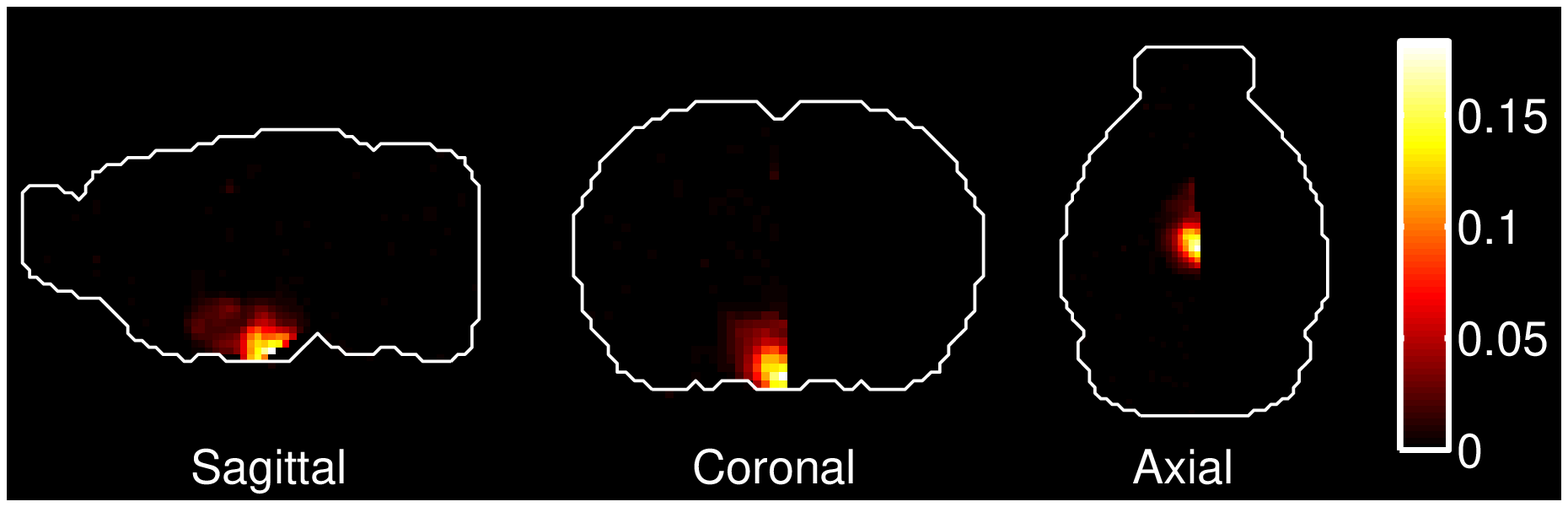}&\includegraphics[width=2in,keepaspectratio]{regionProfile9.eps}\\\hline
Midbrain&\includegraphics[width=2in,keepaspectratio]{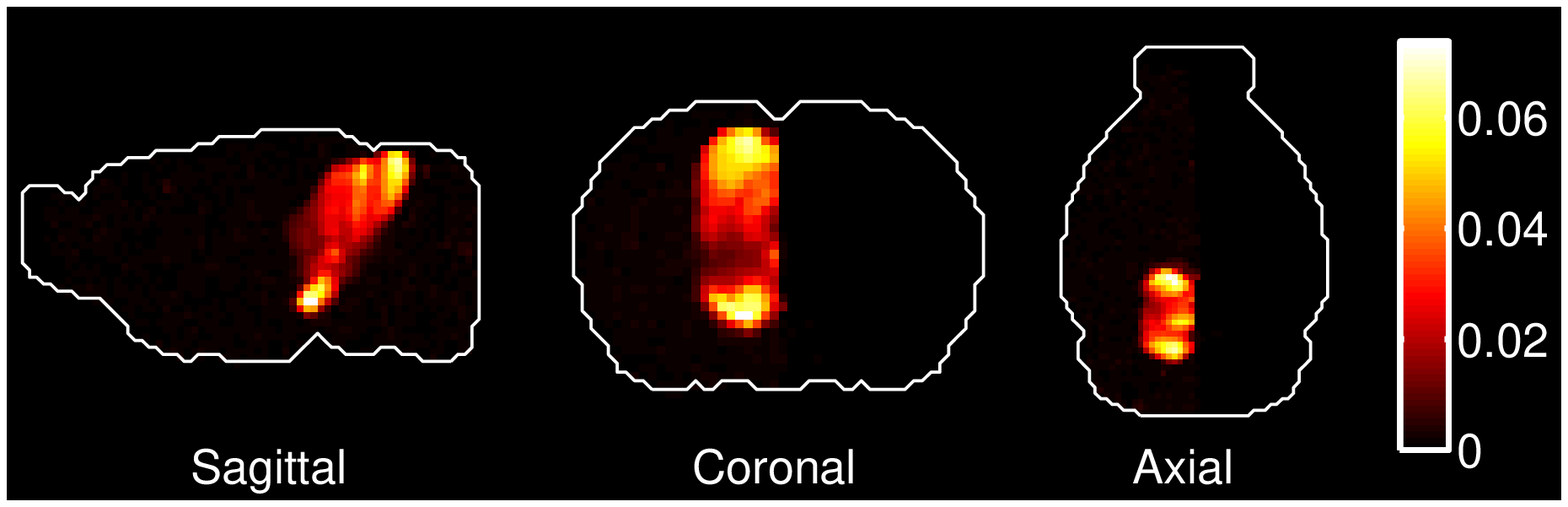}&\includegraphics[width=2in,keepaspectratio]{regionProfile10.eps}\\\hline
Pons&\includegraphics[width=2in,keepaspectratio]{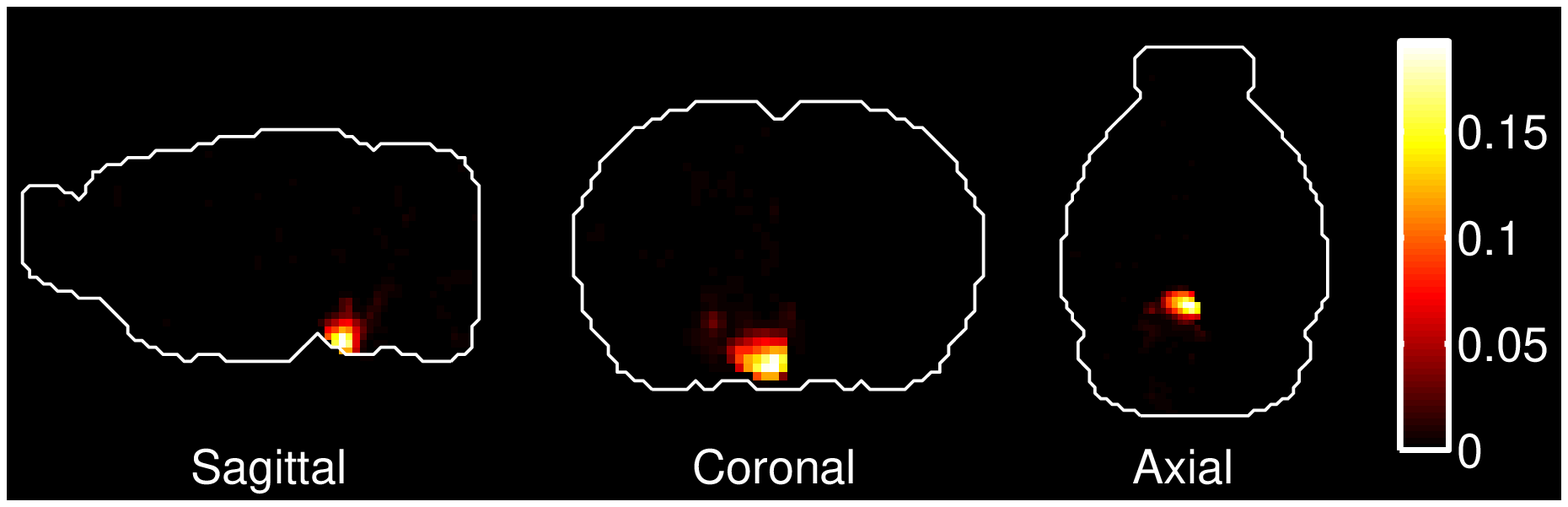}&\includegraphics[width=2in,keepaspectratio]{regionProfile11.eps}\\\hline
Medulla&\includegraphics[width=2in,keepaspectratio]{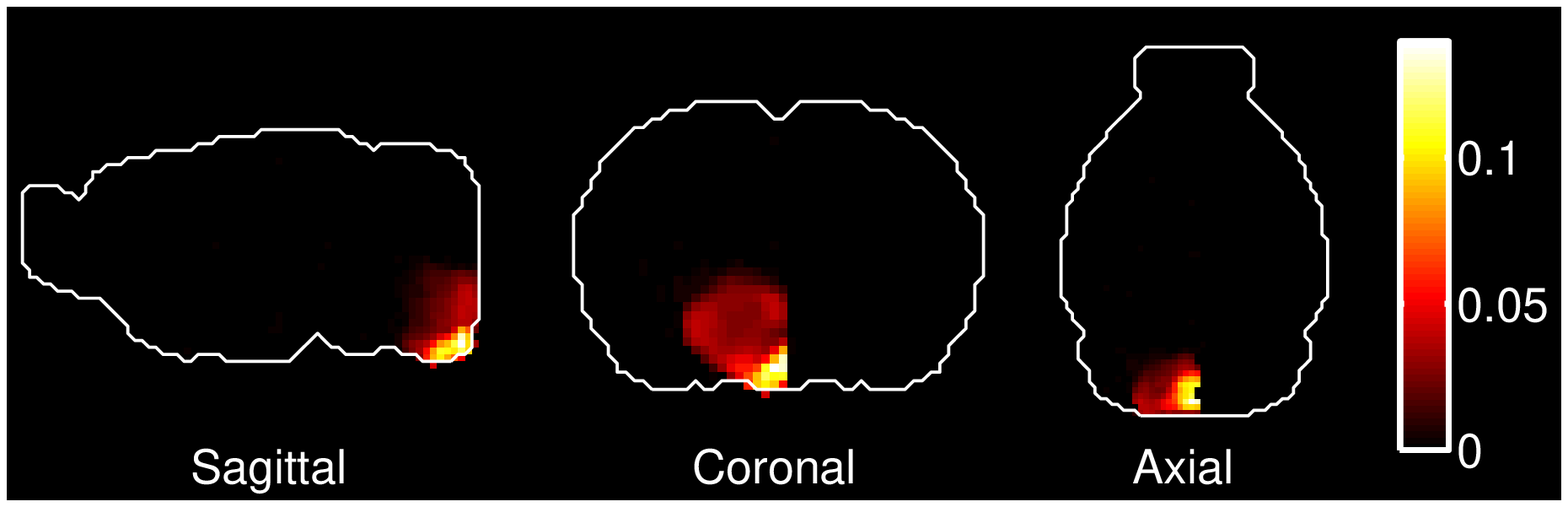}&\includegraphics[width=2in,keepaspectratio]{regionProfile12.eps}\\\hline
Cerebellum&\includegraphics[width=2in,keepaspectratio]{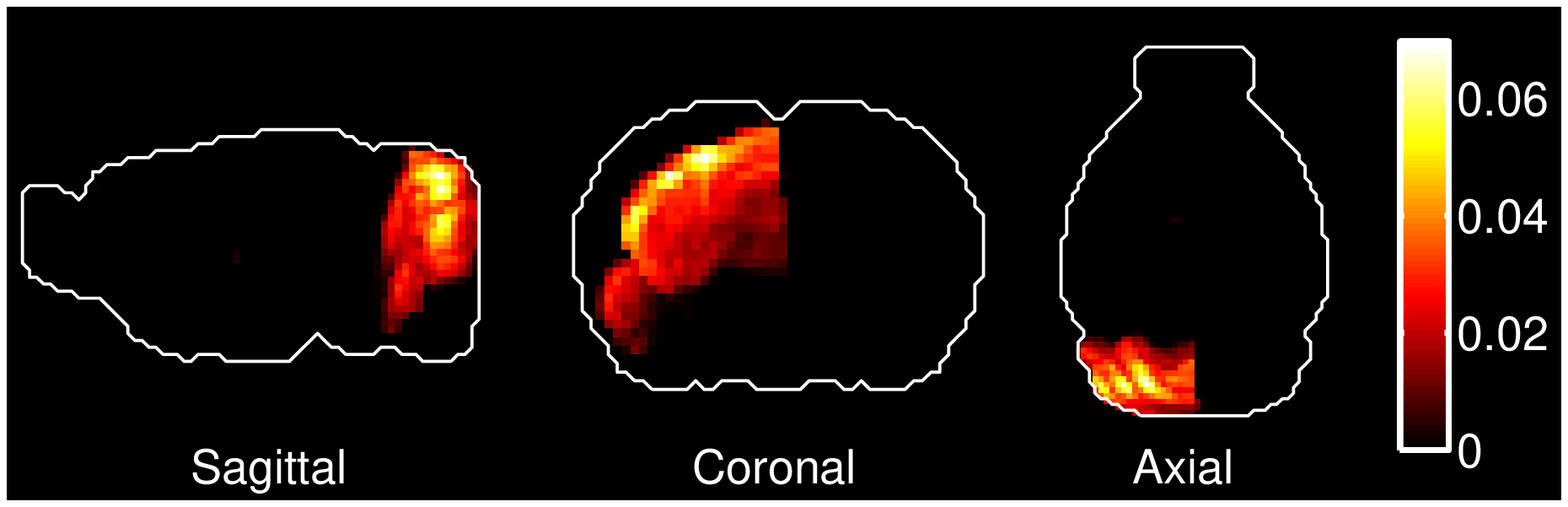}&\includegraphics[width=2in,keepaspectratio]{regionProfile13.eps}\\\hline
\end{tabular}

\caption{Expression profiles of the sets of genes that maximize the localization score
for each of the 12 main regions of the left hemisphere.}
\label{fig:generalizedTableBig12}
\end{figure}
\newpage

\newpage
\begin{figure}
\section{Appendix: Best sets of genes for fitting in the twelve main regions of the left hemisphere} 
\begin{tabular}{|l|l|l|l|}
\hline
\textbf{Brain region}&\textbf{Heat map of best marker}&\textbf{Region profile}\\\hline
Cerebral cortex&\includegraphics[width=2in,keepaspectratio]{genFitProfilebig122.eps}&\includegraphics[width=2in,keepaspectratio]{regionProfile2.eps}\\\hline
Olfactory areas&\includegraphics[width=2in,keepaspectratio]{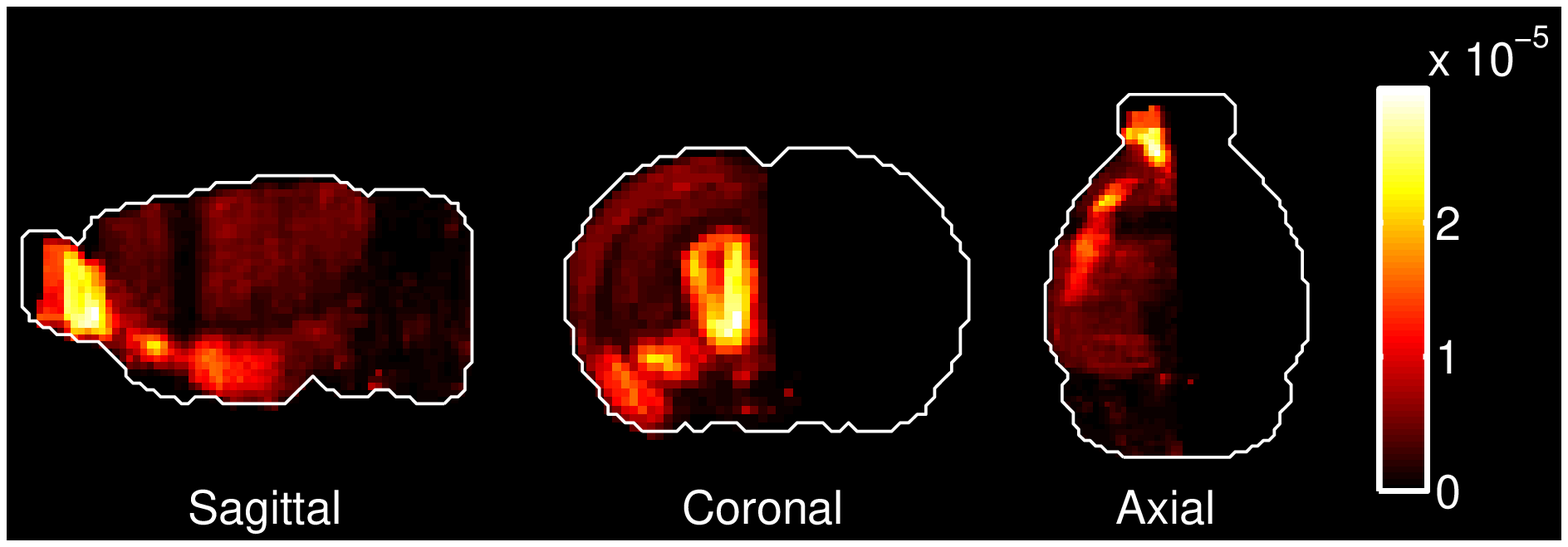}&\includegraphics[width=2in,keepaspectratio]{regionProfile3.eps}\\\hline
Hippocampal region&\includegraphics[width=2in,keepaspectratio]{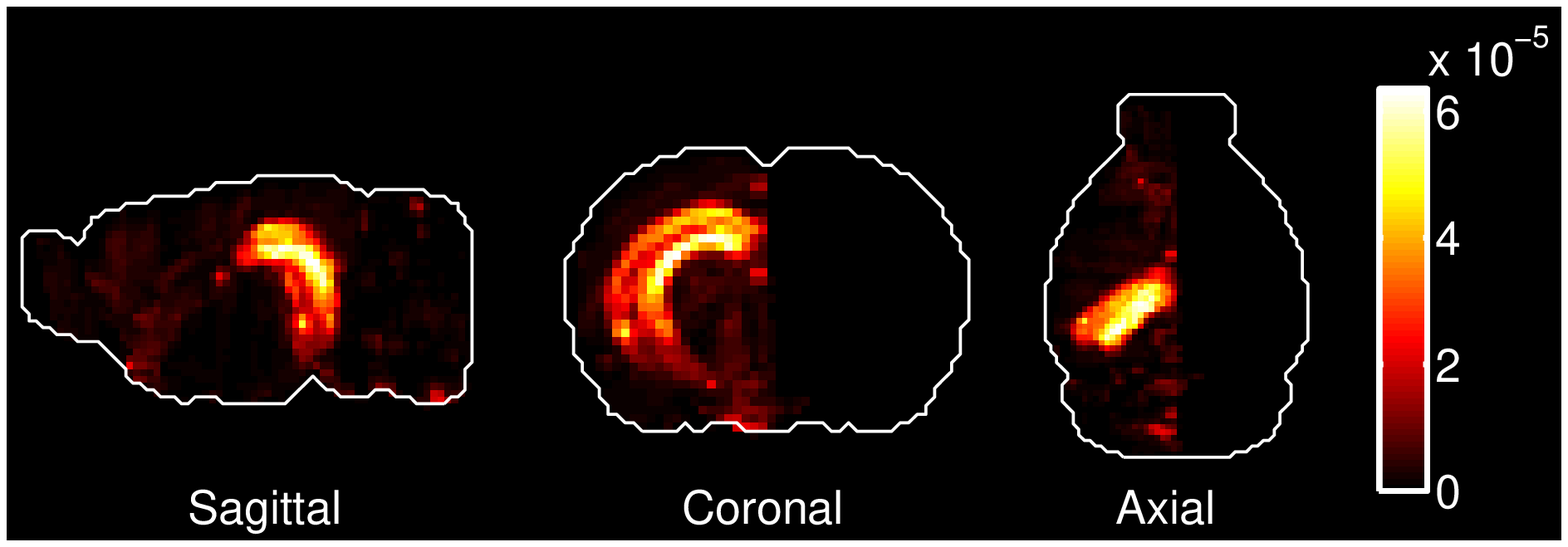}&\includegraphics[width=2in,keepaspectratio]{regionProfile4.eps}\\\hline
Retrohippocampal region&\includegraphics[width=2in,keepaspectratio]{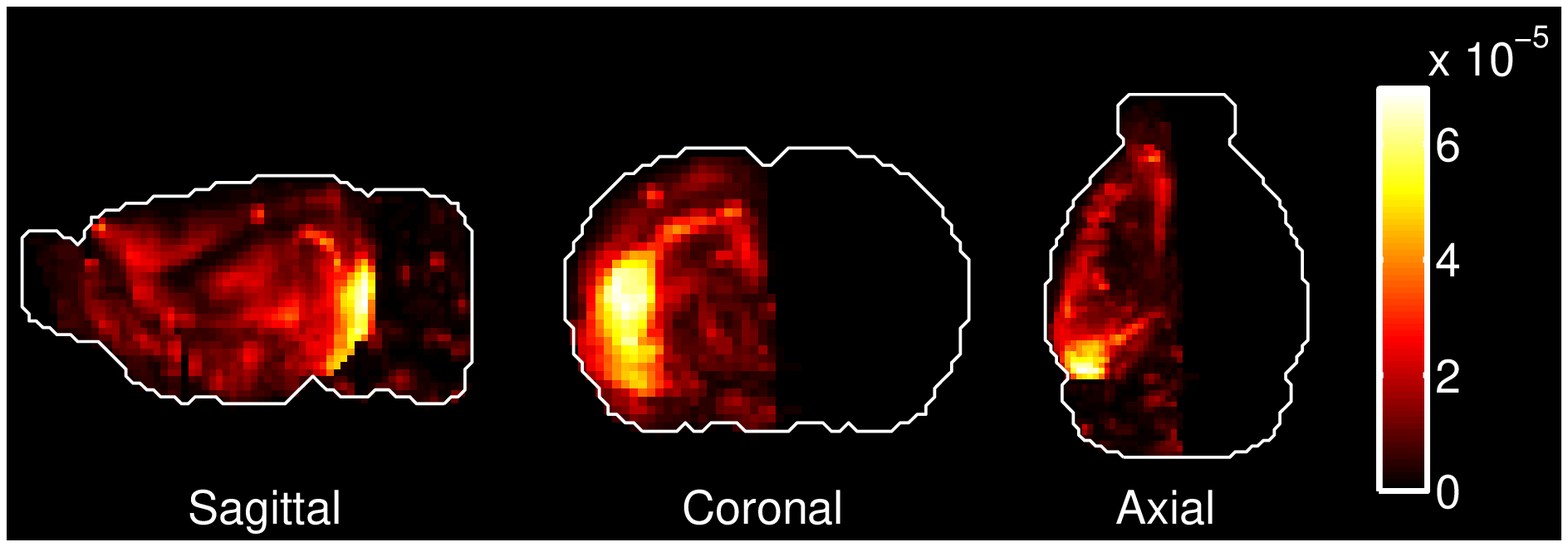}&\includegraphics[width=2in,keepaspectratio]{regionProfile5.eps}\\\hline
Striatum&\includegraphics[width=2in,keepaspectratio]{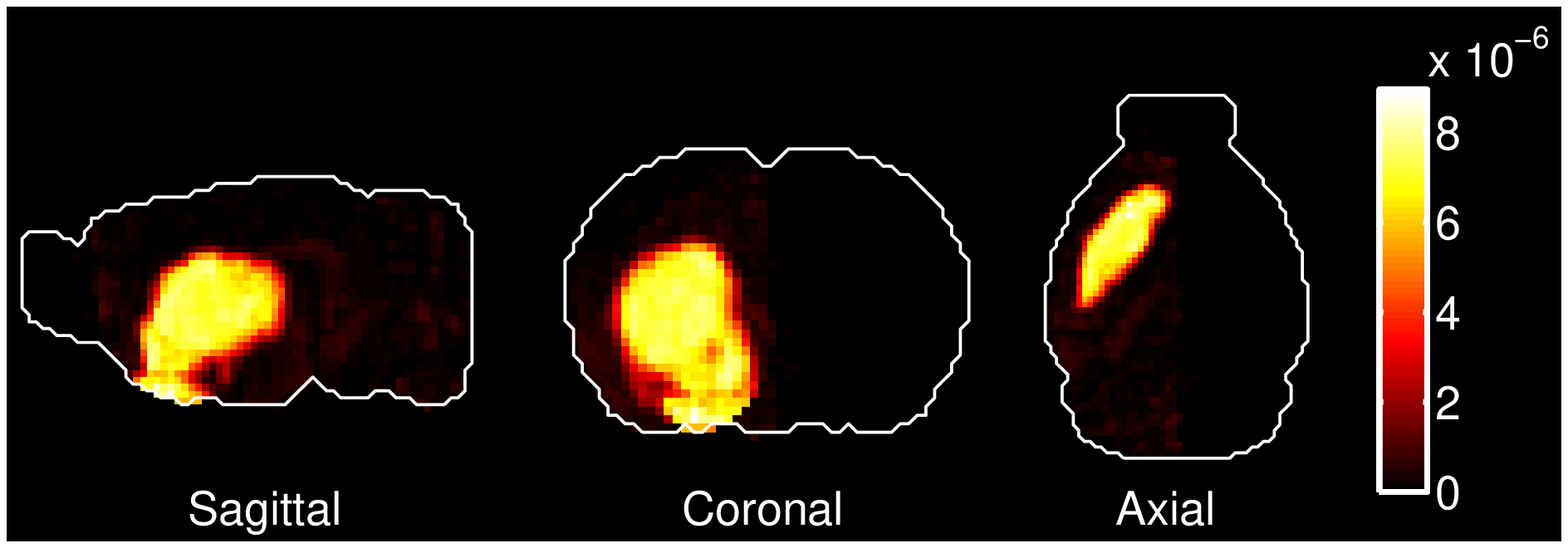}&\includegraphics[width=2in,keepaspectratio]{regionProfile6.eps}\\\hline
Pallidum&\includegraphics[width=2in,keepaspectratio]{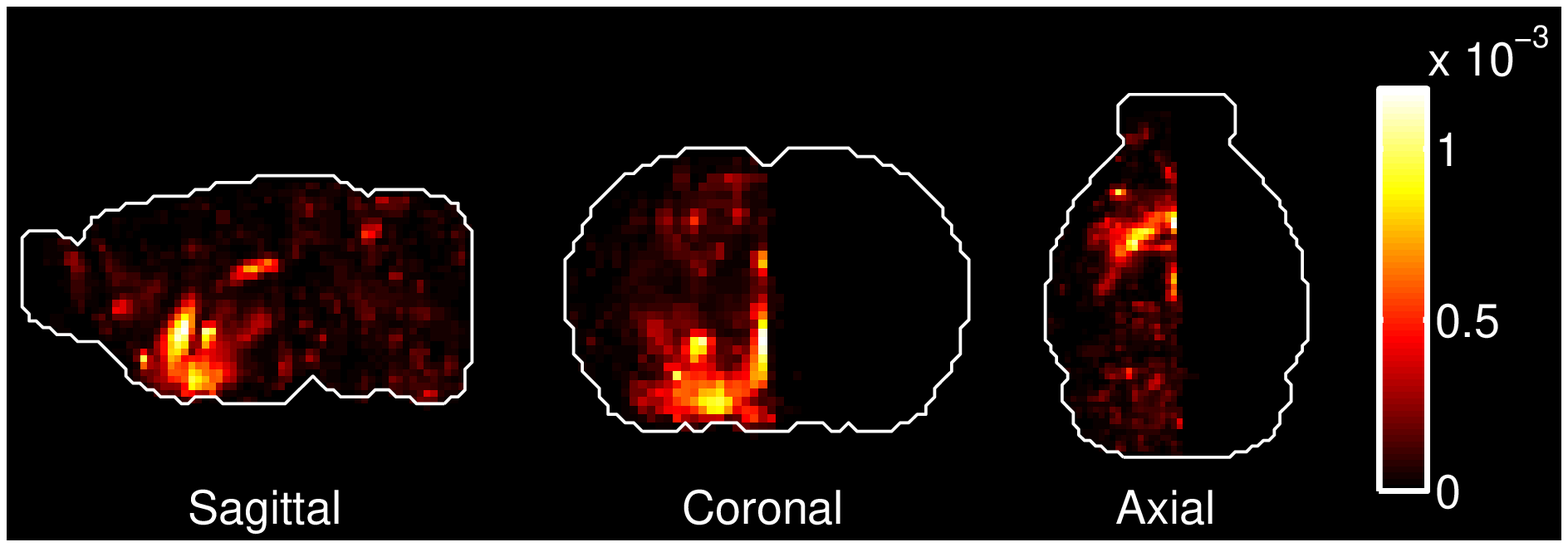}&\includegraphics[width=2in,keepaspectratio]{regionProfile7.eps}\\\hline
Thalamus&\includegraphics[width=2in,keepaspectratio]{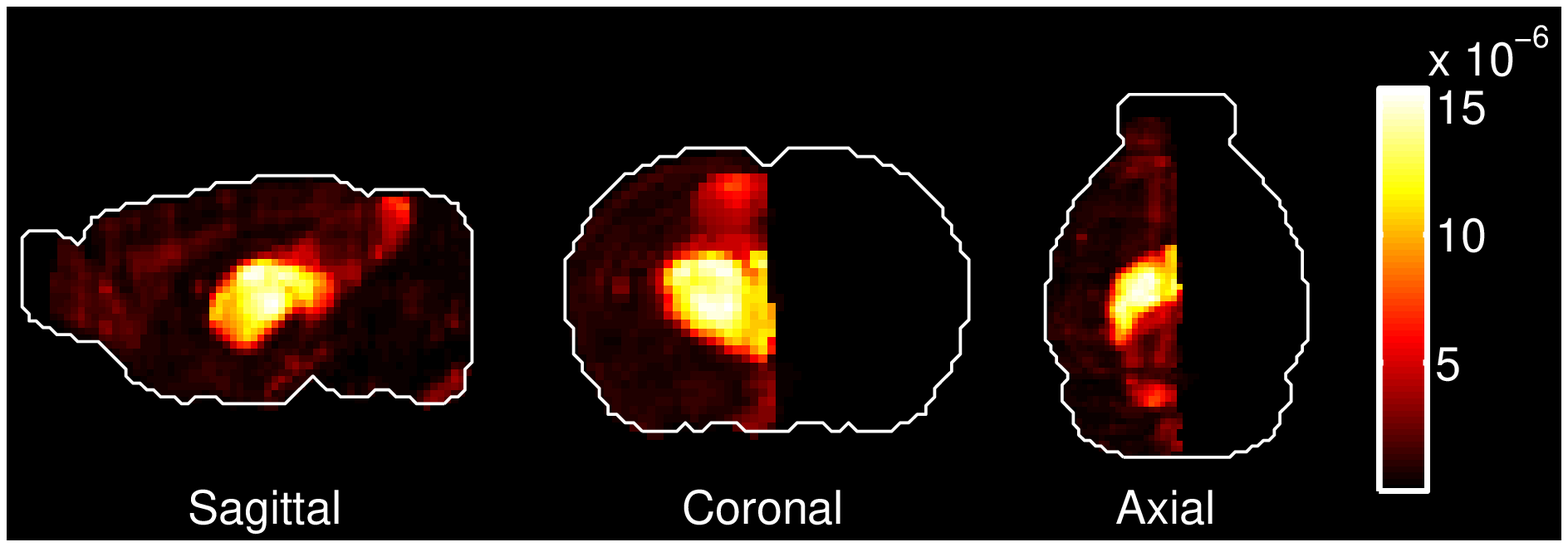}&\includegraphics[width=2in,keepaspectratio]{regionProfile8.eps}\\\hline
Hypothalamus&\includegraphics[width=2in,keepaspectratio]{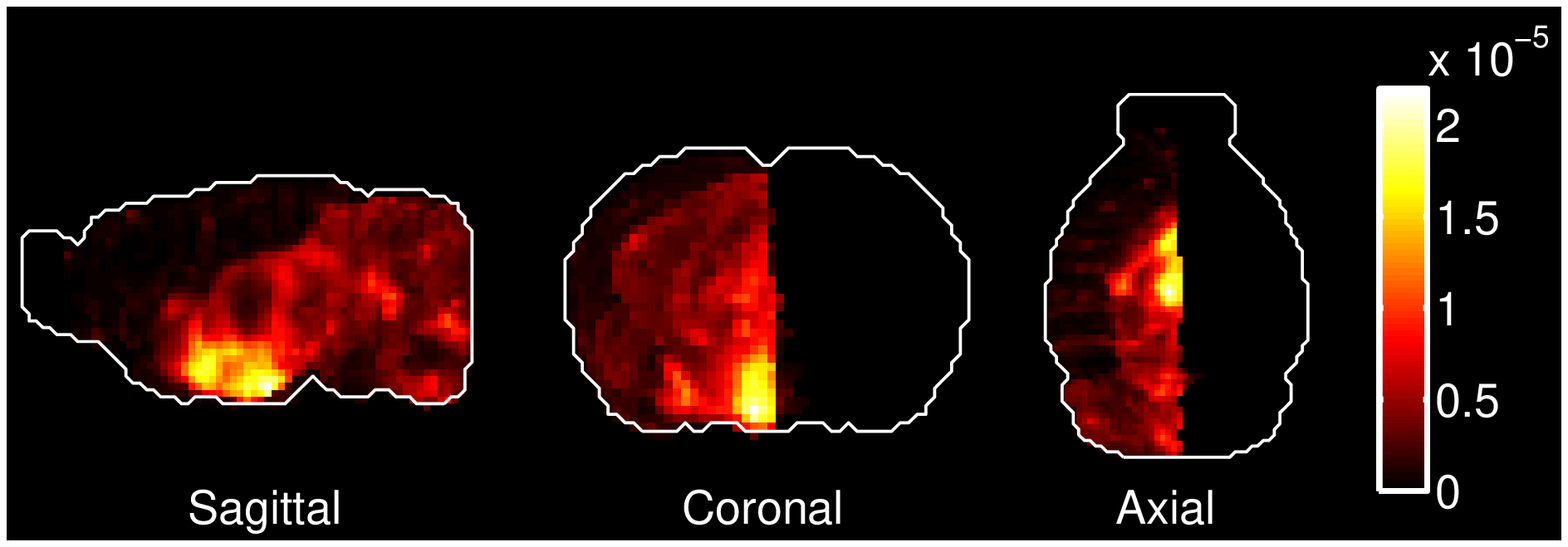}&\includegraphics[width=2in,keepaspectratio]{regionProfile9.eps}\\\hline
Midbrain&\includegraphics[width=2in,keepaspectratio]{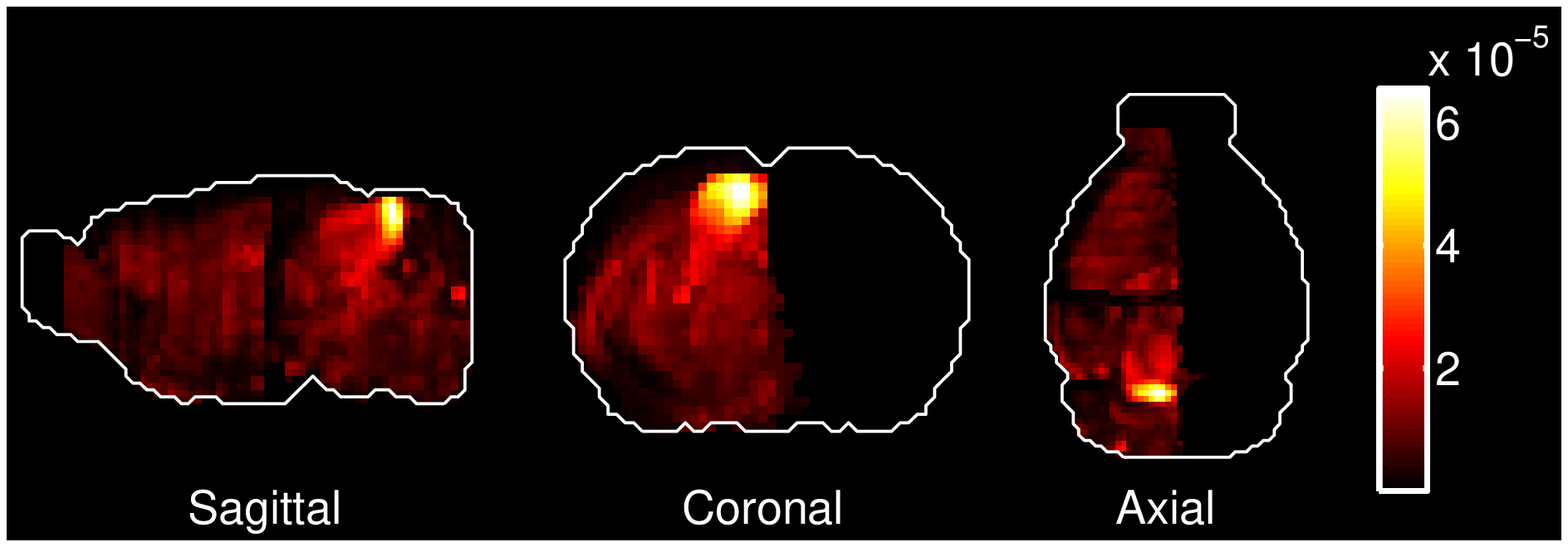}&\includegraphics[width=2in,keepaspectratio]{regionProfile10.eps}\\\hline
Pons&\includegraphics[width=2in,keepaspectratio]{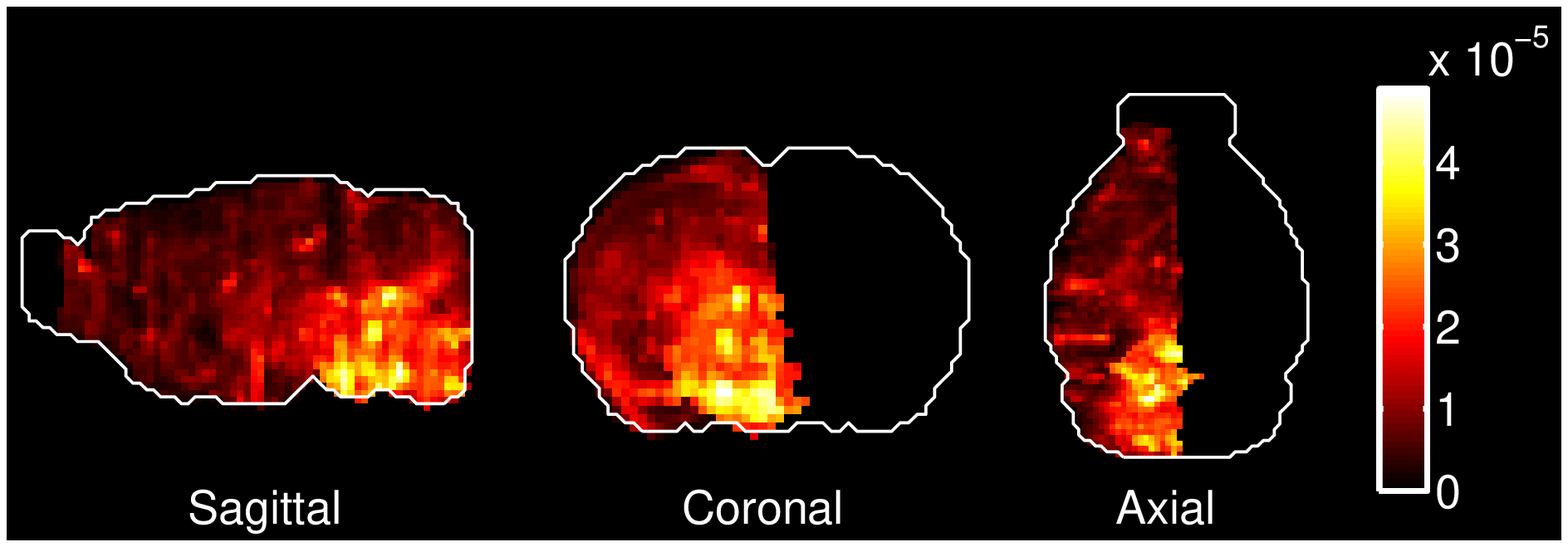}&\includegraphics[width=2in,keepaspectratio]{regionProfile11.eps}\\\hline
Medulla&\includegraphics[width=2in,keepaspectratio]{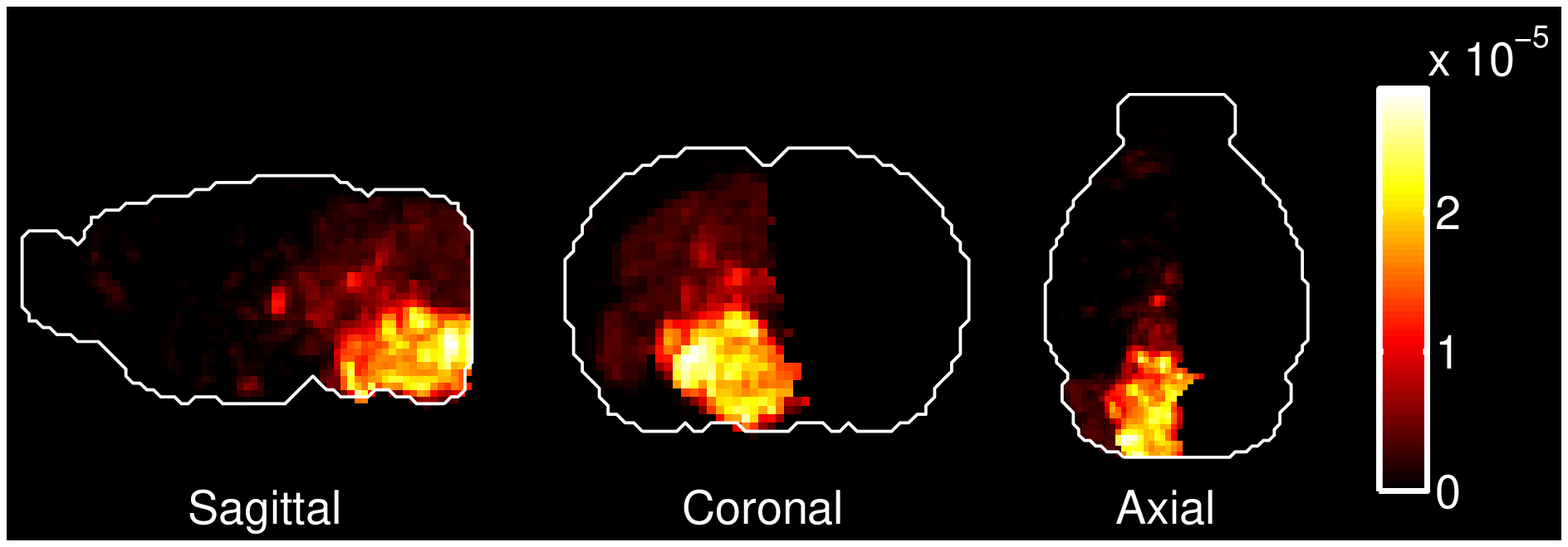}&\includegraphics[width=2in,keepaspectratio]{regionProfile12.eps}\\\hline
Cerebellum&\includegraphics[width=2in,keepaspectratio]{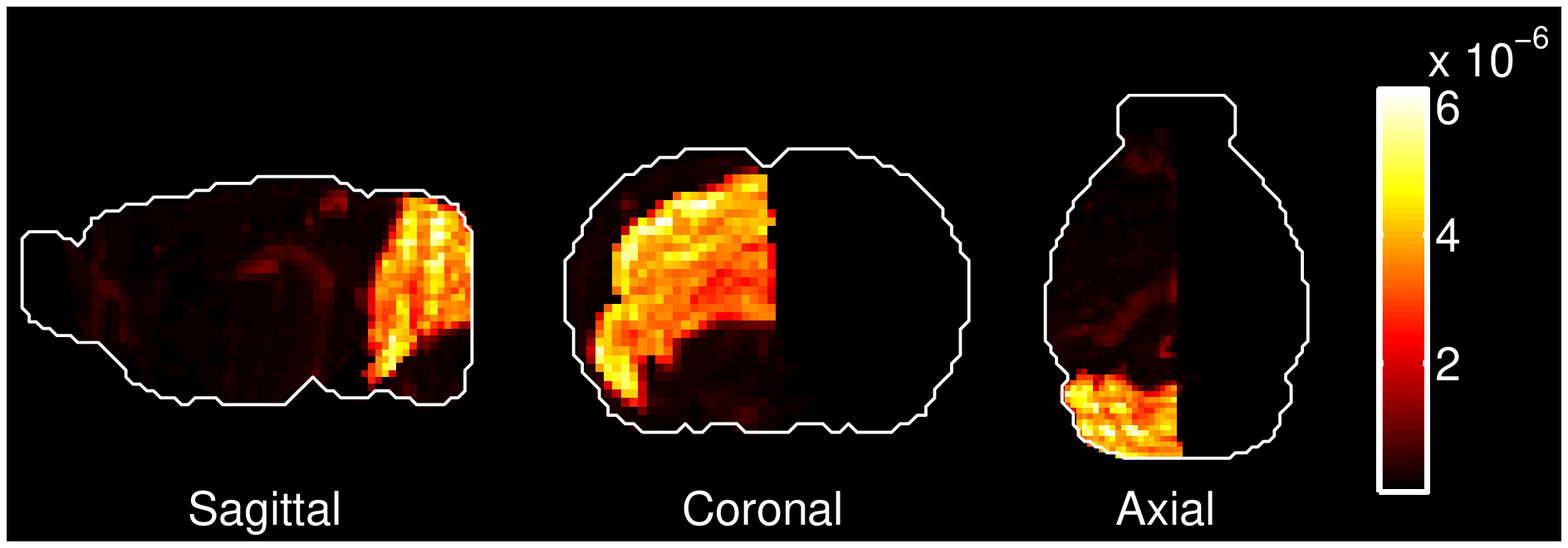}&\includegraphics[width=2in,keepaspectratio]{regionProfile13.eps}\\\hline
\end{tabular}

\caption{Expression profiles of the sums of genes that maximize the fitting score
for each of the 12 main regions of the left hemisphere.}
\label{fig:generalizedTableBig12}
\end{figure}
\newpage

\newpage

\end{document}